\documentclass[11pt,a4paper]{article}\pdfoutput=1

\usepackage{jheppub}
\usepackage{hyperref}
\usepackage{booktabs}
\usepackage{xspace}
\usepackage{xcolor}
\usepackage[toc,page]{appendix}
\usepackage{mathtools}
\usepackage[utf8]{inputenc}
\usepackage{multirow}

\usepackage{slashed}
\usepackage{cancel}

\makeatletter
\g@addto@macro\bfseries{\boldmath}
\makeatother

\newcommand{\noun}[1]{{\sc #1}}
\newcommand{\sss}{\rm}
\newcommand{\MADGRAPH}{\noun{MadGraph v4}}
\newcommand{\GOSAM}{\noun{GoSam 2.0}}
\newcommand{\POWHEG}{\noun{POWHEG}}
\newcommand{\POWHEGBOX}{\noun{POWHEG BOX}}

\newcommand{\MINLO}{\noun{MiNLO}}

\newcommand{\WWJMINLO}{\noun{WWJ-MiNLO}}

\newcommand{\OpenLoops}{{\sc OpenLoops}}

\newcommand{\Matrix}{\scalebox{1.0}{\textsc{Matrix}}}
\newcommand{\Munich}{\scalebox{1.0}{\textsc{Munich}}}
\newcommand{\Collier}{\scalebox{1.0}{\textsc{Collier}}}
\newcommand{\CutTools}{\scalebox{1.0}{\textsc{CutTools}}}
\newcommand{\OneLOop}{\scalebox{1.0}{\textsc{OneLOop}}}

\newcommand{\NNLOPS}{\noun{NNLOPS}}

\newcommand{\PYTHIA}[1]{\noun{Pythia{#1}}}
\newcommand{\MCatNLO}{\noun{MC@NLO}}

\newcommand{\FASTJET}{\noun{FastJet}}
\newcommand{\as}{\alpha_{\rm s}}

\newcommand{\mur}{\mu_{\mathrm{R}}}
\newcommand{\muf}{\mu_{\mathrm{F}}}

\newcommand{\ptlep}{\ensuremath{p_{T,{\ell}}}}
\newcommand{\ptlepone}{\ensuremath{p_{T,{\ell_1}}}}
\newcommand{\etalep}{\ensuremath{|\eta_{\ell}|}}

\newcommand{\mll}{\ensuremath{m_{\ell^-\ell^+}}}
\newcommand{\ptmiss}{\ensuremath{p_{T}^{\text{miss}}}}
\newcommand{\ptmissrel}{\ensuremath{p_{T}^{\text{miss,rel}}}}

\newcommand{\mww}{\ensuremath{m_{WW}}}

\newcommand{\mtww}{\ensuremath{m_{T,WW}}}

\newcommand{\yww}{\ensuremath{y_{WW}}}
\newcommand{\dywpwm}{\ensuremath{\Delta y_{W^+W^-}}}
\newcommand{\ptww}{\ensuremath{p_{T,WW}}}
\newcommand{\ptwm}{\ensuremath{p_{T,W^-}}}
\newcommand{\ptwp}{\ensuremath{p_{T,W^+}}}

\newcommand{\ywp}{\ensuremath{y_{W^+}}}
\newcommand{\mwp}{\ensuremath{m_{W^+}}}
\newcommand{\mwm}{\ensuremath{m_{W^-}}}
\newcommand{\thetap}{\ensuremath{\theta^{\rm\scalebox{0.6}{CS}}_{W^+}}}
\newcommand{\costhetap}{\ensuremath{{\rm cos}\,\thetap}}
\newcommand{\phip}{\ensuremath{\phi^{\rm\scalebox{0.6}{CS}}_{W^+}}}
\newcommand{\thetam}{\ensuremath{\theta^{\rm\scalebox{0.6}{CS}}_{W^-}}}
\newcommand{\costhetam}{\ensuremath{{\rm cos}\,\thetam}}
\newcommand{\phim}{\ensuremath{\phi^{\rm\scalebox{0.6}{CS}}_{W^-}}}

\newcommand{\dyww}{\ensuremath{\dd\yww}}
\newcommand{\ddywpwm}{\ensuremath{\dd\dywpwm}}

\newcommand{\dptwm}{\ensuremath{\dd\ptwm}}

\newcommand{\dmwp}{\ensuremath{\dd\mwp}}
\newcommand{\dmwm}{\ensuremath{\dd\mwm}}
\newcommand{\dcosthetap}{\ensuremath{\dd\costhetap}}
\newcommand{\dphip}{\ensuremath{\dd\phip}}
\newcommand{\dcosthetam}{\ensuremath{\dd\costhetam}}
\newcommand{\dphim}{\ensuremath{\dd\phim}}

\newcommand{\Etlone}{\ensuremath{E_{T,\ell_1}}}
\newcommand{\Etltwo}{\ensuremath{E_{T,\ell_2}}}

\newcommand{\ptll}{\ensuremath{p_{T,\ell\ell}}}

\newcommand{\dphillnunu}{\ensuremath{\Delta\phi_{\ell\ell,\nu\nu}}}
\newcommand{\dphill}{\ensuremath{\Delta\phi_{\ell\ell}}}

\newcommand{\etal}{\ensuremath{\eta_{\ell}}}

\newcommand{\ptjet}{\ensuremath{p_{T,j}}}
\newcommand{\ptjetone}{\ensuremath{p_{T,j_1}}}
\newcommand{\ptjetveto}{\ensuremath{p_{T,j_1}^{\rm veto}}}

\newcommand{\elle}{\ensuremath{\ell}}

\newcommand{\pt}{\ensuremath{p_{T}}}
\newcommand{\qt}{\ensuremath{q_{T}}}

\definecolor{darkgreen}{rgb}{0,0.6,0}
\definecolor{darkpurple}{rgb}{0,0.5,0.5}
\definecolor{darkblue}{rgb}{0,0,0.7}
\definecolor{darkred}{rgb}{0.5,0,0.0}
\definecolor{darkorange}{rgb}{0.8,0.4,0.0}
\definecolor{green}{rgb}{0.0,0.8,0.4}

\newcommand{\thetacs}{\theta^*}
\newcommand{\phics}{\phi^*}

\newcommand{\born}{\Phi_{\sss B}}

\newcommand{\dd}{{\rm d}}
\newcommand{\ww}{\ensuremath{W^+W^-}}
\newcommand{\wz}{\ensuremath{W^\pm Z}}
\newcommand{\zz}{\ensuremath{ZZ}}
\newcommand{\muenn}{\ensuremath{e^- \bar \nu_e\;\mu^+\nu_\mu}}
\newcommand{\emunn}{\ensuremath{e^+ \nu_e\;\mu^- \bar\nu_\mu}}

\newcommand{\citere}[1]{Ref.\,\cite{#1}}
\newcommand{\citeres}[1]{Refs.\,\cite{#1}}

\newcommand{\eqn}[1]{Eq.\,(\ref{#1})}

\newcommand{\fig}[1]{Fig.\,\ref{#1}}
\newcommand{\figs}[1]{Figs.\,\ref{#1}}
\newcommand{\tab}[1]{Tab.\,\ref{#1}}
\newcommand{\sct}[1]{Sec.~\ref{#1}}

\usepackage[mathscr]{euscript}

\preprint{CERN-TH/2018-114, LAPTH-018/18}

\title{NNLOPS accurate predictions for \ww{} production}

\author[a,b]{Emanuele Re,}
\author[a]{Marius Wiesemann}
\author[a]{and Giulia Zanderighi\footnote{On
    leave from Rudolf Peierls Centre for Theoretical Physics,
    University of Oxford, 1 Keble Road, UK.}}

\affiliation[a]{Theoretical Physics Department, CERN, Geneva, Switzerland}
\affiliation[b]{LAPTh, CNRS, Université Savoie Mont Blanc, 74940 Annecy, France}

\emailAdd{emanuele.re@lapth.cnrs.fr}
\emailAdd{marius.wiesemann@cern.ch}
\emailAdd{giulia.zanderighi@cern.ch}

\abstract{We present novel predictions for the production of \ww{}
pairs in hadron collisions that are next-to-next-to-leading order
accurate and consistently matched to a parton shower (NNLOPS).  All diagrams
that lead to the process $pp\to\muenn+X$ are taken into
account, thereby including spin correlations and off-shell effects.
For the first time full NNLOPS accuracy is achieved for a 
$2\rightarrow 4$ process.
We find good agreement, at the 1$\sigma$ level, with the \ww{} rates
measured by ATLAS and CMS.
The importance of NNLOPS predictions is evident from differential
distributions sensitive to soft-gluon effects and from the large 
impact 
($10$\% and more) 
of including next-to-next-to-leading order corrections on top of \MINLO{}.
We define a charge asymmetry for the $W$ bosons and the leptons in \ww{}
production at the LHC, which is sensitive to the $W$ polarizations
and hence can be used as a probe of new physics.}


\begin{document}
\maketitle \flushbottom

\section{Introduction}
\label{sec:intro}

In the rich physics programme of LHC Run II major attention is given
to measurements of Higgs-boson properties, the direct and indirect
search for signals of new-physics phenomena, precision measurements
and the extraction of Standard Model (SM) parameters.  The production of \ww{} pairs is
among the most important LHC processes to study the gauge symmetry
structure of electroweak (EW) interactions and of the mechanism of EW
symmetry breaking in the SM.  In particular, with the
lack of clear signs of new physics, precision measurements have become
of foremost importance to search for small deviations of SM
predictions. They translate into indirect bounds on high-scale beyond-SM (BSM)
models, whose effects manifest themselves in small deformations of SM
predictions at lower energies.  Most important in that respect are
constraints on the allowed size of anomalous trilinear gauge couplings
(aTGCs), which appear already in the leading perturbative
contributions to \ww{} production.  In addition, \ww{} final states
are irreducible background to Higgs-boson measurements and to direct
searches for BSM particles decaying into leptons, missing
energy, and/or jets.

The \ww{} cross section has been measured at both the
Tevatron~\cite{Aaltonen:2009aa,Abazov:2011cb} and the LHC (at
7\,TeV~\cite{ATLAS:2012mec,Chatrchyan:2013yaa},
8\,TeV~\cite{ATLAS-CONF-2014-033,Aad:2016wpd,Chatrchyan:2013oev,Khachatryan:2015sga}
and 13\,TeV~\cite{Aaboud:2017qkn,CMS:2016vww}).  \ww{} measurements,
in particular with new data becoming continuously available in Run II
and beyond, play a major role as SM precision tests and in
constraining BSM physics, as any small deviation from the SM
predictions for the production rate or the shape of distributions
could be a signal of new physics.  The high sensitivity to aTGCs of
the \ww{} process renders \ww{} measurements a powerful tool for
indirect BSM
searches~\cite{ATLAS:2012mec,Chatrchyan:2013yaa,Wang:2014uea,Khachatryan:2015sga,Aad:2016wpd}.\footnote{See
  also
  \citeres{Frye:2015rba,Butter:2016cvz,Zhang:2016zsp,Green:2016trm,Baglio:2017bfe,Falkowski:2016cxu,Panico:2017frx,Franceschini:2017xkh,Liu:2018pkg}
  as examples of theory ideas to exploit precision in diboson processes to
  constrain BSM physics.}
In the context of Higgs-boson measurements in the $H\to \ww$ channel,
the irreducible \ww{} background has been extensively studied in
\citeres{Aad:2012tfa,ATLAS:2014aga,Aad:2016lvc,Chatrchyan:2012xdj,Chatrchyan:2013lba,Chatrchyan:2013iaa,Aad:2015mxa,Khachatryan:2016vnn}.

Measurements of continuum production of \ww{} pairs are not the only
case for which accurate predictions for this process are needed: since
a complete reconstruction of the $W$-boson momenta is prevented by the
presence of two neutrinos in the \ww{} signature, any experimental
study which features \ww{} production as an irreducible background
requires a proper modelling of the \ww{} signal.  In particular, this
affects the sensitivity to $H\to \ww$ and to any BSM resonance
decaying into \ww{} pairs. Apart from that, experimental analyses for
both continuum \ww{} production and Higgs-boson production in the
$H\rightarrow \ww$ channel organize their measurements in categories
according to jet multiplicities. A rather strict veto against jet
radiation is particularly important in that respect to limit the
severe signal contamination due to backgrounds involving top-quarks
($t\bar{t}$ and $tW$).  The fact that the fiducial phase-space
definition involves cuts on the presence of the associated jet
activity induces an increased sensitivity to higher-order QCD effects
due to potentially large logarithms.  Such terms challenge the
reliability of fixed-order predictions in QCD perturbation theory and
cause a significant increase of the uncertainty related to the
extrapolation from the fiducial to the total phase space in
measurements of the inclusive \ww{} cross section.  These issues show
the relevance of fully flexible, hadron-level Monte Carlo predictions
with state-of-the-art perturbative precision for the \ww{} production
process.

An enormous effort has been put into the computation of highly
accurate predictions for \ww{} production in the past years.  Leading
order (LO) \cite{Brown:1978mq} 
and next-to-LO (NLO) \cite{Ohnemus:1991kk,Frixione:1993yp}  predictions
for stable $W$ bosons have been evaluated a long time ago.
More sophisticated parton-level computations at NLO have
become available incorporating leptonic $W$ decays with off-shell
effects and spin
correlations~\cite{Campbell:1999ah,Dixon:1999di,Dixon:1998py,Campbell:2011bn}.
Recently, also NLO electroweak (EW) corrections have been computed in
both the on-shell
approximation~\cite{Bierweiler:2012kw,Baglio:2013toa,Billoni:2013aba}
and including the full off-shell treatment of the $W$
bosons~\cite{Biedermann:2016guo}.  Although EW effects have a minor
impact on the inclusive \ww{} rate, they can be significantly enhanced
up to several tens of percent at transverse momenta of about 1\,TeV.

In light of sizable ${\cal O}(\as)$ effects, higher-order QCD
corrections to \ww{} production are indispensable to ensure highly
accurate theoretical predictions for this process.  \ww{} production
in association with one, two, and three jets has been computed at NLO
QCD in
\citeres{Dittmaier:2007th,Campbell:2007ev,Dittmaier:2009un,Campbell:2015hya},
\citeres{Melia:2011dw,Greiner:2012im}, and~\citere{Cordero:2015hem},
respectively. The simplest ${\cal O}(\as^2)$ contribution to the \ww{}
cross section constitutes the loop-induced $gg\to\ww+X$ subprocess,
which receives an enhancement from the gluon luminosities and is an
important part of the full next-to-NLO (NNLO) QCD corrections. $gg\to
\ww$ predictions at LO have been extensively studied
in~\citeres{Dicus:1987dj,Glover:1988fe,Binoth:2005ua,Binoth:2006mf,Campbell:2011bn},
while the Higgs-interference contribution has been considered in
\citere{Campbell:2011cu}. The corresponding calculation for loop-induced $gg\to \ww$+$1$-jet production has 
been presented in~\citere{Melia:2012zg}.

Employing the two-loop helicity amplitudes
for $gg\to VV'$ \cite{Caola:2015ila,vonManteuffel:2015msa}, NLO QCD
corrections to this subprocess keeping only contributions with $gg$
initial states were computed in \citere{Caola:2015rqy} and have been
extended by the inclusion of the Higgs-boson interference in
\citere{Caola:2016trd}. The complete NLO QCD corrections for $gg\to
\ww$ including also the $gq$ channel are still unknown.

The full NNLO corrections to \ww{} production have been calculated for
both the inclusive cross section in the on-shell approximation
\cite{Gehrmann:2014fva} and the fully differential cross section
incorporating leptonic $W$-boson decays with off-shell effects and
spin correlations \cite{Grazzini:2016ctr}. These computations employed
the two loop helicity amplitudes of
\citeres{Gehrmann:2014bfa,Caola:2014iua,Gehrmann:2015ora}.  It was
found that NNLO QCD corrections have a significant impact on the
inclusive cross section of roughly $10\%$. Contrary to what was widely
expected, the dominant correction is given by the NNLO corrections to
the quark-initiated process, with the size of the loop-induced $gg$
contribution being only about $30\%$ of the ${\cal O}(\as^2)$
terms. This highlights the importance of including the full NNLO
corrections to this process.

Several Monte Carlo predictions have been obtained in the past years:
\ww{} production was part of the original proof-of-concept publication
of the \MCatNLO{} formalism \cite{Frixione:2002ik} to match NLO QCD
predictions with parton showers (NLO+PS); it was followed by
independent NLO+PS computations in
\noun{Herwig++}~\cite{Hamilton:2010mb},
\noun{Sherpa}~\cite{Hoche:2010pf} and
\noun{Powheg-Box}~\cite{Melia:2011tj,Nason:2013ydw}. The recent
\noun{Herwig7} implementation \cite{Bellm:2016cks,Bellm:2015jjp}
includes also single-resonant and gluon-induced contributions, and
supersedes the previous \noun{Herwig++} prediction.  More recently,
also merged computations for $\ww{}$+$0,1$\,jets at NLO+PS have become
available\footnote{For a combination of fixed-order NLO predictions of
  $\ww{}$+$0,1$\,jets see \citere{Campanario:2013wta}.} in the
\noun{MEPS@NLO} approach~\cite{Gehrmann:2012yg,Hoeche:2012yf} within
\noun{OpenLoops+Sherpa} \cite{Cascioli:2013gfa}, in the \noun{FxFx}
scheme \cite{Frederix:2012ps} within \noun{MadGraph5\_aMC@NLO}
\cite{Alwall:2014hca}, and in the \MINLO{}
procedure~\cite{Hamilton:2012np,Hamilton:2012rf} within \POWHEGBOX{}
\cite{Nason:2004rx,Frixione:2007vw,Alioli:2010xd} through the
\WWJMINLO{} generator \cite{Hamilton:2016bfu}. The \MINLO{}
computation has the advantage of being NLO accurate in both 0- and
1-jet quantities simultaneously, while other multi-jet merging
simulations partition the phase space into different jet bins
according to some merging scale, which spoils NLO accuracy in certain
phase space regions.

State-of-the-art resummation techniques have been used to compute all
higher-order logarithmic contributions up to NNLL+NLO for threshold
logarithms \cite{Dawson:2013lya} and up to NNLL+NNLO for the
transverse momentum (\pt{}) of the \ww{} pair~\cite{Grazzini:2015wpa}
as well as the jet-vetoed cross section
\cite{Dawson:2016ysj}.\footnote{See also
  \citeres{Grazzini:2005vw,Wang:2013qua,Meade:2014fca} for earlier,
  less accurate results.} The latter results show that high
theoretical control on the cross section with a veto on the \pt{} of
the \ww{} pair or on the jets can be obtained only by combining both
NNLO accuracy at fixed order and resummation of large logarithmic
terms. Indeed, some tension observed in earlier \ww{} measurements
\cite{Chatrchyan:2013oev,ATLAS-CONF-2014-033} triggered a discussion
on the proper modelling of the jet-vetoed cross sections
\cite{Jaiswal:2014yba,Becher:2014aya,Monni:2014zra,Dawson:2016ysj} and
challenged the validity of lower-order Monte Carlo predictions.
Hence, a combination of parton-shower resummation with
state-of-the-art perturbative precision is crucial to obtain highly
accurate predictions for the production of \ww{} pairs at the LHC.

In this paper we present a novel computation of NNLO-accurate
predictions matched to parton showers (NNLOPS) for \ww{} production at
hadron colliders.  More precisely, we consider all topologies which
lead to two opposite-charge leptons and two neutrinos in the final
state ($\ell\nu_\ell\ell'\nu_{\ell'}$), thereby taking into account
off-shell effects and spin correlations.
This is the first time full NNLOPS accuracy is achieved for a $2\to 4$ process.
Our computation is based on the combination of two earlier
computations: we start from the \WWJMINLO{} implementation
\cite{Hamilton:2016bfu} within the \POWHEGBOX{}
framework~\cite{Nason:2004rx,Frixione:2007vw,Alioli:2010xd} and
combine it with the NNLO predictions of \citere{Grazzini:2016ctr}
which are publicly available within the \Matrix{}
code~\cite{Grazzini:2017mhc,MATRIX}.  To obtain NNLOPS accuracy from
these two ingredients we follow the reweighting procedure used in
\citeres{Hamilton:2013fea,Hamilton:2015nsa,Karlberg:2014qua,Astill:2016hpa}. To
handle the significantly increased complexity inherent to an off-shell
diboson process with four final-state leptons we devise a
parametrization of the Born-level phase space which allows us to
reduce the number of degrees of freedom. In particular, we derive a
formula to describe the angular dependence of the decay products of
the two vector bosons in terms of spherical harmonics, which is
deduced from the known expression for the decay of a single vector
boson \cite{Collins:1977iv}.

Our NNLOPS computation is implemented and will be made publicly
available within the \POWHEGBOX{}
framework~\cite{Nason:2004rx,Frixione:2007vw,Alioli:2010xd}.\footnote{Instructions
  to download the code can be obtained at
  \url{http://powhegbox.mib.infn.it}.}  All-order, higher-twist,
and non-perturbative QCD effects can be approximated through the
interface to a parton shower using hadronization and underlying event
models, which render a complete and realistic event simulation
feasible. Such corrections can have a non-negligible impact on certain
observables. For instance, exclusive jet cross sections can be
considerably modified because of migration effects.
In our implementation and throughout this paper we omit the
loop-induced $gg$ component, since it is already known to higher-order
in QCD in the pure $gg$ channel and can be added at LO+PS
through known tools, such as the \noun{gg2ww} event
generator~\cite{Binoth:2006mf,Kauer:2012hd} (as used by ATLAS and
CMS).\footnote{A NLO+PS generator for $gg\to\ww{}$ production could be obtained along
the lines of~\citere{Alioli:2016xab}.}
Furthermore, in order to define \ww{} signal events free of top-quark
contamination we employ the four-flavour scheme with massive bottom
quarks and drop all contributions with bottom quarks in the final
state.
It has been shown in \citeres{Gehrmann:2014fva,Grazzini:2016ctr} for
both total and fiducial rates at NNLO that this approach is in very
good agreement ($\sim 1$--$2\%$) with an alternative procedure to
obtain top-free \ww{} predictions in the five-flavour scheme. The
latter exploits the resonance structure of top-quark contributions to
fit the part of the cross section independent of the top-quark width.

Besides an extensive validation of our NNLOPS results, we study the
impact of the parton shower on NNLO predictions and show the
importance of including NNLO corrections on top of the \MINLO{}
computation. In particular, the NNLOPS predictions provide new
insights on the size of jet-veto logarithms at higher orders.  We also
compare our predictions against measurements of the total and fiducial
cross sections as measured by ATLAS and CMS, and present distributions
in the presence of experimental selection cuts in the fiducial volume
of \ww{} measurements. We finally use differences observed in the
rapidities of the two $W$ bosons to define a charge asymmetry for
\ww{} production and study to what extent such asymmetry remains when
considering the rapidities of the two charged leptons instead.

The manuscript is organized as follows.  In \sct{sec:calculation} we
describe technical aspects of the computation, including a discussion
of the top-quark contamination (\sct{sec:top}), the reweighting method
to obtain NNLOPS predictions (\sct{sec:method}), its practical
implementation (\sct{sec:practical}), and a validation of our NNLOPS
results (\sct{sec:validation}).  Phenomenological results are
presented in \sct{sec:results}: we first outline input parameters and
fiducial cuts (\sct{sec:input}); report cross-section predictions and
compare them to data (\sct{sec:rates}); study the impact of jet-veto
logarithms at NNLOPS (\sct{sec:jetveto}); demonstrate the importance
of NNLOPS predictions in the fiducial phase space
(\sct{sec:fiducial}); and finally define a charge asymmetry for \ww{}
production (\sct{sec:chargeasymmetry}).  We summarize our findings in
\sct{sec:summary}.

\section{Description of the calculation}
\label{sec:calculation}

\begin{figure}
\begin{center}
\begin{tabular}{ccccc}
\includegraphics[width=.25\textwidth]{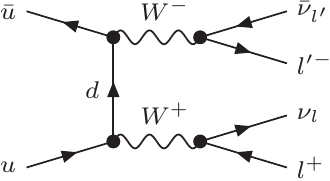} & &
\includegraphics[width=.25\textwidth]{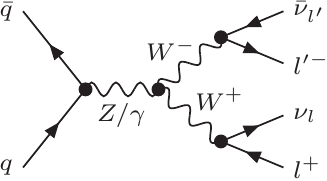} & &
\includegraphics[width=.25\textwidth]{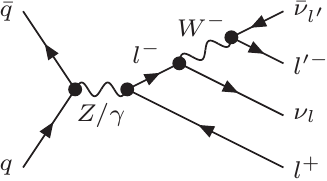} \\[0ex]
(a) & & (b) & & (c)
\end{tabular}
\end{center}
\vspace*{.5ex}
\begin{center}
\begin{tabular}{ccc}
\includegraphics[width=.25\textwidth]{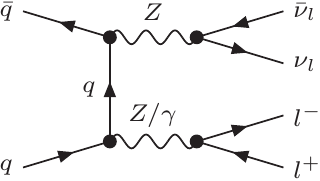} & &
\includegraphics[width=.25\textwidth]{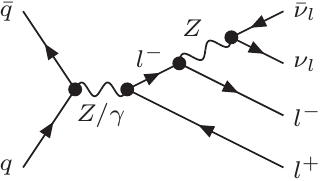} \\[0ex]
(d) & & (e)
\end{tabular}
\caption[]{\label{fig:diag}{Born-level Feynman diagrams for \ww{}
    production: (a-c) contribute to both the DF channel ($\elle\neq
    \elle^\prime$) and the SF channel ($\elle=\elle^\prime$); (d-e)
    only contribute in the SF case.}}
\end{center}
\end{figure}

We consider the production of two opposite-charge leptons and two
neutrinos in proton--proton collisions
\begin{equation}
\label{eq:process}
pp\to \elle^-\bar\nu_{\elle}\;\elle^{\prime\, +}{\nu}_{\elle^\prime}+X,
\end{equation}
where the two leptons are of different flavour ($\ell \neq \ell'$).
Our computation includes off-shell effects and spin correlations by
taking into account all the resonant and non-resonant topologies
leading to this process. For convenience, we simply refer to it as
\ww{} production in the upcoming sections. Already at LO these
topologies involve different combinations of vector-boson resonances,
such as double-resonant $t$-channel \ww{} production; double-resonant
$s$-channel \ww{} production via $Z$ or $\gamma^\ast$; and
single-resonant DY-like topologies with subsequent decay. The relevant
Born-level diagrams are shown in \fig{fig:diag}\,(a-c).

While the same type of diagrams contribute also to the same-flavour
(SF) case ($\ell = \ell'$), this channel involves additional
topologies depicted in \fig{fig:diag}\,(d-e): double-resonant
$t$-channel \zz{} production; and single-resonant DY-like
topologies. The SF channel therefore mixes double-resonant \zz{} and
\ww{} contributions. It was shown in
\citeres{Melia:2011tj,Kallweit:2017khh}, however, that interference
effects between \zz{} and \ww{} topologies are generally small, so
that the two processes can be computed separately and added
incoherently.  Thus, we focus on \ww{} production in the
different-flavour (DF) channel in what follows. More precisely, while
our computation is applicable to any combination
$\elle,\elle^\prime\in \{e,\mu,\tau\}$ of two massless leptons of
different flavour, for the sake of simplicity, we will study the
process $pp\to\muenn+X$ and its charge conjugate in \sct{sec:results}.

\subsection{Top-quark contamination in \ww{} production}
\label{sec:top}

\ww{} production is subject to a severe contamination from top-quark
contributions with $t\rightarrow Wb$ decays, which enter radiative
corrections through interference with real-emission diagrams featuring
final-state bottom quarks.  Such contributions ought to be removed to
define a top-free \ww{} cross section. Without a consistent removal of
the top-quark contamination, the \ww{} cross section, in particular in
the inclusive phase space, can be increased by even an order of
magnitude upon inclusion of radiative corrections, thereby corrupting
the convergence of the perturbative expansion. Two approaches have
been followed in the literature to compute the top-subtracted \ww{}
cross section, which will be described below. The two methods have
been shown to agree within $\sim 1$--$2\%$ for both the inclusive case
\cite{Gehrmann:2014fva} and with fiducial cuts
\cite{Grazzini:2016ctr}. 

In the five-flavour scheme (5FS) bottom
quarks are treated as massless and appear as both initial and
final-state particles.  In this scheme, the presence of real
bottom-quark emission is inevitably tied to $g\to b\bar b$ splittings
in the virtual corrections through collinear singularities. Hence,
such contributions must not be separated to guarantee infrared (IR)
safety.  Instead, the scaling behaviour of the cross sections in the
limit of a vanishing top-quark width can be exploited to determine all
contributions free from top-quark resonances. This approach requires
the repeated computation of the cross section for varying top-quark
widths in the limit $\Gamma_{t}\to0$ in order to fit the resonance
structure and isolate double-resonant (single-resonant) contributions,
which depend quadratically (linearly) on $1/\Gamma_t$, while top-free
\ww{} contributions have no enhancement at small $\Gamma_t$.  In the
four-flavour scheme (4FS), on the other hand, bottom quarks are
treated as massive and bottom quarks appear only as final-state
particles. The bottom mass renders all partonic subprocesses with
bottom quarks in the final state separately finite.  The top-quark
contamination in the 4FS can simply be avoided by dropping all such
contributions from the computation, which are then considered part of
the (off-shell) top-pair background. For convenience, we employ this
approach in the calculation and throughout this paper.

\subsection{NNLOPS method}\label{sec:method}
Our computation of NNLO-accurate parton shower predictions for \ww{}
production builds upon two recent computations for this process: the
fully differential NNLO corrections for \ww{} production, which were
calculated in \citere{Grazzini:2016ctr} and have become available in
the \Matrix{} framework \cite{Grazzini:2017mhc,MATRIX}, and a \MINLO{}
computation for \ww+jet production in the \POWHEGBOX{}
\cite{Nason:2004rx,Frixione:2007vw,Alioli:2010xd} (\WWJMINLO{}), which
was presented in \citere{Hamilton:2016bfu}.

\Matrix{} is a computational framework, which features NNLO QCD
corrections to a large number of hadron-collider processes with
color-neutral final states. This code (and earlier versions of it) has
been used to obtain several state-of-the-art NNLO predictions, in
particular for $Z\gamma$~\cite{Grazzini:2013bna,Grazzini:2015nwa},
$W^\pm\gamma$~\cite{Grazzini:2015nwa},
$ZZ$~\cite{Cascioli:2014yka,Grazzini:2015hta},
\ww{}~\cite{Gehrmann:2014fva,Grazzini:2016ctr},
\wz{}~\cite{Grazzini:2016swo,Grazzini:2017ckn} and $HH$
\cite{deFlorian:2016uhr,Grazzini:2018bsd} production.\footnote{It was
  also used to compute the resummed transverse-momentum spectra for
  $ZZ$ and \ww{} pairs at NNLL+NNLO in \citere{Grazzini:2015wpa}.}
\Matrix{} uses a fully general implementation of the \qt{}-subtraction
formalism \cite{Catani:2007vq} to achieve NNLO accuracy, in
combination with an automated implementation of the Catani--Seymour
dipole subtraction method \cite{Catani:1996jh,Catani:1996vz} within
the Monte Carlo program \Munich{}\footnote{The Monte
  Carlo program \Munich{} features a general implementation of an
  efficient, multi-channel based phase-space integration and computes
  both QCD and EW~\cite{Kallweit:2014xda,Kallweit:2015dum} corrections
  to NLO accuracy for arbitrary SM processes.}~\cite{munich}.  All (spin- and
colour-correlated) tree-level and one-loop amplitudes are obtained
from \OpenLoops{}\footnote{\OpenLoops{} relies on the fast and stable
  tensor reduction of \Collier{}~\cite{Denner:2014gla,Denner:2016kdg},
  supported by a rescue system based on quad-precision
  \CutTools\cite{Ossola:2007ax} with \OneLOop\cite{vanHameren:2010cp}
  to deal with exceptional phase-space
  points.}~\cite{Cascioli:2011va,Buccioni:2017yxi,hepforge:OpenLoops},
while dedicated computations of the two-loop amplitudes are employed
\cite{Anastasiou:2002zn,Gehrmann:2011ab,Gehrmann:2015ora,deFlorian:2013uza}.
Most importantly, the two-loop amplitudes for the production of a pair
of off-shell massive vector bosons \cite{Gehrmann:2015ora} are taken
from the publicly available code \textsc{VVamp}~\cite{hepforge:VVamp},
which enters our computation for \ww{} production.

The \WWJMINLO{} computation of \citere{Hamilton:2016bfu} implements
\ww{}+jet production within the \POWHEGBOX{} framework and upgrades it
by the \MINLO{} procedure.
As described in~\citere{Hamilton:2016bfu}, all tree-level matrix elements
have been obtained using the \POWHEGBOX{} interface
to \MADGRAPH{}~\cite{Alwall:2007st,Campbell:2012am}, while the one-loop amplitudes 
have been
obtained using \GOSAM{}~\cite{Cullen:2014yla}.
The \MINLO{} procedure merges $\ww{}$+$0,1$-jet
multiplicities to obtain fully exclusive hadron-level events with NLO
accuracy. In particular, the inclusion of a numerical implementation
of the $B_2$ resummation coefficient, ensures that observables 
inclusive over the extra jet are also NLO accurate. In fact, the
\WWJMINLO{} computation of \citere{Hamilton:2016bfu} was the first to
implement this approach for a genuine $2\to2$ process, with non-trivial virtual corrections. 

To obtain NNLOPS accurate predictions from these two ingredients we
follow closely the method which has already been successfully applied
in the computations of Higgs \cite{Hamilton:2013fea}, Drell-Yan
\cite{Karlberg:2014qua}, $HW^\pm$ \cite{Astill:2016hpa} and
$HZ$~\cite{Astill:2018ivh} production: the Les Houches events (LHE)
produced with the \WWJMINLO{} generator are reweighted to the correct
NNLO prediction fully differentially in the Born phase space.  This is
done by means of a multidifferential reweighting covering the entire
phase-space of the colourless system ($\muenn$) at LO.
In its simplest form, the reweighting proceeds as follows: for each
\MINLO{} event, the reweighting factor is computed as the ratio of the
NNLO cross section in the given configuration of the Born-level
variables and the original \MINLO{} weight associated with the
Born-level variables of the respective event:
\begin{align}
  \label{eq:W}
  {\cal W}(\born{}) = \frac{\dd\sigma^{\textrm{NNLO}}/\dd\born{}}{\dd\sigma^{\text{\MINLO}}/\dd\born{}}\,.
\end{align}
$\dd\sigma^{\textrm{NNLO}}/\dd\born{}$ is a multi-differential
distribution obtained from the NNLO computation, while
$\dd\sigma^{\text{\MINLO}}/\dd\born{}$ is the same multi-differential
distribution, but determined from the \WWJMINLO{} events.  The
observables defining the multi-differential cross section are to a
large extent arbitrary as long as they form a basis of the Born-level
phase space ($\born{}$).  Our specific choice for \ww{} production
will be discussed in detail in \sct{sec:practical}.

By construction, this procedure promotes the \WWJMINLO{} events to be
NNLO accurate in all Born-level variables. As proven in
\citeres{Hamilton:2013fea,Karlberg:2014qua} the reweighting does not
spoil the NLO accuracy of the \WWJMINLO{} computation.  Also the
parton shower does not interfere with the NNLO accuracy of the \ww{}
sample, which is obvious considering the fact that the second emission
is generated from the \POWHEG{} prescription keeping NLO accuracy of
the \ww{}+jet process.  Only starting from the third one, parton
emissions are generated by the shower, whose impact is beyond NNLO as
it affects terms from ${\cal O}(\alpha_s^3)$ onwards.  In conclusion,
the reweighting procedure under consideration allows us to obtain
fully differential hadron-level events, while retaining NNLO accuracy
for \ww{} production.

One should bear in mind that \eqn{eq:W} reflects the reweighting
factor only in its simplest form.  As pointed out in
\citere{Hamilton:2013fea} it has the disadvantage of spreading the
$\textrm{NNLO/NLO}$ $K$-factor uniformly for observables which are
non-trivial starting from the \ww{}+$1$-jet phase space only, such as
the transverse momentum of $W$-boson pair (\ptww{}). Away from the
singular region, such observables are described at the same formal
accuracy (effectively NLO) by the \ww{} NNLO computation and the
\WWJMINLO{} generator. Hence, no improvement can be obtained for them
through the reweighting procedure. On the contrary, given that the only observables 
that are formally NNLO accurate are those that are non-trivial at Born level, where $\ptww{}=0$, it appears
to be more natural to limit the range in \ptww{} in which the
reweighting takes effect to small values of \ptww{}.  Indeed, this is
much closer to what is done in the matching between fixed order and
analytic transverse-momentum resummation of the \ww{} system
\cite{Grazzini:2005vw,Wang:2013qua,Meade:2014fca,Grazzini:2015wpa}. In
fact, in analytic resummation all logarithmic terms are unambiguously
matched between the two contributions upon truncation at a given order
in $\as{}$. For the analytically resummed \ptww{} spectrum at
NNLL+NNLO, see \citere{Grazzini:2015wpa}, the NNLO contribution from
the two-loop virtual corrections is, roughly speaking, distributed in
$\ptww{}$ between zero and the respective resummation scale.  As a
consequence, the NLO transverse momentum distribution is recovered at
large \ptww{}.

Following this idea it was suggested in \citere{Hamilton:2013fea} to
introduce a reweighing factor that evolves smoothly to one in regions
where the NNLO computation is formally only NLO accurate and thus does
not improve the NLO accuracy of the \MINLO{} event sample:
\begin{align} 
\begin{split}\label{eq:NNLOPS-overall-rwgt-factor-1}
  \mathcal{W}\left(\born,\,\pt{}\right)&=h\left(\pt\right)\,\frac{\smallint
    \dd\sigma^{{\sss
        \mathrm{NNLO\phantom{i}}}}\,\delta\left(\born-\born\left(\Phi\right)\right)-\smallint
    \dd\sigma_{\sss B}^{{\sss
        \text{\MINLO}}}\,\delta\left(\born-\born\left(\Phi\right)\right)}{\smallint
    \dd\sigma_{\sss A}^{{\sss
        \text{\MINLO}}}\,\delta\left(\born-\born\left(\Phi\right)\right)}\\ &+\left(1-h\left(\pt\right)\right)\,, 
\end{split}
\end{align}
where the function $h(p_{\sss T})$ has the property that it is one at
$\pt=0$ and vanishes at infinity. This function is used in
\eqn{eq:NNLOPS-overall-rwgt-factor-1} to split the cross section into
\begin{align}
\dd\sigma_{\sss A} = \dd\sigma\cdot h(p_{\sss T})\,,\qquad  
\dd\sigma_{\sss B} = \dd\sigma\cdot (1-h(p_{\sss T}))\,.
\end{align}
Here we use the following smoothing function:
\begin{align}
h(p_{\sss T}) = \frac{(2\,M_{W})^2}{(2\,M_{W})^2+p_{\sss T}^{\; 2}}\,. \label{eq:h_pt}
\end{align}
It is trivial to see that the exact value of the NNLO differential
cross-section in the Born-level phase space is preserved using this
reweighting factor:
\begin{align}
  \left(\frac{d\sigma}{d\born}\right)^{{\sss
      \mathrm{NNLOPS}}} & =
  \left(\frac{d\sigma}{d\born}\right)^{{\sss
      \mathrm{NNLO}}}\,.\label{eq:NNLOPS-NNLOPS-eq-NNLO_0+MINLO_1-1}
\end{align}

We have not yet specified what $p_{\sss T}$ exactly stands
for. Between the two natural choices, the transverse momentum of the
colourless system or of the leading jet, we refrain from using the
former, and have chosen the transverse momentum of the leading jet
instead.  This choice is motivated by the fact that only the latter is
a direct indicator of whether QCD radiation is present in a given
event or not. This ensures that $h(p_T)$ goes to one only for
Born-like configurations, while it tends to zero in the presence of
hard radiation, with ${\cal W}(\born,p_{\sss T})$ going to one
accordingly.  To define jets in $h(p_T)$ we employ the inclusive
$k_T$-algorithm with $R=0.4$~\cite{Catani:1993hr,Ellis:1993tq} as
implemented in \FASTJET{}~\cite{Cacciari:2011ma}.

\subsection{Practical implementation}
\label{sec:practical}

We now turn to discussing practical details on the implementation of
the reweighting procedure for \ww{} production sketched in the
previous section.  First we have to find a parametrization of the Born
phase space. To this end, we select a set of nine independent
observables, with nine being the degrees of freedom of the
$4$-particle (\muenn{}) phase space we have at LO, after removing an
overall azimuthal angle. This defines our basis for the
multidimensional reweighting. We choose the variables $\Phi=\{\ptwm$,
$\yww$, $\dywpwm$, $\costhetap$, $\phip$, $\costhetam$, $\phim$,
$\mwp$, $\mwm\}$, which correspond to the transverse momentum of $W^-$
(that is equal and in the opposite direction to the one of $W^+$ at
LO), the rapidity of the \ww{} pair, the rapidity difference between
the two $W$ bosons ($\dywpwm=y_{W^+}-y_{W^-}$), the Collins-Soper (CS)
angles for $W^+$ and $W^-$ as introduced in \citere{Collins:1977iv},
and the invariant masses of the two $W$ bosons, respectively.  The
differential cross section in the Born phase space is then defined as
\begin{align}\label{eq:initialphasespace}
\frac{\dd\sigma}{\dd\born}=\frac{\dd^9\sigma}{\dptwm\dyww\ddywpwm\dcosthetap\dphip\dcosthetam\dphim\dmwp\dmwm}\,.
\end{align}
Given the high complexity of both the NNLO and the \MINLO{}
computation for \ww{} production the computation of a nine-dimensional
cross section is virtually impossible with current technology.
However, we can make use of two facts: first of all, we can drop the
invariant $W$-boson masses by realizing that their differential K
factor is practically flat over the whole phase space. This is true
especially in the peak region of the $W^\pm$ resonances, where the
majority of the events originate from, but even applies in the region
where the $W$ bosons become off-shell. Validation plots confirming
this approximation are discussed in \sct{sec:validation}.  We
therefore reduce the number of free parameters from nine to
seven. Secondly, the angular dependence of each $W$-boson decay is
fully described by the corresponding two CS angles and we exploit the
fact that one can parametrize the dependence of the cross section on
the CS angles for the decay of each of the $W$ bosons in terms of the nine spherical harmonic
functions $Y_{lm}(\thetacs,\phics)$ with $l\le 2$ and $|m|\le l$
\cite{Collins:1977iv}.  This allows us to significantly simplify the
calculation of the cross section in the Born phase space by expressing
the sevenfold-differential distribution through the evaluation of 81
triple-differential distributions of the cross section multiplied by
functions depending on the CS angles, which renders a numerical
evaluation feasible.

Strictly speaking, the parametrization through CS angles is fully
applicable only to double-resonant \ww{} topologies. However, they
provide by far the dominant contribution to the cross section. Indeed,
the validation in \sct{sec:validation} reveals no remnants of using
this procedure as an approximation in the single- and non-resonant
contributions.

Before demonstrating how to express the cross section in terms of
spherical harmonics of the CS angles, we briefly describe our choices
of the bin sizes. For the three remaining variables\footnote{The star
  indicates that the CS angles of the respective $W$ decays are
  integrated out.} $\Phi_{W^+_*W^-_*}=\{\ptwm$, $\yww$, $\dywpwm\}$,
we choose bin edges:
\begin{equation}
  \begin{split}
\ptwm{}\,:\quad [&0.,17.5,25.,30.,35.,40.,47.5,57.5,72.5,100.,200.,350.,600.,1000.,1500.,\infty]\,;\\
\yww{}\,:\quad [&-\infty,-3.5,-2.5,-2.0,-1.5,-1.0,-0.5,0.0,0.5,1.0,1.5,2.0,2.5,3.5,\infty]\,;\\ 
\dywpwm{}\,:\quad [&-\infty,-5.2,-4.8,-4.4,-4.0,-3.6,-3.2,-2.8,-2.4,-2.0,-1.6,-1.2, \\ 
  &-0.8,-0.4,0.0,0.4,0.8,1.2,1.6,2.0,2.4,2.8,3.2,3.6,4.0,4.4,4.8,5.2,\infty]\,. 
\label{eq:bins}
  \end{split}
  \end{equation}
These bins have been selected following two criteria: firstly, the
bins should be sufficiently populated by events to ensure statistical
robustness. Secondly, we tried to cover regions of phase space with
finer binnings where the NNLO K factor features large shape
effects. Not in all cases both criteria are fully compatible, in
particular when there are shape effects far in the tail of
distributions. The present choice constitutes a judicious compromise
in these phase space regions. We will show in \sct{sec:validation}
that the chosen bin edges are sufficient to obtain NNLO-accurate
parton-shower predictions in all distributions and phase space regions
of phenomenological interest.

We now turn to deriving a novel expression for the expansion of the
cross section in spherical harmonic functions of the CS angles for a
process involving the decay of two vector bosons. We start from the
well-known formula for the decay of a single vector boson~\cite{Collins:1977iv}:

\begin{align}\label{eq:single}
\frac{\dd\sigma}{\dd\born} &= \frac{\dd^7\sigma}{\dptwm\dyww\ddywpwm\dcosthetap\dphip\dcosthetam\dphim}\\
&=\frac{3}{16\pi}\,\sum\limits_{i=0}^8 
A_i\,f_i(\thetam,\phim)\nonumber
=\frac{3}{16\pi}\,\sum\limits_{i=0}^8 B_i\,f_i(\thetap,\phip)\nonumber,
\end{align}
where the first expansion (with $A_i$) corresponds to the
parametrization of the $W^-$ decay in terms of two CS angles and the
second one (with $B_i$) is the same, but for the $W^+$ decay.  The
functions $f_i(\theta, \phi)$ are given by
\begin{equation}
\label{eq:f}
\begin{aligned}
f_0(\theta,\phi) &= \left(1-3\cos^2\theta\right)/2\,,\quad \\
f_3(\theta,\phi) &= \sin\theta \cos\phi\,, \\ 
f_6(\theta,\phi) &= \sin 2\theta \sin \phi\,, \\
\end{aligned}
\begin{aligned}
f_1(\theta,\phi) &= \sin2\theta \cos\phi\,, \quad\\
f_4(\theta,\phi) &= \cos\theta\,, \\ 
f_7(\theta,\phi) &= \sin^2\theta \sin 2\phi\,,
\end{aligned}
\begin{aligned}
f_2(\theta,\phi) &= (\sin^2\theta \cos2\phi)/2\,, \\
f_5(\theta,\phi) &= \sin\theta \sin \phi\,, \\ 
f_8(\theta,\phi) &= 1+\cos^2\theta\,.
\end{aligned}
\end{equation}
For $i\in\{0,...,7\}$ they have the property that their integral
vanishes when integrating over $\dd{\rm cos}\,\theta\,\dd\phi$. The
coefficients $A_i$ and $B_i$ are defined as moments of the
differential cross section integrated over the respective CS angles:
\begin{align}
\begin{split}\label{eq:coeffsingle}
A_i &= \int \frac{\dd\sigma}{\dd\Phi_B}\,g_i(\thetam,\phim)\,\dcosthetam\dphim\,,\\
B_i &= \int \frac{\dd\sigma}{\dd\Phi_B}\,g_i(\thetap,\phip)\,\dcosthetap\dphip.
\end{split}
\end{align}
The functions $g_i(\theta, \phi)$ are defined as
\begin{equation}
\begin{aligned}
g_0(\theta,\phi) &= 4- 10 \cos^2 \theta\,, \quad\\
g_3(\theta,\phi) &= 4 \sin\theta \cos\phi\,, \\ 
g_6(\theta,\phi) &= \sin2\theta \sin\phi\,, \\
\end{aligned}
\begin{aligned}
g_1(\theta,\phi) &= \sin 2\theta \cos\phi\,,\\
g_4(\theta,\phi) &= 4 \cos\theta\,, \\ 
g_7(\theta,\phi) &= 5 \sin^2\theta \sin2\phi\,,\quad
\end{aligned}
\begin{aligned}
g_2(\theta,\phi) &= 10 \sin^2\theta \cos 2\phi\,, \\
g_5(\theta,\phi) &= 4 \sin\theta \sin\phi\,, \\ 
g_8(\theta,\phi) &= 1\,.
\end{aligned}
\end{equation}
Note that $A_8$ and $B_8$ are actually not moments, but correspond to
the differential cross section itself integrated over the respective
CS angles.

With the notation that we have introduced to write \eqn{eq:single} in
such a compact form, it is straightforward to deduce the combined
formula including both decays by inserting the expression of
\eqn{eq:single} for the $W^-$ decay into the $B_i$ coefficient of the
$W^+$ decay in \eqn{eq:coeffsingle}, or vice versa. Hence, our
generalization to the decay of both vector bosons for the expansion of
the cross section in all four CS angles can be cast into the following
form:
\begin{align}\label{eq:double}
\frac{\dd\sigma}{\dd\born} &=\frac{9}{256\pi^2}\,\sum\limits_{i=0}^8 \sum\limits_{j=0}^8 
AB_{ij}\,f_i(\thetam,\phim)\,f_j(\thetap,\phip)\,,
\end{align}
with coefficients 
\begin{align}\label{eq:coeffdouble}
AB_{ij} &= \int \frac{\dd\sigma}{\dd\Phi_B}\,g_i(\thetam,\phim)\,g_j(\thetap,\phip)\,\dcosthetam\dphim\dcosthetap\dphip\,.
\end{align}
These 81 coefficients are simply computed as triple-differential
distributions of the variables $\{\ptwm$, $\yww$, $\dywpwm\}$ in the
Monte Carlo integration via moments of the cross section.  In
particular, the coefficient $AB_{88}$ corresponds to the
triple-differential cross section itself:
\begin{align}
AB_{88} = \frac{\dd^3\sigma}{\dptwm\dyww\ddywpwm}\,.
\end{align}
In conclusion, the computation of 81 triple-differential distributions
allows us to determine the fully differential cross section in the
Born phase space and is feasible within both the NNLO code and the
\MINLO{} generator.

\subsection{Validation}
\label{sec:validation}

As detailed in the previous sections, the NNLOPS procedure under
consideration reweights the \MINLO{} events to the NNLO cross section
using a set of observables spanning the Born-level phase
space. Therefore, the un-showered LHE files after reweighting should
match the NNLO distributions for Born-like observables up to
differences caused by limited numerics. In particular, the
normalization of the event sample should reproduce the inclusive NNLO
cross section. In this section we provide an extensive validation of
our computation for \ww{} production by comparing LHE-level results
with the nominal NNLO predictions.  The NNLO results have been obtain
from a statistically independent computation with respect to the one
employed for the reweighting.
In order to obtain all results of this paper, we have used the input
parameters specified in Sec.~\ref{sec:input}. For the 
validation plots presented here we consider the 
process $pp\to\muenn+X$ in the 
inclusive phase space with no fiducial cuts. 

We first point out that the inclusive NNLO cross section is reproduced
to about two permille, which is at the level of the statistical
uncertainties. This level of agreement can be appreciated by examining
the cross-section numbers shown in \tab{tab:rates}. We will come back
to the discussion and interpretation of these numbers later.

Instead, we now turn to the discussion of differential
observables. The figures of this section are organized according to
the following pattern: there is a main frame, where NNLOPS (blue,
solid) and \MINLO{} (black, dotted) results at LHE level as well as
NNLO predictions (red, dashed) are shown with their absolute
normalisation, and as cross section per bin (namely, the sum of the
contents of the bins is equal to the total cross section, possibly
within cuts). In an inset we display the bin-by-bin ratio of all the
histograms which appear in the main frame over the NNLOPS curve,
chosen as a reference.  The bands correspond to the residual
uncertainties due to scale variations, which we compute as follows:
the uncertainty of the NNLO and the \MINLO{} distributions are
obtained by performing a variation of the renormalization ($\mur$) and
factorization ($\muf$) scales by a factor two around the central
choice subject to the restriction $1/2 \le \mur/\muf \le 2$. In the
case of \MINLO{} the central scale choice is dictated by the \MINLO{}
procedure:
the transverse-momentum of the \ww{} system is chosen as a scale on a
point-by-point basis, and, upon integration over radiation, one
recovers the inclusive cross-section with renormalization and
factorization scales that scale as \mww{}. In
the case of the NNLO the central scale is chosen to be the average of
the transverse masses of the two $W$ bosons, see
\eqn{eq:dynscale_nnlo}.
In order to assess the uncertainty of our \NNLOPS{} predictions,
computed using \eqn{eq:NNLOPS-overall-rwgt-factor-1}, we have
evaluated the NNLO and \MINLO{} differential cross-sections using the
same scale-variation factors. As a result, the uncertainty of the
\NNLOPS{} is also the envelope of a 7-point scale variation.

\begin{figure}[tp]
\begin{center}
\begin{tabular}{ccc}
\hspace*{-0.22cm}
\includegraphics[trim = 7mm -7mm 0mm 0mm, width=.226\textheight]{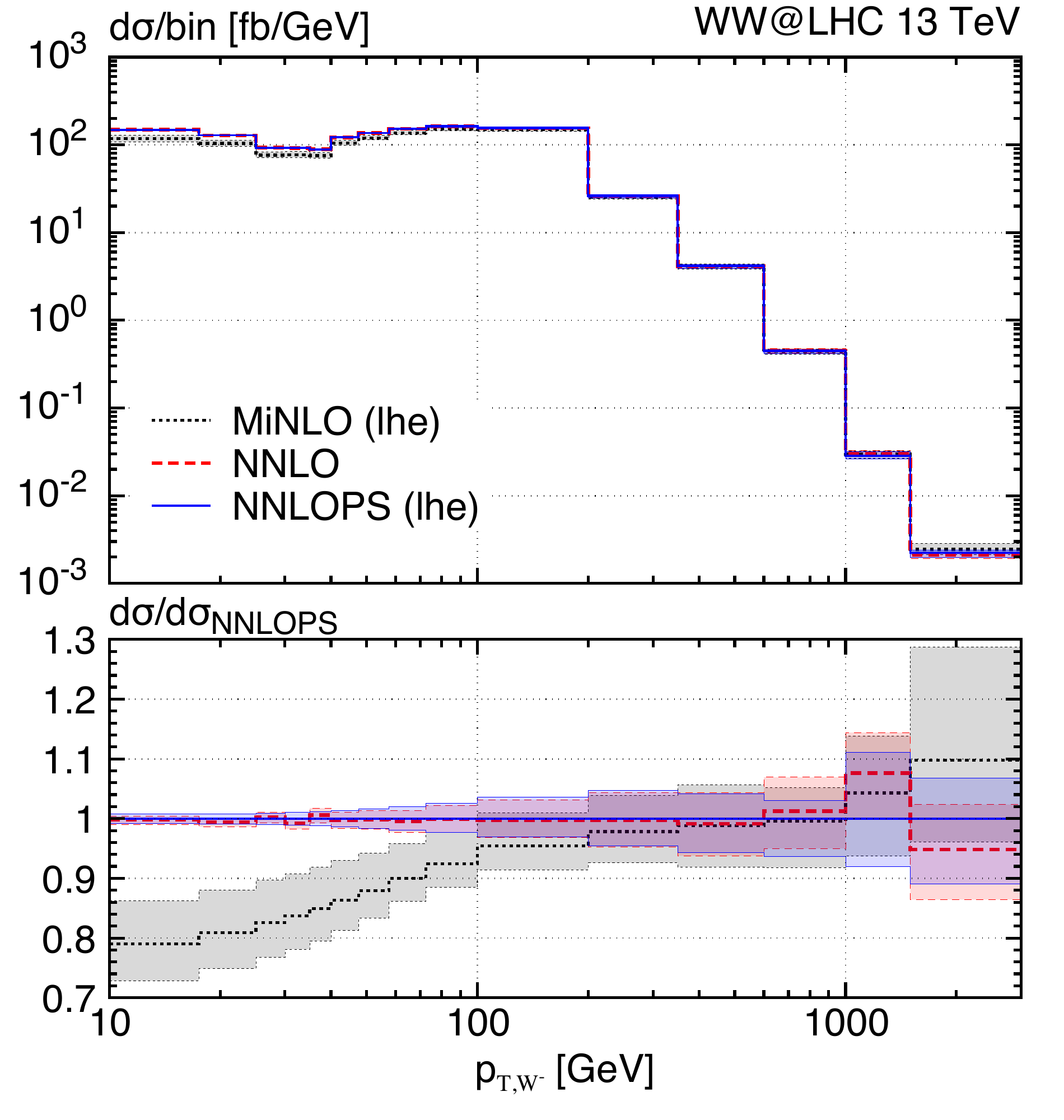} &\hspace{-0.6cm}
\includegraphics[trim = 7mm -7mm 0mm 0mm, width=.226\textheight]{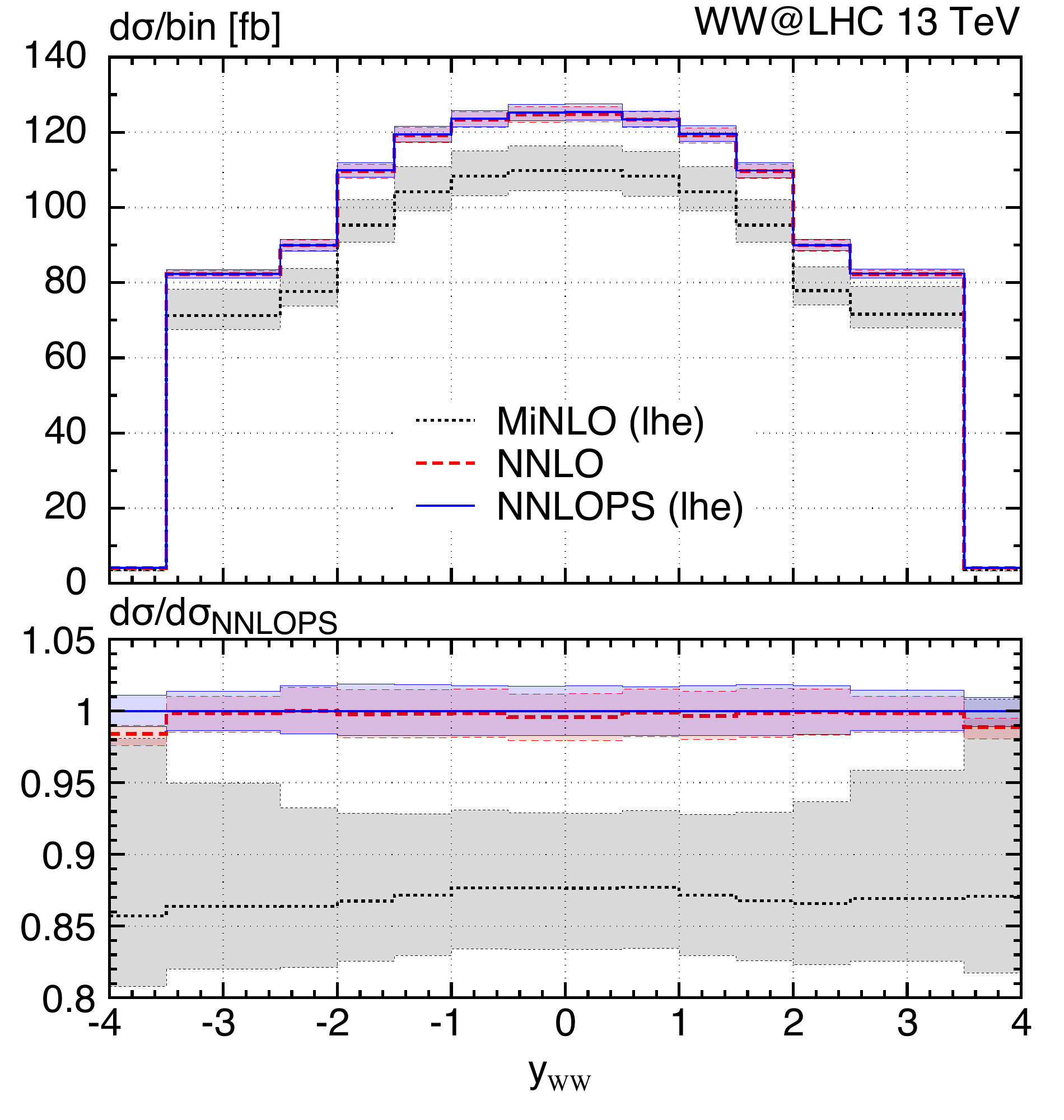} &\hspace{-0.6cm}  \includegraphics[trim = 7mm -7mm 0mm 0mm, width=.226\textheight]{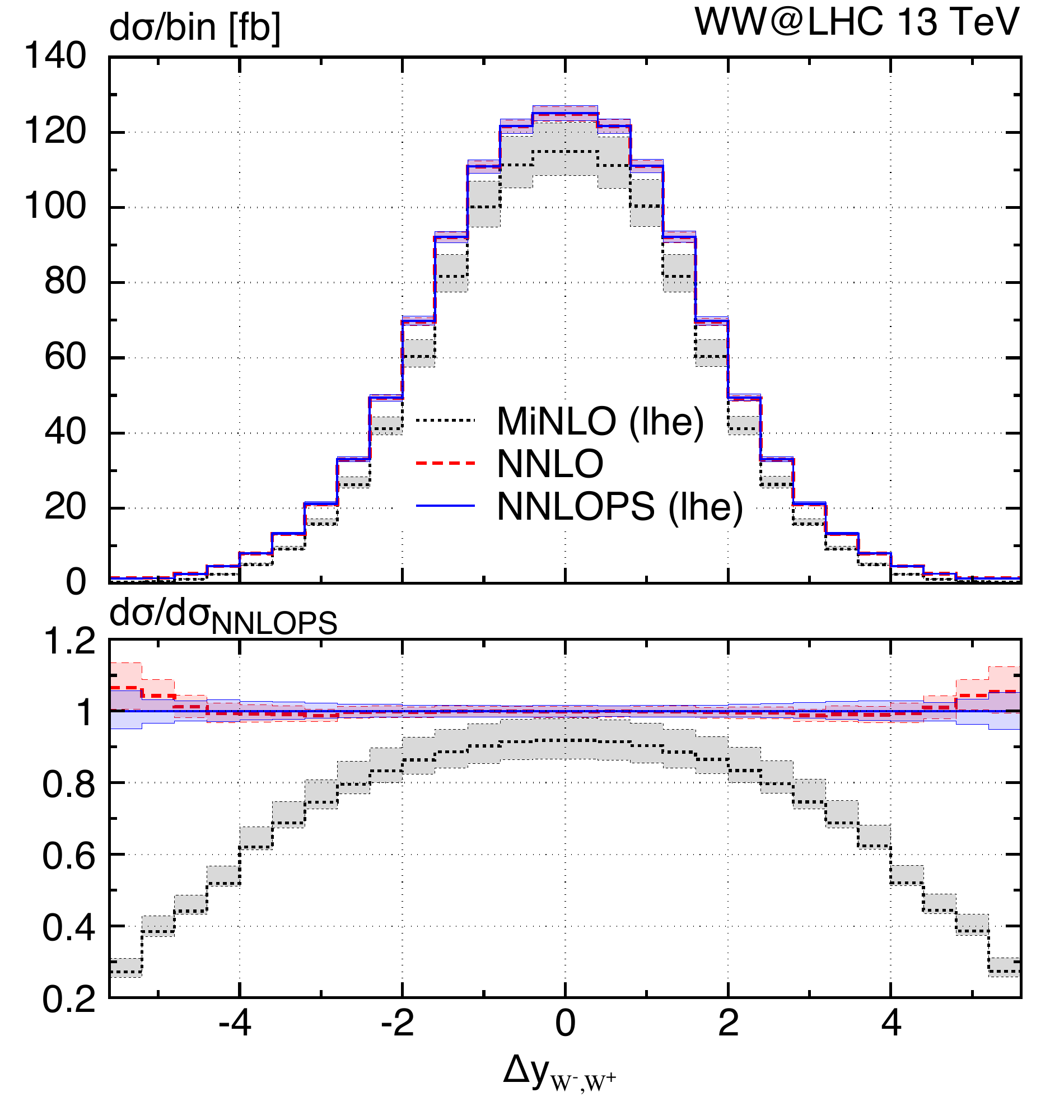} \\[-1em]
 (a) & \hspace{-1em}(b) & \hspace{-1em}(c)
\end{tabular}
\caption[]{\label{fig:ptwmywwdywpwm}{Comparison at LHE level of our
    NNLOPS results (solid, blue) with the nominal NNLO predictions
    (red, dashed) for the three distributions used in the reweighting,
    with the binning of \eqn{eq:bins}: (a) $\ptwm$, (b) $\yww$ and (c)
    $\dywpwm$; \MINLO{} results at the LHE level (black, dotted) are
    shown for reference; see text for details.}}
\end{center}
\end{figure}

We begin by showing in \fig{fig:ptwmywwdywpwm} the three observables
applied in our NNLOPS reweighting with the binning as in
\eqn{eq:bins}. We see that the NNLO distributions are nicely
reproduced by the NNLOPS computation. Differences are below one
percent in the bulk region of the distributions and increase to the
few-percent level in regions with limited numerics only. This
validates the three-dimensional reweighting we used to obtain our
NNLOPS results.

\begin{figure}[tp]
\begin{center}
\begin{tabular}{cc}
\hspace*{-0.17cm}
\includegraphics[trim = 7mm -7mm 0mm 0mm, width=.33\textheight]{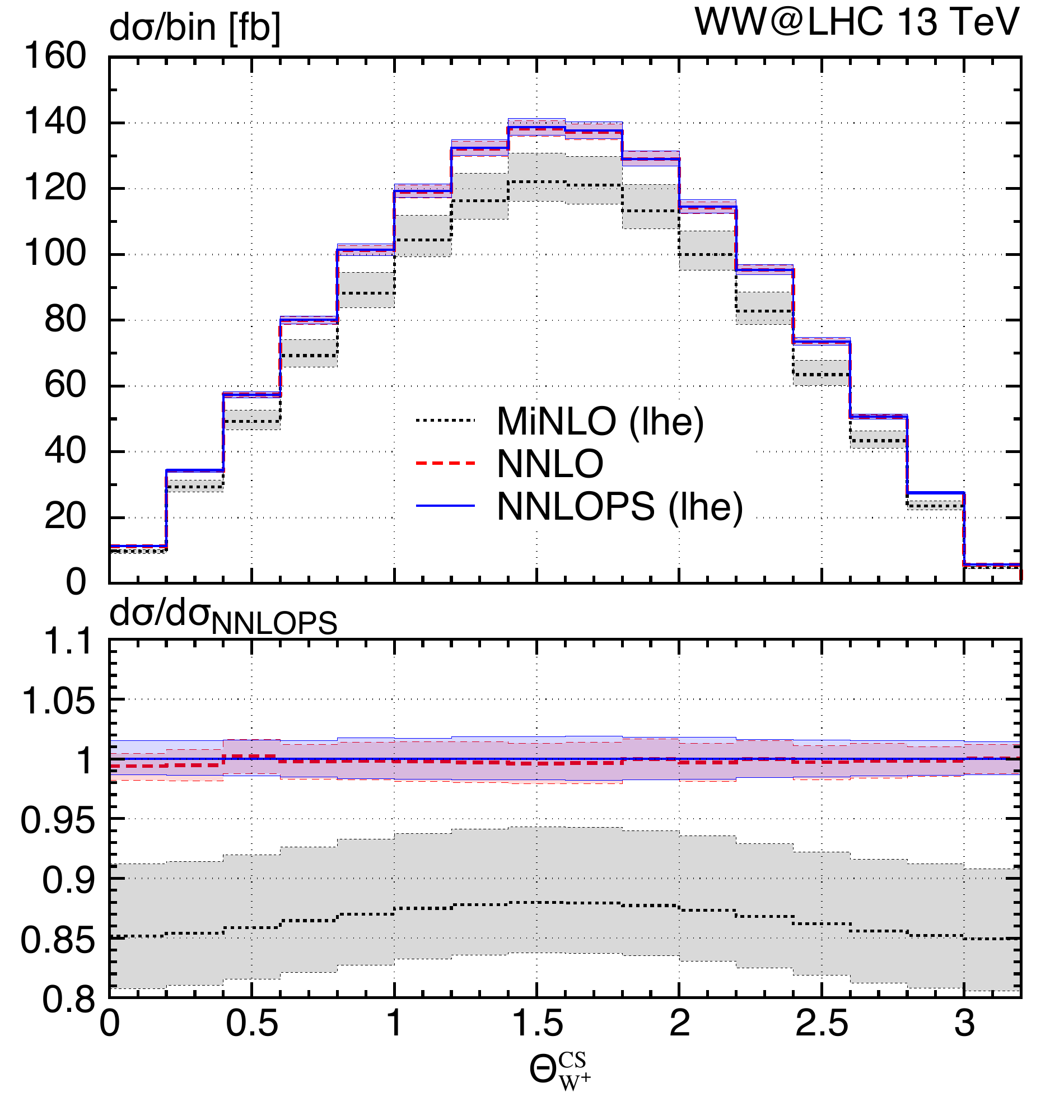} &
\includegraphics[trim = 7mm -7mm 0mm 0mm, width=.33\textheight]{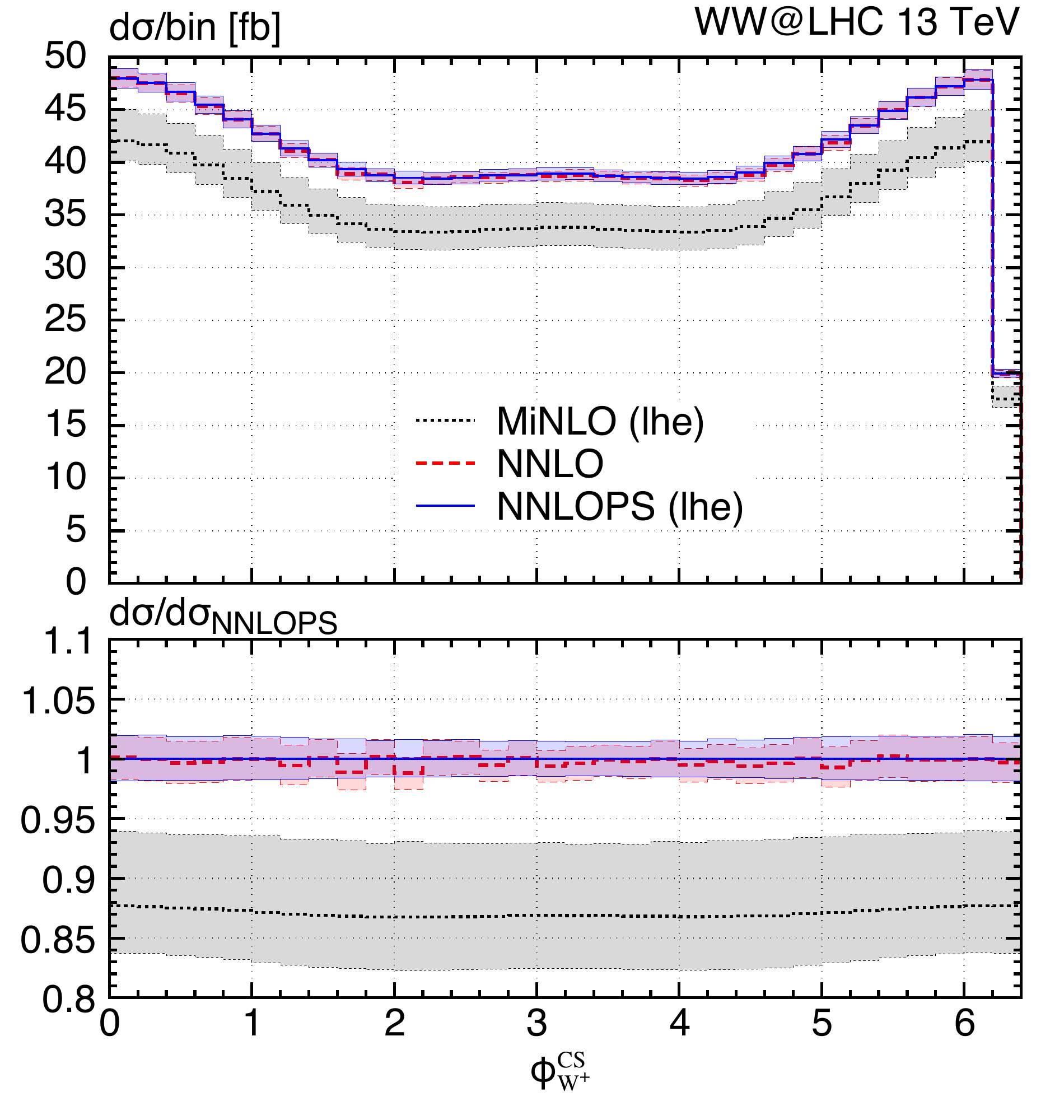} \\[-1em]
\hspace{0.6em} (a) & \hspace{1em}(b)
\end{tabular}
\caption[]{\label{fig:CSangles}{Same as \fig{fig:ptwmywwdywpwm}, but for the CS angles of the $W^+$ decay: (a) $\thetap$ and (b) $\phip$.}}
\end{center}\vspace{0.5cm}
\end{figure}

We next consider the CS angles of the $W^+$ decay. The corresponding
results for the $W^-$ decay are practically identical which is why we
refrain from discussing them separately. \fig{fig:CSangles} shows that
the distributions in \thetap{} and \phip{} are in perfect agreement
between NNLOPS and NNLO, which demonstrates the validity of our
procedure to describe the $W$ decays via CS angles.  In fact, we have
checked explicitly at NNLO level that \eqn{eq:double} reproduces the
correct cross section when being differential in any two of the four
CS angles at the same time.

\begin{figure}[tp]
\begin{center}
\begin{tabular}{cc}
\hspace*{-0.17cm}
\includegraphics[trim = 7mm -7mm 0mm 0mm, width=.33\textheight]{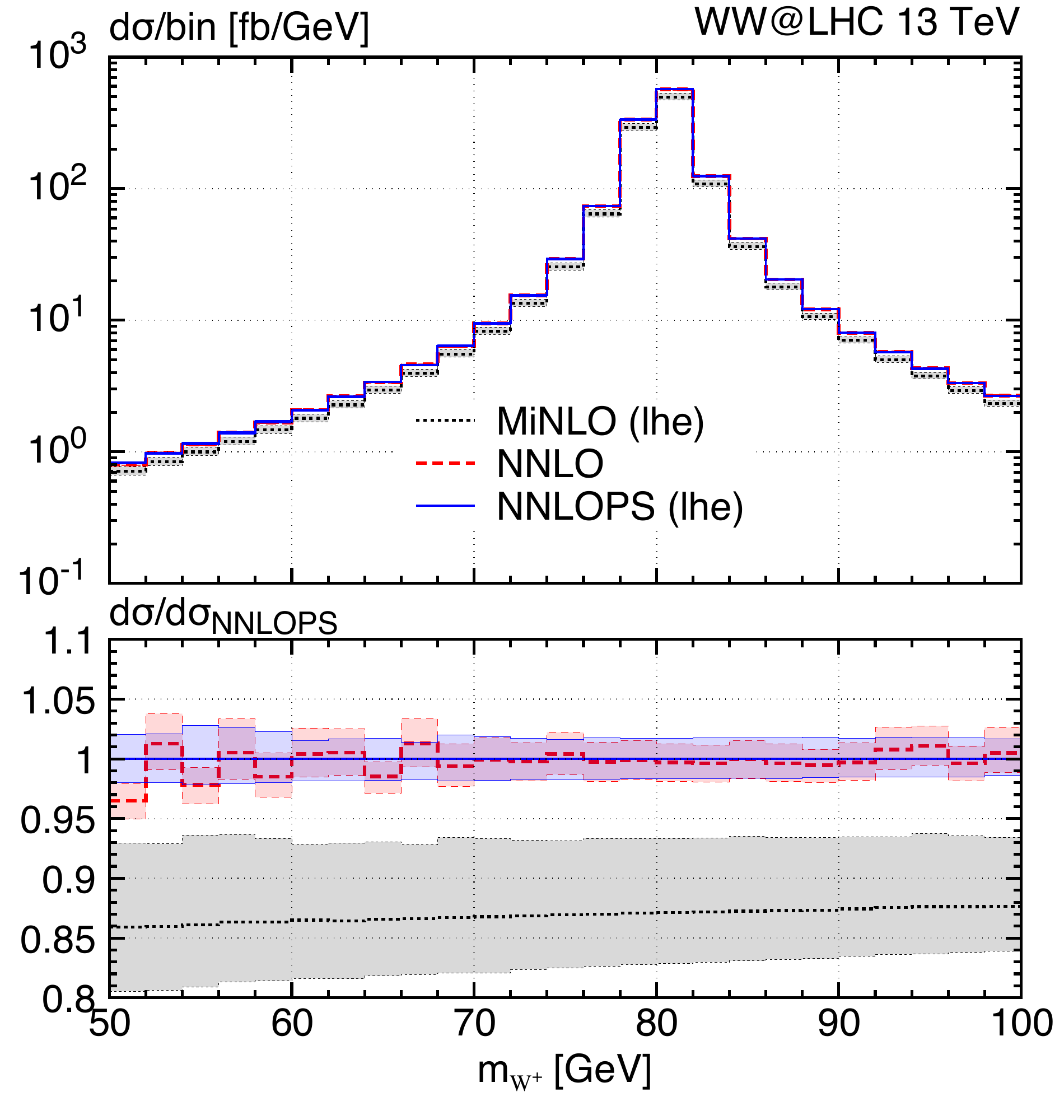} &
\includegraphics[trim = 7mm -7mm 0mm 0mm, width=.33\textheight]{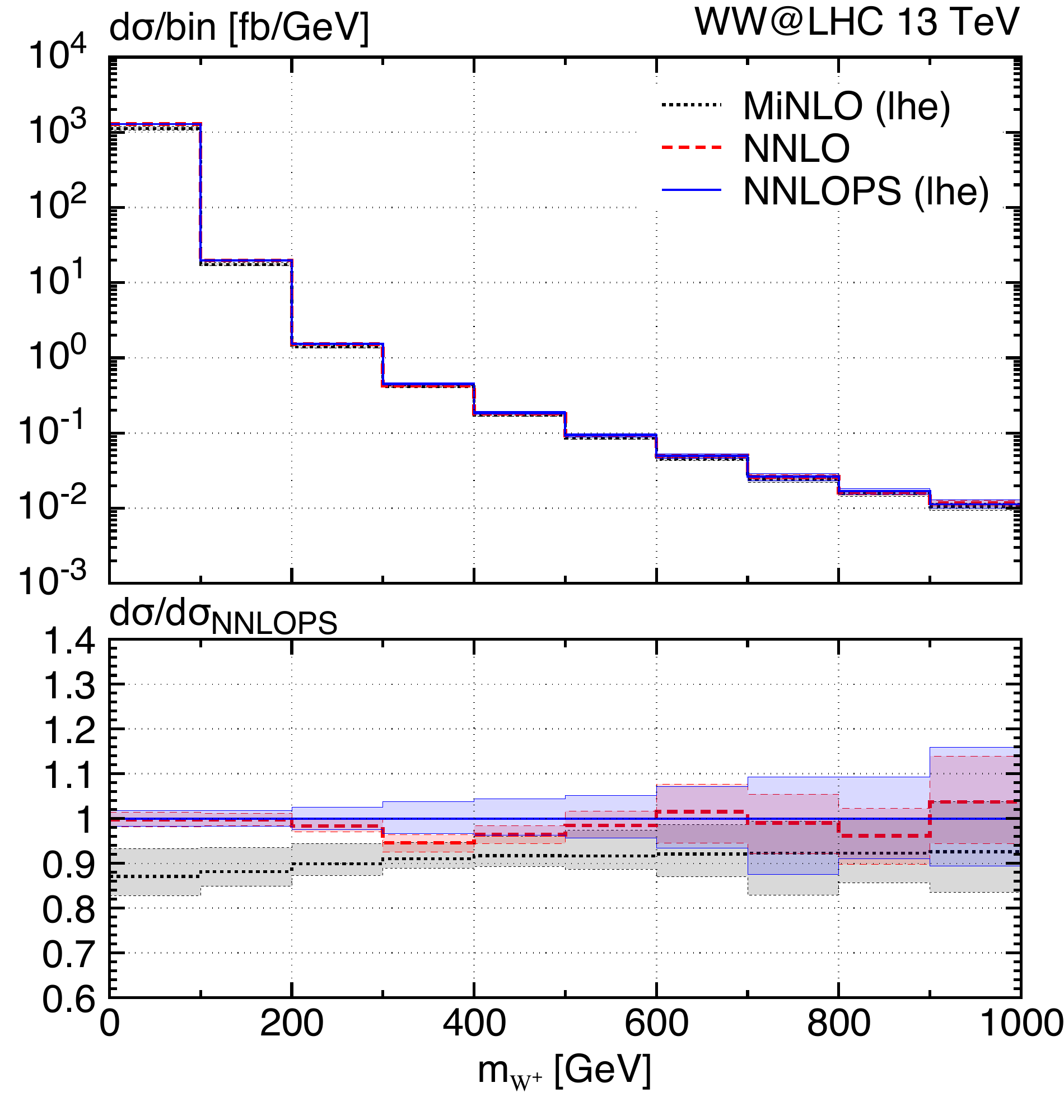} \\[-1em]
\hspace{0.6em} (a) & \hspace{1em}(b)
\end{tabular}
\caption[]{\label{fig:mwp}{Same as \fig{fig:ptwmywwdywpwm}, but for the invariant mass of the $W^+$ boson $\mwp$ in two different regions: (a) around the $W$-mass peak, $\mwp\in[50,100]$\,GeV, and (b) including off-shell regions, $\mwp\in[0,1000]$\,GeV.}}
\end{center}
\end{figure}

\begin{figure}[tp]
\begin{center}
\hspace*{-0.15cm}
\begin{tabular}{ccc}
\includegraphics[trim = 7mm -7mm 0mm 0mm, width=.226\textheight]{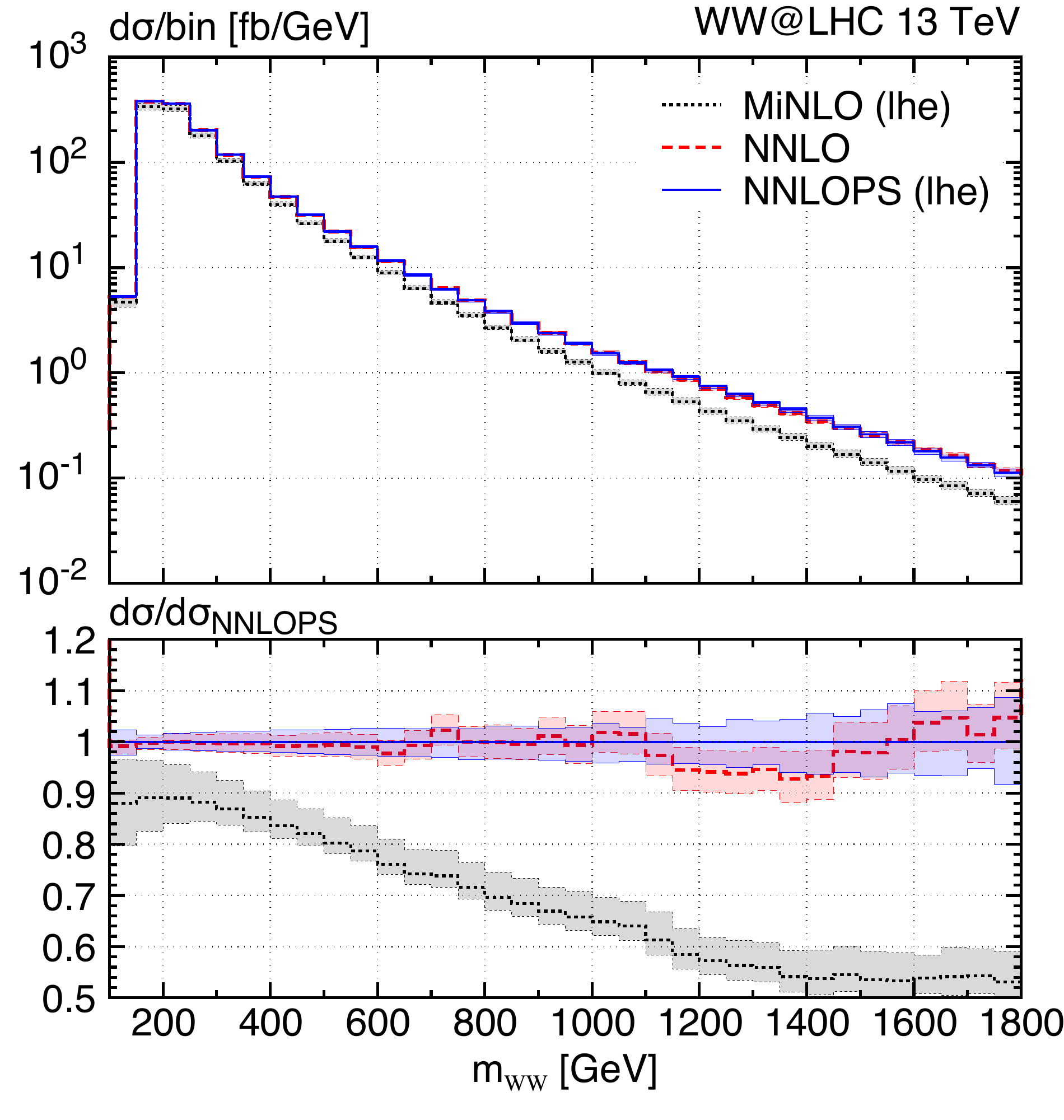} &\hspace{-0.6cm}
\includegraphics[trim = 7mm -7mm 0mm 0mm, width=.226\textheight]{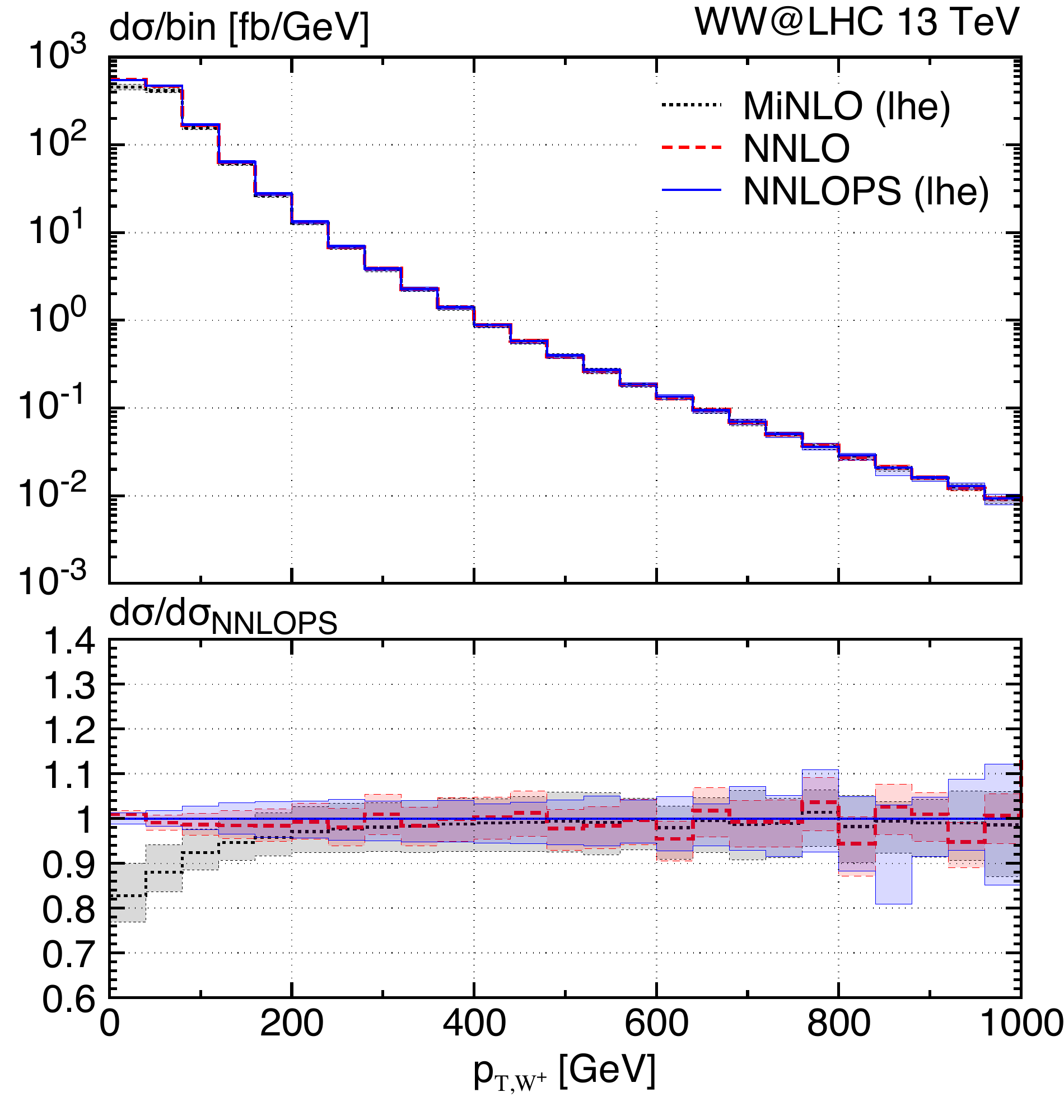} &\hspace{-0.6cm}  \includegraphics[trim = 7mm -7mm 0mm 0mm, width=.226\textheight]{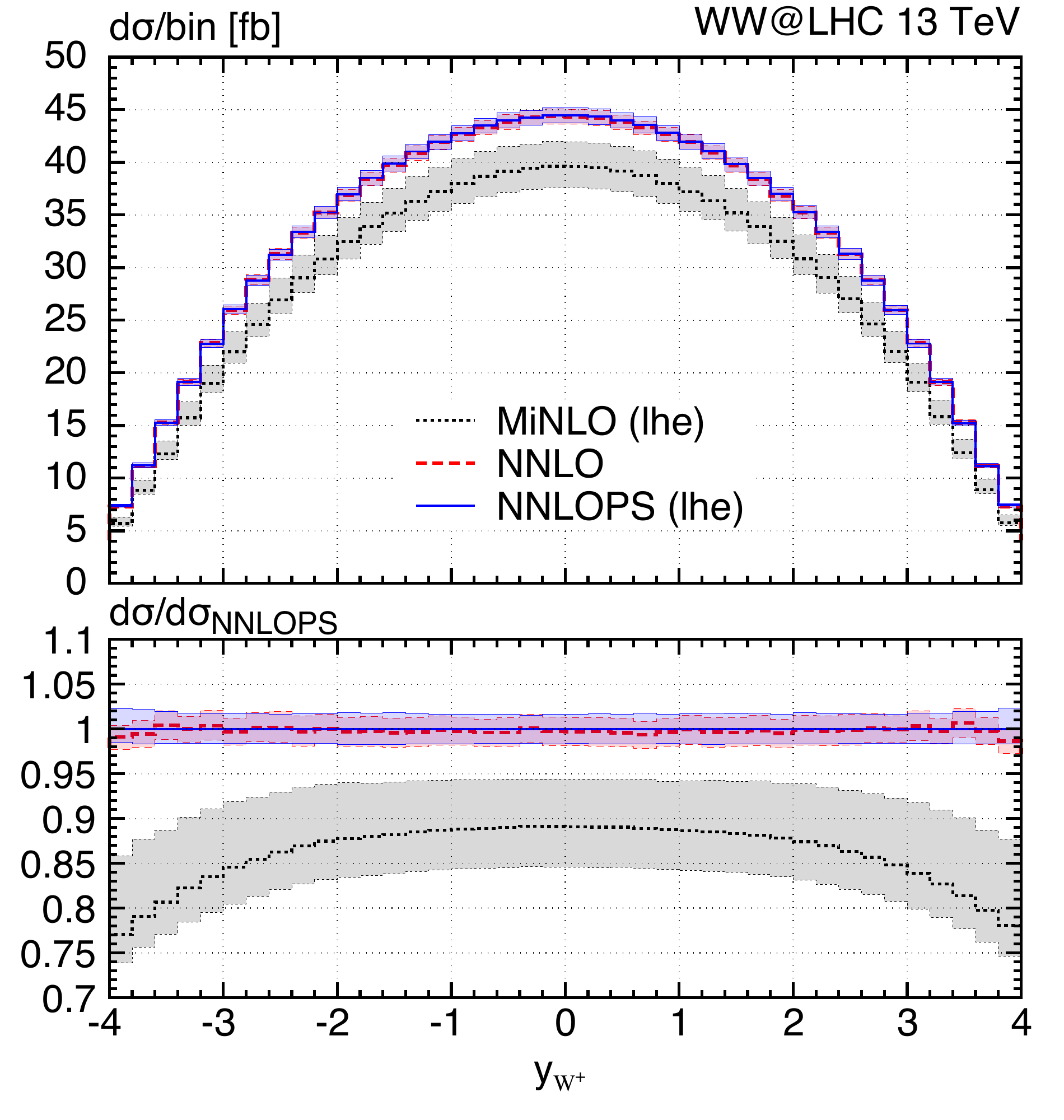} \\[-1em]
 (a) & \hspace{-1em}(b) & \hspace{-1em}(c)\\[0.3cm]
 \includegraphics[trim = 7mm -7mm 0mm 0mm, width=.226\textheight]{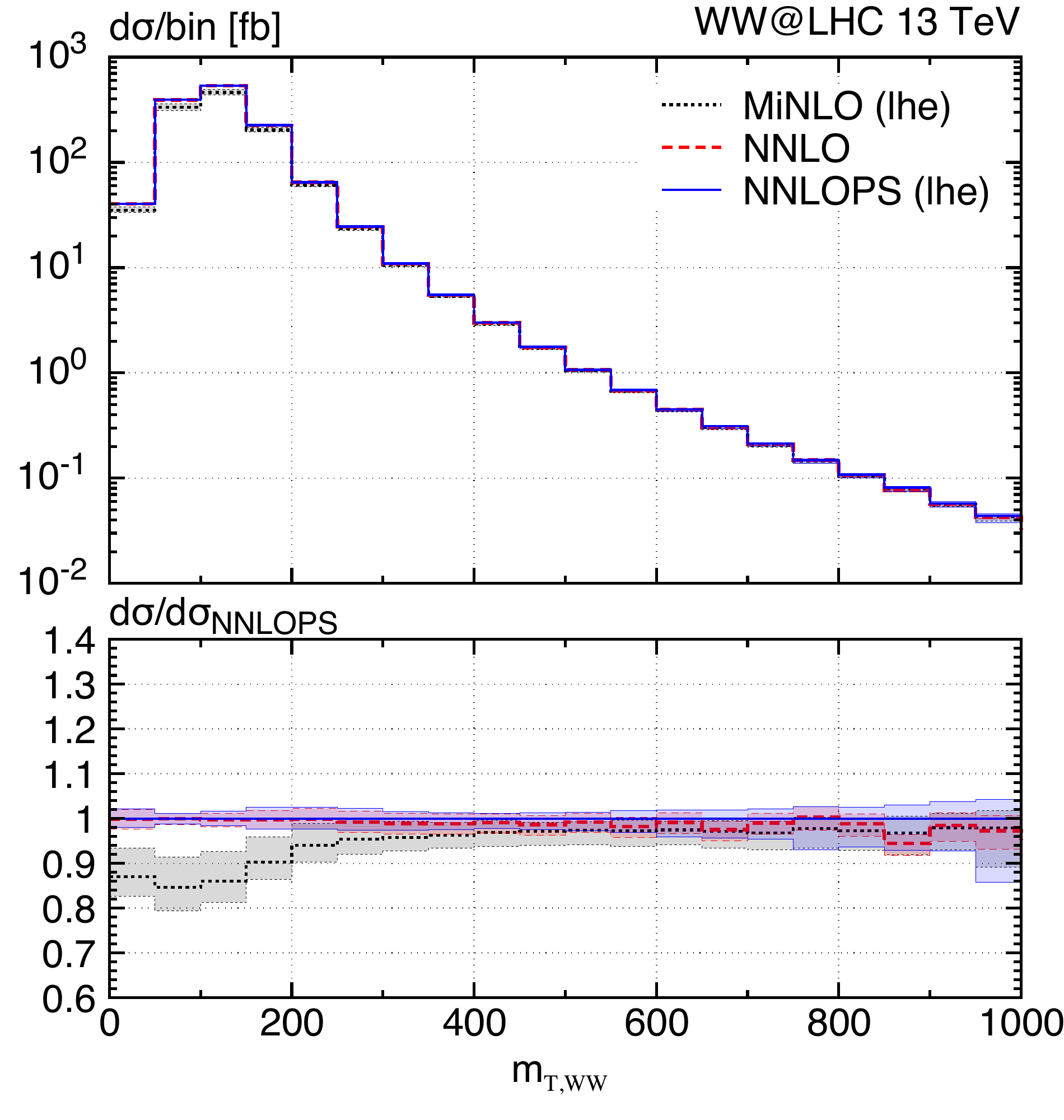} &\hspace{-0.6cm}
\includegraphics[trim = 7mm -7mm 0mm 0mm, width=.226\textheight]{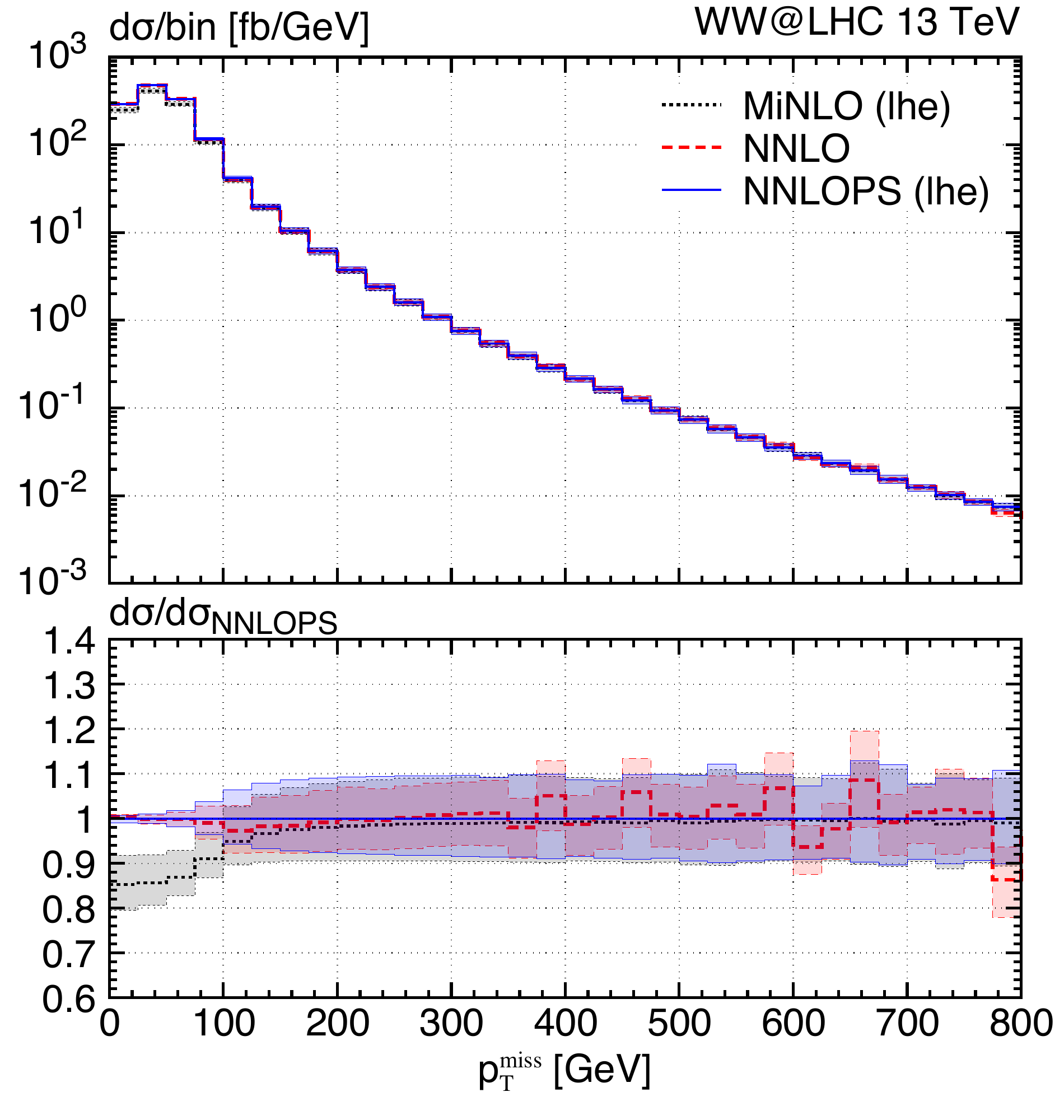} &\hspace{-0.6cm}  \includegraphics[trim = 7mm -7mm 0mm 0mm, width=.226\textheight]{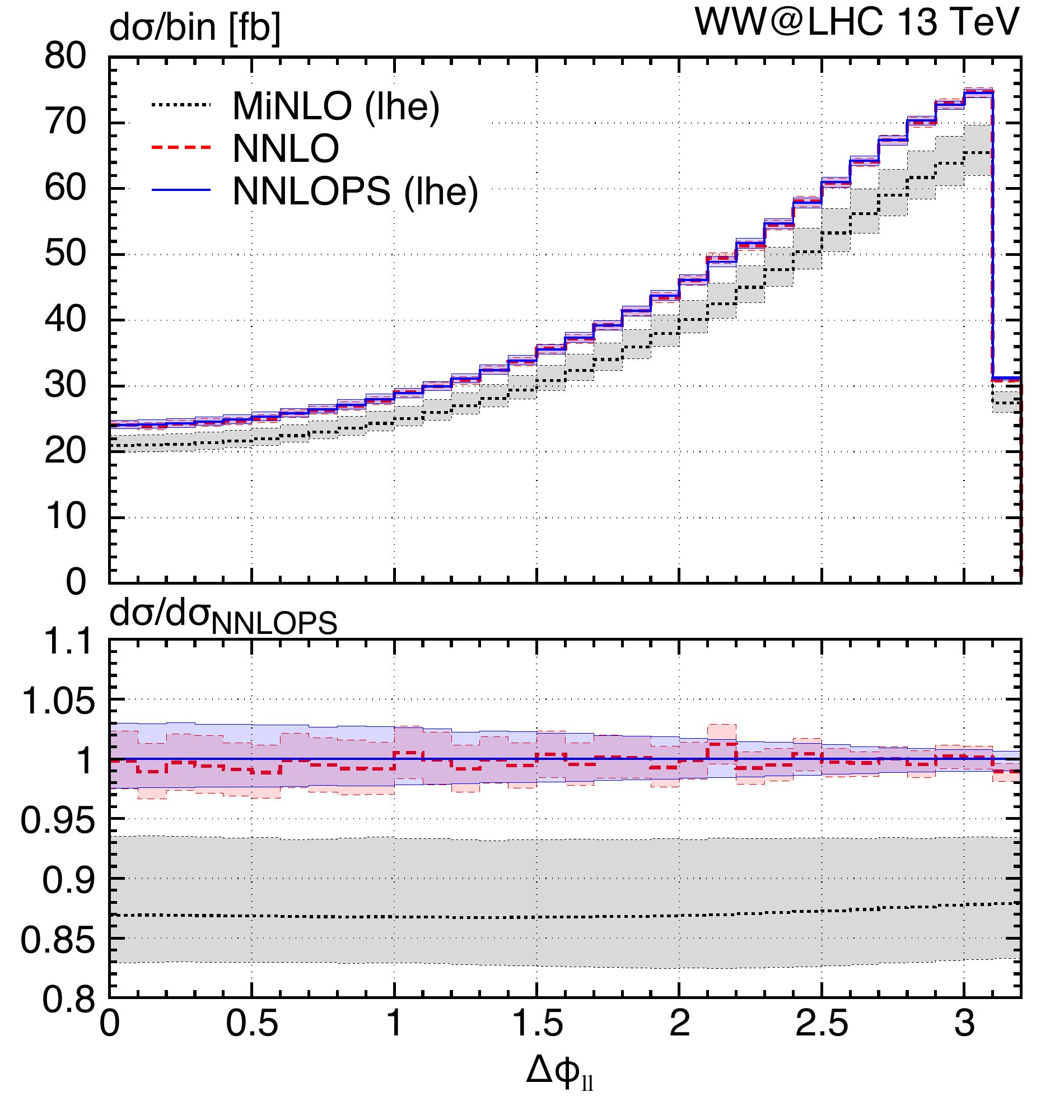} \\[-1em]
 (d) & \hspace{-1em}(e) & \hspace{-1em}(f)
\end{tabular}
\caption[]{\label{fig:allother}{Same as \fig{fig:ptwmywwdywpwm}, but
    for observables which have not been used in
    \eqn{eq:initialphasespace} to define a basis of the Born-level
    phase space: (a) invariant mass of the \ww{} pair $\mww$, (b)
    transverse momentum $\ptwp$ and (c) rapidity $\ywp$ of $W^+$, (d)
    transverse mass of the \ww{} pair{} $\mtww$ defined \eqn{eq:mTWW},
    (e) missing transverse momentum $\ptmiss$ and (f) lepton
    separation $\dphill$.}}
\end{center}
\end{figure}

Let us add at this point that we have also tried to only use the
three-dimensional reweighting in $\dd\Phi_{W^+_*W^-_*}$ without
using the CS angles by replacing
\begin{align}\label{eq:initialphasespace3D}
\frac{\dd\sigma}{\dd\born}\equiv\frac{\dd\sigma}{\dd\Phi_{W^+_*W^-_*}}=\frac{\dd^3\sigma}{\dptwm\dyww\ddywpwm}
\end{align}
in Eq.~\eqref{eq:W}. As expected, this reduces
some statistical fluctuations. In fact, we found that excluding the CS
angles the NNLO distributions are still very well reproduced by the
NNLOPS sample. Of all one-dimensional distributions we considered,
only \thetap{} and \thetam{} show a mildly different shape (at the
few-percent level) in this case.  We therefore provide the reweighting
without CS angles as an option in our code, while keeping the
application of the full expression in \eqn{eq:double} the default in
the code and throughout this paper.  One must bear in mind that as
soon as double differential distributions in angular observables of
the leptons are considered the validity of the application of the
reweighting without CS angles may be limited.

The only observables in our definition of the Born phase space, see
\eqn{eq:initialphasespace}, which remain to be validated are the
invariant masses of the two $W$ bosons. We first recall that for
reasons of complexity we excluded them from the Born-level variables
in the reweighting procedure by assuming them to feature flat
higher-order corrections. Indeed, \fig{fig:mwp}\,(a) confirms this to
be an appropriate assumption in the peak-region of the spectrum, where
the bulk of events is situated and the agreement of the NNLO with the
NNLOPS distributions is close to perfect. Even in the phase-space
areas where the two $W$ bosons become far off-shell the NNLOPS result
deviates by less than 5\% from the NNLO curve, see
\fig{fig:mwp}\,(b). This discrepancy is at the level of the
statistical uncertainty in these regions.  We note that we only show
the $\mwp$ distribution in that figure, because the $\mwm$ results are
practically identical.

We conclude this section by studying distributions which have not been
used in the parametrization of our phase-space definition in
\eqn{eq:initialphasespace}.  This is important in order to convince
oneself that, beyond the observables used for the reweighting, our
procedure reproduces correctly the NNLO cross section for other
distributions.  \fig{fig:allother} shows corresponding plots for the
invariant mass of the \ww{} pair, the transverse momentum and the
rapidity of $W^+$, the transverse mass of the \ww{} pair{} defined as
\begin{equation}\label{eq:mTWW}
  \mtww{} = \sqrt{\left(\Etlone+\Etltwo+\ptmiss\right)^2-\left({\bf p}_{T,l_1}+{\bf p}_{T,l_2}+
    {\mathbf\ptmiss}\right)^2}\,,
\end{equation}
the missing transverse momentum, as well as the separation in the
azimuthal angle $\Delta \phi_{ll}$ of the two charged leptons. We
stress that we have considered a large number of relevant observables
and that we have picked a representative set of distributions here.
In particular, $\mww{}$ is the one we found most sensitive to
statistical effects.  Looking at \fig{fig:allother} it is clear that
our reweighting procedure allows us to promote the \MINLO{} sample to
have NNLO accuracy for Born-level observables: even in regions where
the \MINLO{} and NNLO curves are far apart the NNLOPS predictions are
perfectly consistent with the NNLO ones.
This is particularly evident in the region of large $m_{WW}$, where
the \MINLO{} result is almost a factor 2 below the NNLO one. 
Not only the central predictions are in reasonable agreement, also the NNLOPS
uncertainty bands are of the expected size, being close to the NNLO
ones.
The fact that upon reweighting NNLOPS agrees with the NNLO
for Born-level observables verifies that the parametrization through CS 
angles, which, strictly speaking, is applicable only to the double-resonant 
\ww{} contributions, is an excellent approximation in general.
 
One might wonder what is the reason for the large difference between
\MINLO{} and NNLO at large $m_{WW}$. It can be easily traced back to
the different scale choice in the NNLO calculation and in \MINLO{}. In
fact, \MINLO{} uses effectively $m_{WW}$ as a primary scale, which
becomes a very hard scale in the tail of the $m_{WW}$ distribution. On
the contrary, the dynamic scale choice in the NNLO calculation, see
\eqn{eq:dynscale_nnlo}, is not sensitive to the rapidity
distance between the two $W$-bosons. Hence, it can be much smaller
than $m_{WW}$ when there is a large rapidity separation between the
$W$ bosons.  As expected, similar effects can be observed also in the
rapidity distribution of the $W^+$ boson and the rapidity difference
between the two $W$ bosons, see \fig{fig:allother}\,(c) and
\fig{fig:ptwmywwdywpwm}\,(c), respectively.
We note that the difference between NNLO and \MINLO{} would be
about $20$\% smaller in the tail of the $\dywpwm$ and $\mww{}$ 
distributions if $\mu=\mww{}$ was used.
It is not clear which scale choice is more appropriate for \ww{}
production.  For instance, if the $W$ bosons originate from the
$s$-channel decay of an off-shell $Z$ boson, the invariant mass
(possibly transverse mass) of the $W$-boson pair is the natural
choice.  However, through the $t$-channel diagrams it is also possible
to emit one $W$ boson as initial-state (soft, large rapidity)
radiation and a second $W$ boson as a standard hard Drell Yan
interaction. In this case, the choice done in the NNLO calculation
would be more appropriate. Since of course all topologies interfere,
it seems hard to argue in favour of any of the two scale
choices.

\section{Results}
\label{sec:results}

In this section we present NNLOPS-accurate predictions for the
processes $pp\to e^\mp \nu_e\, \mu^\pm \nu_\mu +X$.  Hence, we
consider the production of two different-flavour leptons together with 
the two corresponding neutrinos and its charge-conjugated
process. After defining our general setup, we discuss rates and
distributions both in the inclusive phase space and in presence of
fiducial cuts.

\subsection{Input parameters and fiducial cuts}\label{sec:input}

We study predictions at the 13\,TeV LHC. The $G_\mu$ scheme is
employed for EW parameters and the complex-mass
scheme~\cite{Denner:2005fg} for EW decays of the $W$ bosons. Thus,
complex $W$- and $Z$-boson masses are used and the EW mixing angles
are defined as
$\cos\theta_W^2=(m_W^2-i\Gamma_W\,m_W)/(m_Z^2-i\Gamma_Z\,m_Z)$ and
$\alpha=\sqrt{2}\,G_\mu m_W^2\sin^2\theta_W/\pi$.  The input
parameters are set to the PDG~\cite{Olive:2016xmw} values: $G_F =
1.16639\times 10^{-5}$\,GeV$^{-2}$, $m_W=80.385$\,GeV,
$\Gamma_W=2.0854$\,GeV, $m_Z = 91.1876$\,GeV, $\Gamma_Z=2.4952$\,GeV,
and $m_H = 125.0$\,GeV. We obtain a branching fraction of
$\textrm{BR}(W^\pm \rightarrow \ell^\pm\nu_\ell) = 0.108987$ from
these inputs for the $W$-boson decay into massless leptons.  The CKM
matrix is set to unity, which, because of unitarity and because we
consider only massless external quarks, is almost equivalent to using
the full Cabibbo matrix.\footnote{An approximation is made in the real correction
when the two $W$ bosons are emitted from two different fermion lines,
one in the initial state and one in the final state. First of all,
these contributions are very small, as they contain a gluon propagator
in the $s$-channel which is pushed far off-shell by the $W$ boson
emitted off the final state fermion line. Additionally, 
these effects are further suppressed by the heavy-flavour PDFs.
Hence, replacing the CKM matrix by the unit matrix is a very good approximation.}
As outlined in \sct{sec:top}, we use the 4FS with massive bottom
quarks throughout and consistently remove top-quark contamination by
dropping all partonic subprocesses with real bottom-quark emissions,
which are separately IR finite. The on-shell top- and bottom-quark
masses are set to $m_t = 173.2$\,GeV and $m_b =
4.92$\,GeV.\footnote{We note that the contributions involving massive
fermion loops, which include also the exchange of a Higgs boson 
and appear starting from ${\cal O}(\as^2)$, 
are accounted for through the reweighting to the NNLO.}
We use the NNPDF3.0~\cite{Ball:2014uwa} $n_f=4$ PDF sets with the
corresponding value of the strong coupling constant.\footnote{The
  strong coupling constant of the $n_f=4$ NNPDF set is derived from
  the standard variable-flavour-number PDF set with
  $\alpha_s^{(5\mathrm{F})}(M_Z)=0.118$ using an appropriate backward
  and forward evolution with five and four active flavours,
  respectively. This results in values of 0.1136, 0.1123 and 0.1123
  for $\alpha_s^{(4\mathrm{F})}(M_Z)$ at LO, NLO and NNLO.}  As usual,
for the fixed-order results, we choose N$^n$LO PDF sets in accordance
with the perturbative order under consideration, while the evolution
of $\as$ is done at $(n+1)$-loop order. In the \WWJMINLO{} simulation
NNLO PDFs are used. We use dynamical renormalization ($\mu_R$) and
factorization ($\mu_F$) scales: for the NNLO computation the average
of the transverse masses of the two $W$ bosons is chosen as a central
scale:
\begin{align}
\label{eq:dynscale_nnlo}
\mu_R=\mu_F=\mu_0\equiv \frac12\,\left(\sqrt{m_{e^-\bar\nu_e}^2+p^2_{T,e^-\bar\nu_e}}+\sqrt{m_{\mu^+\nu_\mu}^2+p^2_{T,\mu^+\nu_\mu}}\right)\,,
\end{align}
while, upon integration over all radiation, the scales in the \MINLO{}
generator effectively reduce to an \mww{}-like scale.  As described in
\sct{sec:validation}, uncertainties from missing higher-order
contributions are estimated from the customary $7$-point variation,
while keeping the $\mu_R$ and $\mu_F$ values correlated in the NNLOPS
reweighting factor. All showered results are obtained through matching
to the \PYTHIA{8} parton shower \cite{Sjostrand:2014zea}. Results are
shown at parton level, without hadronization or underlying-event
effects.

\renewcommand\arraystretch{1.5}
\begin{table}[t]
\begin{center}
\begin{tabular}{l | c  }
\toprule
\bf lepton cuts
& $\ptlep > 25$\,GeV, \quad $\etalep<2.4$, \quad $\mll>10$\,GeV\\
\bf lepton dressing & add photon FSR to lepton momenta with $\Delta R_{\ell\gamma}<0.1$\\[-0.2cm]
&{\it (our results do not include photon FSR, see text)}\\
\bf neutrino cuts
& $\ptmiss > 20$\,GeV, \quad $\ptmissrel>15$\,GeV\\
\multirow{3}{*}{\bf jet cuts}
 & anti-$k_T$ jets with $R=0.4$;\\
& $N_{\rm jet} = 0$ for $\ptjet>25$\,GeV, $|\eta_j|<2.4$ and $\Delta R_{ej}<0.3$\\
& $N_{\rm jet} = 0$ for $\ptjet>30$\,GeV, $|\eta_j|<4.5$ and $\Delta R_{ej}<0.3$\\
\bottomrule
\end{tabular}
\end{center}
\renewcommand{\baselinestretch}{1.0}
\caption{\label{tab:cuts} Fiducial cuts used in the \ww{} analysis by
  ATLAS at 13 TeV \cite{Aaboud:2017qkn}. See text for details.}
\vspace{0.75cm}
\end{table}
\renewcommand\arraystretch{1}

In \tab{tab:cuts} we summarize the set of cuts used in the definition
of the fiducial phase space. They involve standard cuts on the
transverse momentum (\ptlep) and pseudo-rapidity (\etal) of the
charged leptons as well as a lower threshold on the invariant-mass of
the dilepton pair (\mll).  Lepton dressing with QED final-state radiation (FSR) 
is not included in the fiducial results shown in this paper. However, we discuss
the general effects of its simulation with \PYTHIA{8} below.
A typical minimal
requirement on the missing transverse momentum (\ptmiss) is
supplemented by a cut on the so-called relative missing transverse
momentum (\ptmissrel), which denotes the component of the \ptmiss{}
vector perpendicular to the direction of the closest lepton in the
azimuthal plane:
\begin{align}
\ptmissrel = \left\{\begin{array}{ll}\ptmiss\cdot \sin|\Delta\phi| & \; \;{\rm for}\; \Delta\phi<\pi/2\,, \\ \ptmiss & \; \;{\rm for}\; \Delta\phi>\pi/2\,,\end{array}\right.
\end{align}
where $\Delta\phi$ denotes the azimuthal angle between the \ptmiss{}
vector and the nearest lepton.  Finally, there is a two-folded
jet-veto requirement: jets are rejected for a softer \ptjet{}
threshold in a narrow pseudo-rapidity ($\eta_j$) range, while slightly
harder jets are vetoed also in a wider pseudo-rapidity range.  This
setup follows precisely the definition of the fiducial volume employed
in the ATLAS 13 TeV \ww{} measurement of \citere{Aaboud:2017qkn},
which we will compare to below. We refer to this  default set of
fiducial cuts, which include the jet-veto requirements, as {\tt
  fiducial-JV}.
In the following, it will be instructive to also consider the same
fiducial setup, but without any restriction on the jet activity, which
we denote as {\tt fiducial-noJV} in the respective figures.

We refrain from showing results including charged leptons dressed with
photon FSR in the following, in order to allow for a more direct
comparison between NNLO and the showered results.  Besides, a proper
treatment is closely tied to the specific choices made by the
experimental collaborations.  Nevertheless, for completeness we have
simulated such effects by generating photon emissions with \PYTHIA{8},
and successively considering dressed leptons, i.e.\ we added to their
momentum all photon momenta in a cone of $\Delta
R_{\ell\gamma}<0.1$. By and large, the impact of photon FSR is at the
level of a few percent. In particular, the cross section in the
fiducial phase space is reduced by about 2\%.  Relatively large
effects ($>10\%$) are found only in distributions where it is
expected, such as the invariant mass of each of the two $W$ bosons or
the charged lepton transverse momenta.  We stress that with our NNLOPS
computation the experimentalists have now a tool to produce NNLO
accurate predictions, and, at the same time, to consistently include
lepton dressing through photon FSR as obtained by a parton shower.

All fiducial results in this section have been obtained for the $pp
\to \muenn$ process, while multiplying them with a factor of two to
account for the charge-conjugated process ($pp \to
\emunn$).\footnote{We have explicitly checked that the minor asymmetry
  in the electron and muon cuts, which appears only in the
  electron-jet separation of the jet-veto definition, has a completely
  negligible impact. Our procedure can therefore be considered to
  provide the exact result for the sum of the two (charge-conjugated)
  processes.} As pointed out in the introduction, contributions which
stem from the loop-induced $gg$ channel and enter the NNLO corrections
to \ww{} production are disregarded throughout this work.  We employ
this simplification to perform a clean study of the newly computed
NNLOPS effects.  For a fully consistent comparison to data all
contributions should be combined with correlated scale variations.

\subsection{Inclusive and fiducial rates}\label{sec:rates}

\renewcommand\arraystretch{1.5}
\begin{table}[ht]
\begin{center}
\resizebox{\columnwidth}{!}{%
\begin{tabular}{l | c c c}
\toprule
$q\bar{q}$ (no loop$^2$ $gg$)
& $\sigma_{\rm incl}(pp\to\ww)$ [pb]
& $\sigma_{\rm fid}(pp\to e^\mp \nu_e\, \mu^\pm \nu_\mu)$ [fb]
& A = $\sigma_{\rm fid}$/$\sigma_{\rm incl}$ [$\%$]\\

\midrule
LO  & $70.66(1)_{-6.2\%}^{+5.1\%}$ & $440.5(0)_{-7.1\%}^{+6.0\%}$ & 0.623\\
NLO & $99.96(3)_{-2.8\%}^{+3.5\%}$ & $411.8(1)_{-2.3\%}^{+2.7\%}$ & 0.412\\
NNLO & $110.0(1)_{-1.6\%}^{+1.6\%}$ & $413.1(2)_{-0.7\%}^{+1.0\%}$ & 0.376\\
\MINLO{} & $96.05(1)_{-4.9\%}^{+7.1\%}$ & $359.6(1)_{-8.3\%}^{+5.4\%}$ & 0.374\\
NNLOPS & $110.2(2)_{-1.6\%}^{+1.7\%}$ & $413.0(2)_{-2.3\%}^{+2.2\%}$ & 0.375\\
\midrule
ATLAS$-gg$ \cite{Aaboud:2017qkn}&$
124.7 \pm 5\,{\rm (stat)} \pm 13\,{\rm (syst)} \pm 3\,{\rm (lumi)}$ & $
473\pm 20\,{\rm (stat)} \pm 50\,{\rm (syst)} \pm 11\,{\rm (lumi)}$ & 0.379 \\
CMS$-gg$ \cite{CMS:2016vww}& $
108.5 \pm 5.8\,{\rm (stat)} \;^{\pm 5.7 (\rm exp.\,syst)}_{\pm 6.4 (\rm theo.\,syst)} \,\pm 3.6\,{\rm (lumi)}$ & --- & ---\\
\bottomrule
\end{tabular}}
\end{center}
\renewcommand{\baselinestretch}{1.0}
\caption{\label{tab:rates} Cross sections for inclusive \ww{}
  production and $e^\mp \nu_e\, \mu^\pm \nu_\mu$ production with
  fiducial cuts in various approximations compared to data. At NNLO,
  all corrections to $q\bar{q}$-bar induced \ww{} production are taken
  into account up to $\mathcal{O}(\alpha_s^2)$, while excluding the
  loop-induced $gg$ contribution. The central values of the
  experimental results have been corrected by subtracting the
  $\mathcal{O}(\alpha_s^3)$ theoretical prediction for the
  (non-resonant) $gg$ component \cite{Caola:2015rqy} as quoted in the
  ATLAS analysis \cite{Aaboud:2017qkn}. In contrast to CMS, ATLAS
  includes resonant Higgs bosons decaying to \ww{} pairs in their
  \ww{} signal measurement.  The theoretical predictions of this
  additional $gg$-initiated contribution in the inclusive
  \cite{deFlorian:2016spz} and fiducial
  \cite{deFlorian:2016spz,Bagnaschi:2011tu} phase space as quoted in
  the ATLAS analysis \cite{Aaboud:2017qkn} have also been removed from
  the central ATLAS result.\protect\footnotemark{}}
\end{table}
\footnotetext{We note that the prediction used for the inclusive Higgs
  results includes the N3LO cross section in the heavy-top limit of
  \citeres{Mistlberger:2018etf,Anastasiou:2016cez,Anastasiou:2015ema}
  and quark-mass effects
  \cite{Djouadi:1991tka,Marzani:2008az,Harlander:2009my,Pak:2009dg,Bagnaschi:2011tu,Mantler:2012bj,Harlander:2012hf,Neumann:2014nha,Bagnaschi:2015bop,Hamilton:2015nsa}. The
  fiducial acceptance for the resonant Higgs contributions in
  \citere{Aaboud:2017qkn} has been computed with the \POWHEG{}
  implementation \cite{Bagnaschi:2011tu}, but equivalent tools using
  the \MCatNLO{} approach \cite{Mantler:2015vba}, or even more
  sophisticated merging \cite{Buschmann:2014sia,Frederix:2016cnl} and
  NNLOPS \cite{Hamilton:2013fea,Hamilton:2015nsa,Hoche:2014dla}
  predictions could have been used to determine the acceptance. Given
  the minor impact ($\sim 2\%$) of resonant Higgs contributions in the
  fiducial phase space, which is due to the applied jet veto, a more
  precise modelling of the Higgs contributions is not required.}
\renewcommand\arraystretch{1}

In \tab{tab:rates}, we report results for integrated cross sections,
both fully inclusive and with fiducial cuts.
The inclusive \ww{} results have been obtained from the full off-shell
computation of the leptonic process in \eqn{eq:process} by simply
dividing out the branching fraction of the $W\to \ell\nu$ decays.
The numbers inside the brackets after the central prediction are the
numerical errors, while the percentages reflect the uncertainties due
to scale variations.  For reference, we also quote the acceptance
obtained from the ratio of the central prediction for the fiducial
cross section over the inclusive one. The predicted rates are provided
in various approximations, with NNLOPS being our best prediction.  All
the available experimental results at 13 TeV by ATLAS and CMS are
quoted in the same table. Since we omit loop-induced $gg$
contributions in the $\mathcal{O}(\alpha_s^2)$ calculations, the
central values of the measured cross sections have been corrected by
removing the respective theory prediction of the $gg$ component to
facilitate a meaningful comparison, as detailed in the caption of the
table.  The main conclusions that can be drawn from the table are the
following:
\begin{enumerate}
\item Radiative corrections on the inclusive cross section are
  large. They amount to $+41.4\%$ at NLO and are still $+10.0\%$ at
  NNLO. The \MINLO{} result is, in accordance with its formal
  accuracy, very close to the inclusive NLO rate. By construction,
  NNLOPS yields the inclusive NNLO cross section up to statistics.
\item In the fiducial phase space the situation is quite
  different. Radiative corrections are much smaller and even negative
  at NLO. They amount to $-6.5\%$ at NLO and $+0.32\%$ at NNLO. In
  fact, looking only at the $\mathcal{O}(\alpha_s^2)$ coefficient,
  i.e.\ comparing against NLO computed with NNLO PDFs (referred to as
  NLO$^\prime$) which yields $\sigma_{\rm fid}^{\rm
    NLO'}=424.6(1)_{-2.1\%}^{+2.5\%}$, one realizes that the
  $\mathcal{O}(\alpha_s^2)$ contributions from the NNLO matrix
  elements are actually also negative.  We stress that these findings
  are caused entirely by the restrictions of jet activity in the
  fiducial phase space.  If we remove the jet-veto requirements,
  i.e.\ consider the {\tt fiducial-noJV} setup, radiative corrections
  are similar to the inclusive case.
\item When comparing \MINLO{} and NLO results, it appears quite
  surprising that the two fiducial cross sections turn out to be so
  different, despite the fact that they are practically identical in  
  the {\tt fiducial-noJV} setup. However, it was pointed out some time
  ago \cite{Monni:2014zra} that the \POWHEG{} generator tends to
  underestimate the jet-vetoed cross section for \ww{} production,
  which seems to persist also in its \MINLO{} extension. When \MINLO{}
  is reweighted to the NNLO this deficit in the fiducial cross section
  disappears. The reasons for why this happens are twofold: first, the
  fiducial cross section without a jet-veto is about 10\% higher at
  NNLO than for \MINLO{}; second, the jet-veto efficiency predicted by
  \MINLO{} is $\sim 5\%$ lower than at NNLOPS for relevant jet-veto
  cuts (see \sct{sec:jetveto}). It shall be noted, however, that the
  reasonable size of the fiducial cross section at NLO is
  accidental. It is caused by the interplay of the small cross section
  without a jet veto and a poor modelling of the jet-veto efficiency.
  This is apparent considering the acceptance in the last column,
  which is rather similar among NNLO, \MINLO{} and NNLOPS, but quite
  higher at NLO.
\item It is interesting to note that the fiducial NNLOPS result is
  identical with the NNLO cross section, despite the fact that its
  description of jet-veto logarithms is more accurate at low jet-veto
  scales. We postpone a detailed discussion to \sct{sec:jetveto},
  where we analyze the cross section as a function of the jet-veto
  cut. We note, however, that the perturbative uncertainties of the
  NNLOPS result are more realistic than at NNLO. As expected they are
  larger in the restricted phase space than in the fully inclusive
  one, while the opposite is the case at NNLO.
\item Despite the rather small QCD corrections in the fiducial phase
  space, only beyond NLO a reliable prediction for the fiducial
  acceptance is obtained.
\item As expected, scale uncertainties successively decrease upon
  inclusion of QCD perturbative corrections. At LO and NLO, they
  underestimate, however, the actual size of missing higher-order
  terms in the case of the inclusive cross section.
\item The agreement between the NNLO(PS) predictions and the measured
  cross sections is quite good.  This is particularly true for the
  inclusive cross section, which is fully consistent with the CMS
  measurement within the statistical uncertainty, and agrees also with
  the ATLAS one as soon as systematics are taken into
  account. Clearly, the tension found in some early 8 TeV \ww{}
  measurements of the inclusive cross section
  \cite{Chatrchyan:2013oev,ATLAS-CONF-2014-033} does not persist at 13
  TeV with state-of-the-art theoretical predictions. For the fiducial
  cross-section measurement by ATLAS the relative difference to the
  NNLOPS result is somewhat higher, but still within the quoted
  uncertainties.
\end{enumerate}

\subsection{Jet-vetoed cross section and impact of the parton shower}\label{sec:jetveto}

We now consider the integrated cross section with a jet-veto as a
function of the jet-veto cut ($\ptjetveto$) defined as
\begin{align}
\sigma(\ptjetone < \ptjetveto) = \int_0^{\ptjetveto}
\dd\ptjetone\,\frac{\dd\sigma}{\dd\ptjetone}= \sigma_{\rm int} -
\int_{\ptjetveto}^\infty
\dd\ptjetone\,\frac{\dd\sigma}{\dd\ptjetone}\,.
\end{align}
$\sigma_{\rm int}$ denotes the cross section integrated over all
\ptjetone{}. In addition to the jet-veto requirement any IR-safe set
of cuts may be imposed on the cross section in the previous
equation. We further define the jet-veto efficiency as
\begin{align}
\varepsilon(\ptjetveto)=\sigma(\ptjetone < \ptjetveto)/\sigma_{\rm
  int}.
\end{align}

\begin{figure}[tp]
\begin{center}
\begin{tabular}{cc}
\hspace*{-0.17cm}
\includegraphics[trim = 7mm -7mm 0mm 0mm, width=.33\textheight]{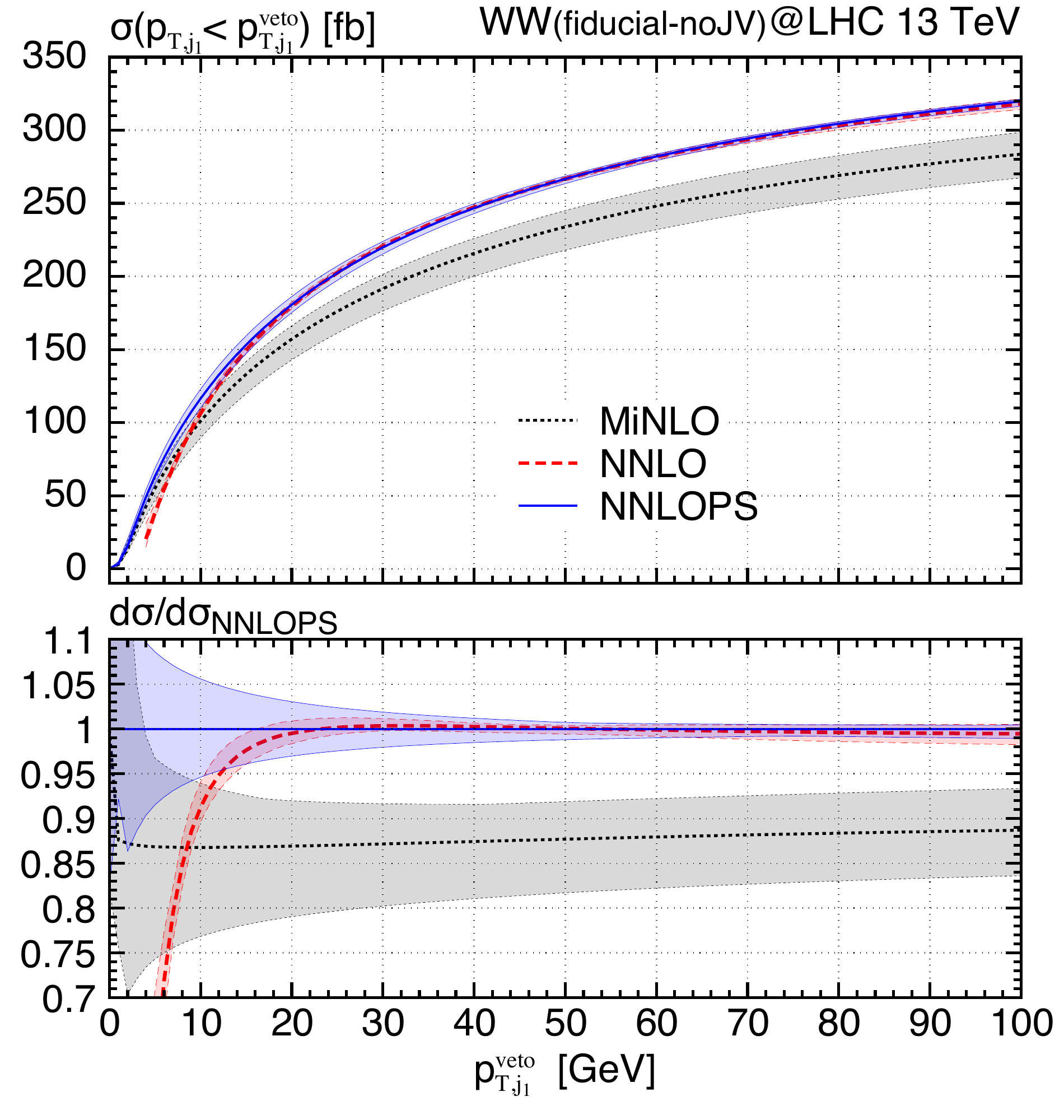} &
\includegraphics[trim = 7mm -7mm 0mm 0mm, width=.33\textheight]{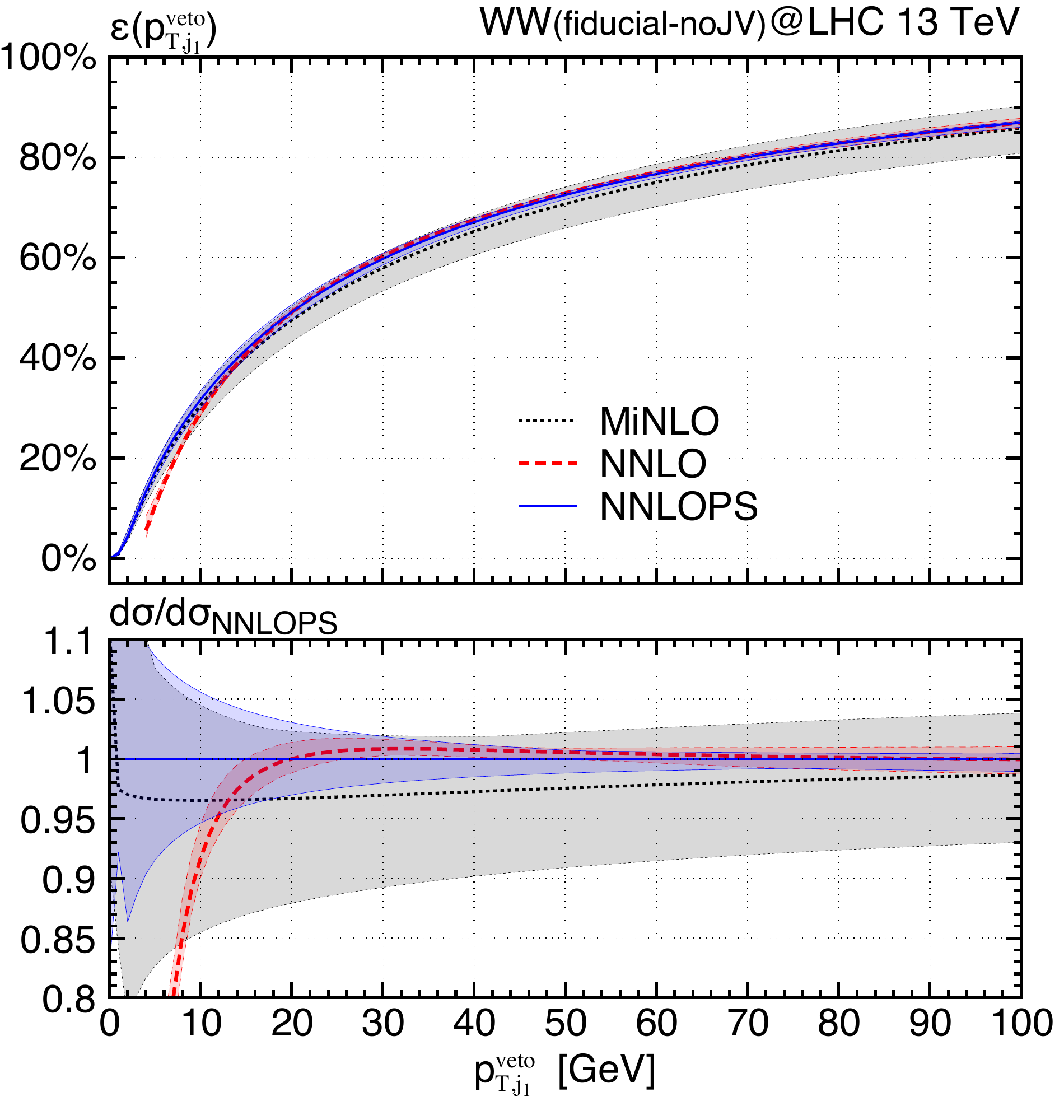} \\[-1.3em]
\hspace{0.6em} (a) & \hspace{1em}(b)
\end{tabular}
\caption[]{\label{fig:jetveto}{Comparison of \MINLO{} (black, dotted),
    NNLO (red, dashed) and NNLOPS (blue, solid) predictions in the
    fiducial phase space as a function of \ptjetveto{} for (a) the
    cross section and (b) the jet-veto efficiency.}}
\end{center}
\end{figure}

\fig{fig:jetveto} depicts both the jet-vetoed cross section and the
jet-veto efficiency in the fiducial phase space. We point out that the
relative behaviour of the curves is practically identical in the fully
inclusive phase space, which is why we refrain from discussing them
separately, and that the general conclusions drawn here also apply in
the inclusive case. The figures throughout this section follow the
same pattern as in \sct{sec:validation}, only that they now show in
the main frame physical results for \MINLO{} and NNLOPS after shower,
and not the ones at LHE level.

Since NNLO (red, dashed) and NNLOPS (blue, solid) have almost
identical cross sections, there is virtually no difference in relative
terms between their jet-vetoed cross sections and the respective
efficiency. \MINLO{} (black, dotted), on the other hand, has a
different normalization in the {\tt fiducial-noJV} phase space of
roughly $-13\%$. Furthermore, the jet-veto efficiency predicted by
\MINLO{} is about $4\%$ below the NNLOPS one for typical jet-veto
cuts applied by the experiments
($20$\,GeV\,$\lesssim\ptjetveto\lesssim30$\,GeV).

The agreement between NNLO and NNLOPS results is remarkable. Even
down to $\ptjetveto=15$\,GeV their difference is within $\sim
2\%$.  Similar results were found in \citere{Dawson:2016ysj}
with resummation effects at high logarithmic accuracy of about $\sim 2$--$3\%$
beyond NNLO for $\ptjetveto=30$\,GeV.
This shows that jet-veto logarithms at typical jet-veto cuts
applied by the experiments are not particularly large and still well
described by a NNLO computation. Clearly, below $\ptjetveto=15$\,GeV
NNLO loses all predictive power and even turns negative at some point.
The scale-uncertainty band completely underestimates the true
uncertainty of the NNLO prediction due to missing higher-order corrections in
this region.
It is nice to see how matching to the parton shower cures the
unphysical behaviour of the NNLO result, so that NNLOPS yields accurate predictions in
the entire range of jet-veto cuts.  Furthermore, the scale uncertainty
band of the NNLOPS curve widens at small $\ptjetveto$, reflecting 
the
fact that higher-order logarithmic terms become important in this
region and degrade the accuracy of the perturbative prediction.

\subsection{Differential distributions in the fiducial phase space}\label{sec:fiducial}

We now turn to discussing differential cross sections.  The figures in
this section have the same layout as before. Additionally, we show the
central NNLOPS result at LHE level, i.e.\ before the shower is applied,
in the ratio frame.
We start by considering observables which are sensitive to soft-gluon
emissions. In phase-space regions where the cross section becomes
sensitive to soft-gluon effects, large logarithmic terms spoil the
validity of fixed-order computations and must be resummed to all
orders to yield a physical description.  This can be done either via
analytic resummation techniques or via a parton-shower approach.
Therefore, the largest and most relevant effects of combining NNLO
predictions with parton showers are expected in regions where
observables are sensitive to soft-gluon radiation.

\begin{figure}[tp]
\begin{center}
\begin{tabular}{cc}
\hspace*{-0.17cm}
\includegraphics[trim = 7mm -7mm 0mm 0mm, width=.33\textheight]{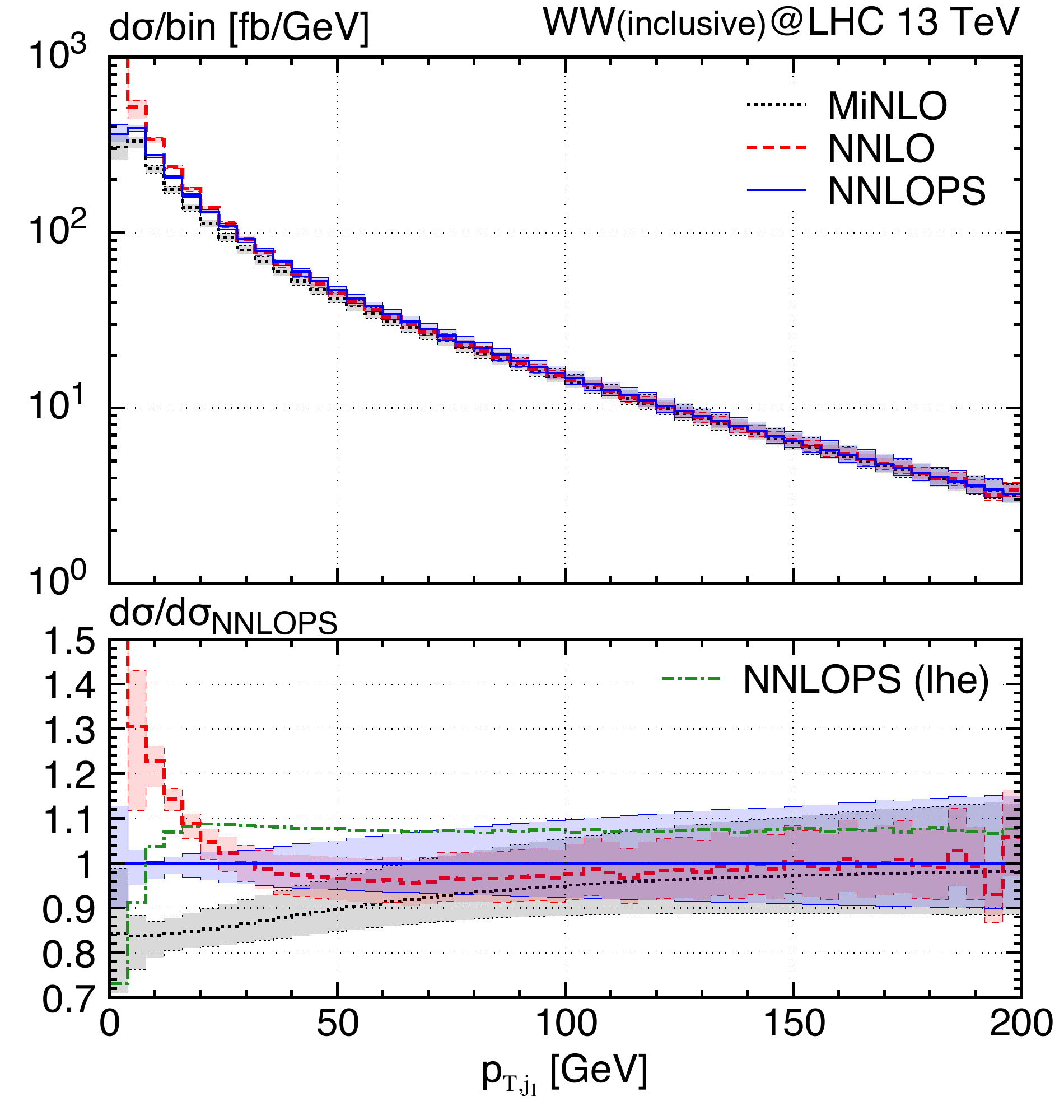} &
\includegraphics[trim = 7mm -7mm 0mm 0mm, width=.33\textheight]{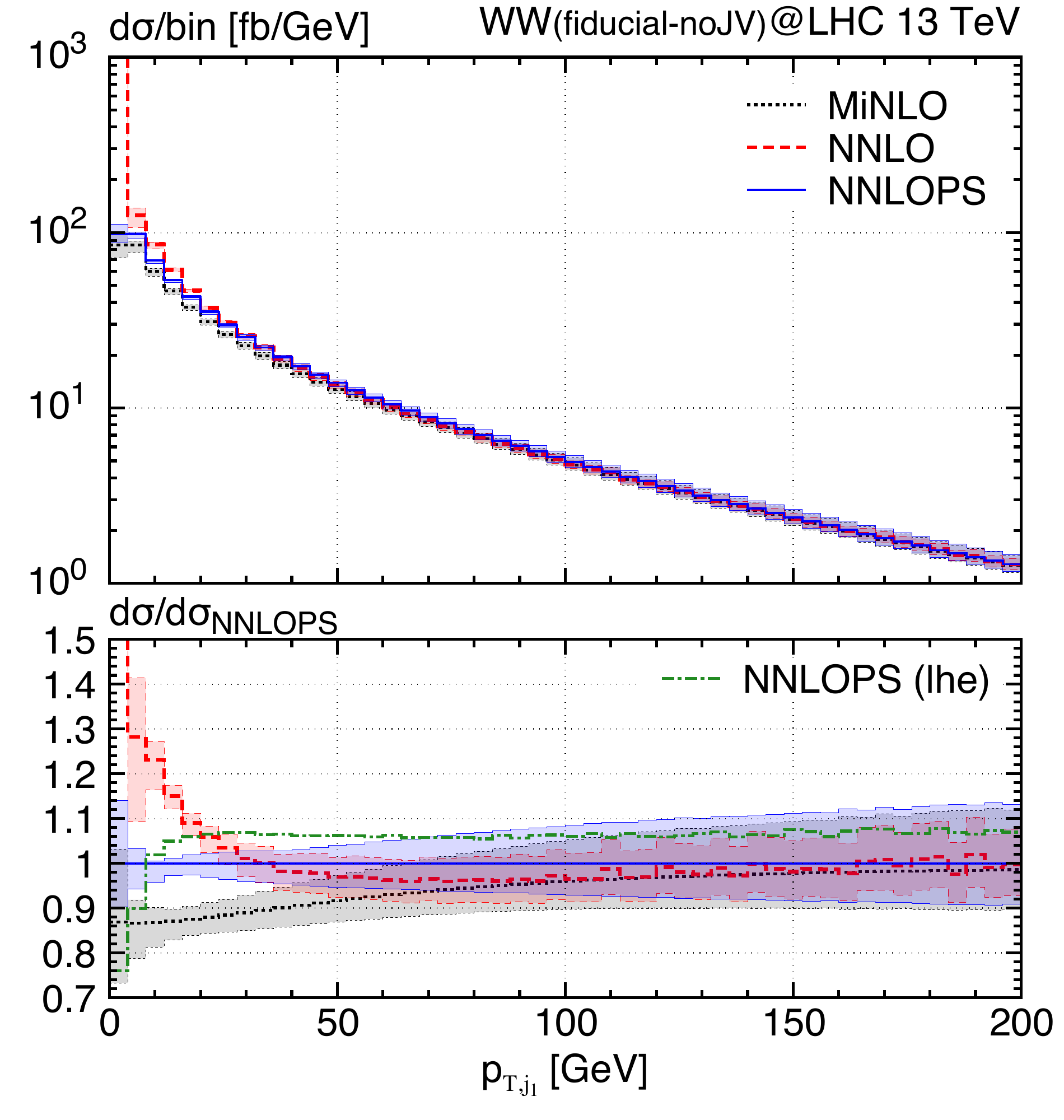} \\[-1em]
\hspace{0.6em} (a) & \hspace{1em}(b) \\[0.3cm]
\end{tabular}
\caption[]{\label{fig:leadingjet}{Distribution of the cross section in
    transverse momentum of the leading jet as predicted by \MINLO{}
    (black, dotted), NNLO (red, dashed) and NNLOPS (blue, solid); for
    reference, central results of NNLOPS at LHE level (green,
    dash-dotted) are also shown in the ratio frame; (a) inclusive phase space
    and (b) {\tt fiducial-noJV} setup.}}
\end{center}
\end{figure}

\fig{fig:leadingjet} depicts the transverse-momentum distribution of
the hardest jet in the inclusive and the {\tt fiducial-noJV} phase
space. The relative behaviour of the results in the two scenarios is
very similar. This remains true also for other distributions which is
why we refrain from showing any other inclusive result and focus
instead only on distributions in the fiducial phase below. As
expected, the NNLO curve diverges as small jet \pt{} due to large
logarithmic terms. On the contrary, both NNLOPS and \MINLO{} remain
finite in the small-\pt{} region. However, the importance of including
NNLO accuracy on top of \MINLO{} is obvious from the $\sim 15\%$
differences to NNLOPS in that region, which also produce a shape
distortion. At high transverse momenta all three predictions have the
same (NLO) perturbative accuracy and are consistent within scale
uncertainties.
The uncertainty band of the NNLOPS result is smallest at around
$\ptjetone = 15$\,GeV. This narrowing of the band is due to the fact
that at high \pt{} the NNLOPS prediction is formally only NLO accurate 
and at very small \pt{} the uncertainty increases due to missing
large logarithmic higher-order terms. As a consequence, the
uncertainty will be smallest in the intermediate-\pt{} region.

\begin{figure}[tp]
\begin{center}
\begin{tabular}{cc}
\hspace*{-0.17cm}
\includegraphics[trim = 7mm -7mm 0mm 0mm, width=.33\textheight]{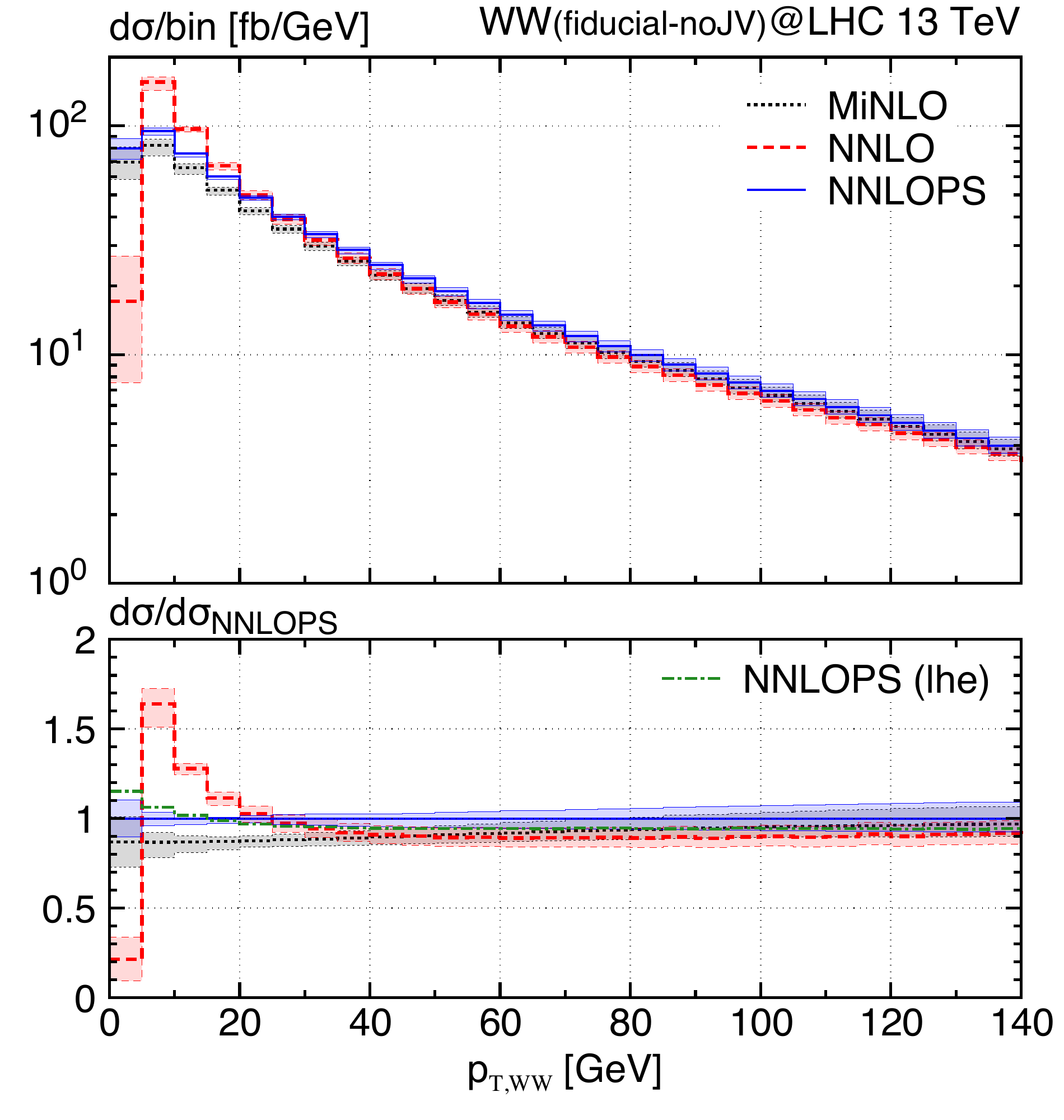} &
\includegraphics[trim = 7mm -7mm 0mm 0mm, width=.33\textheight]{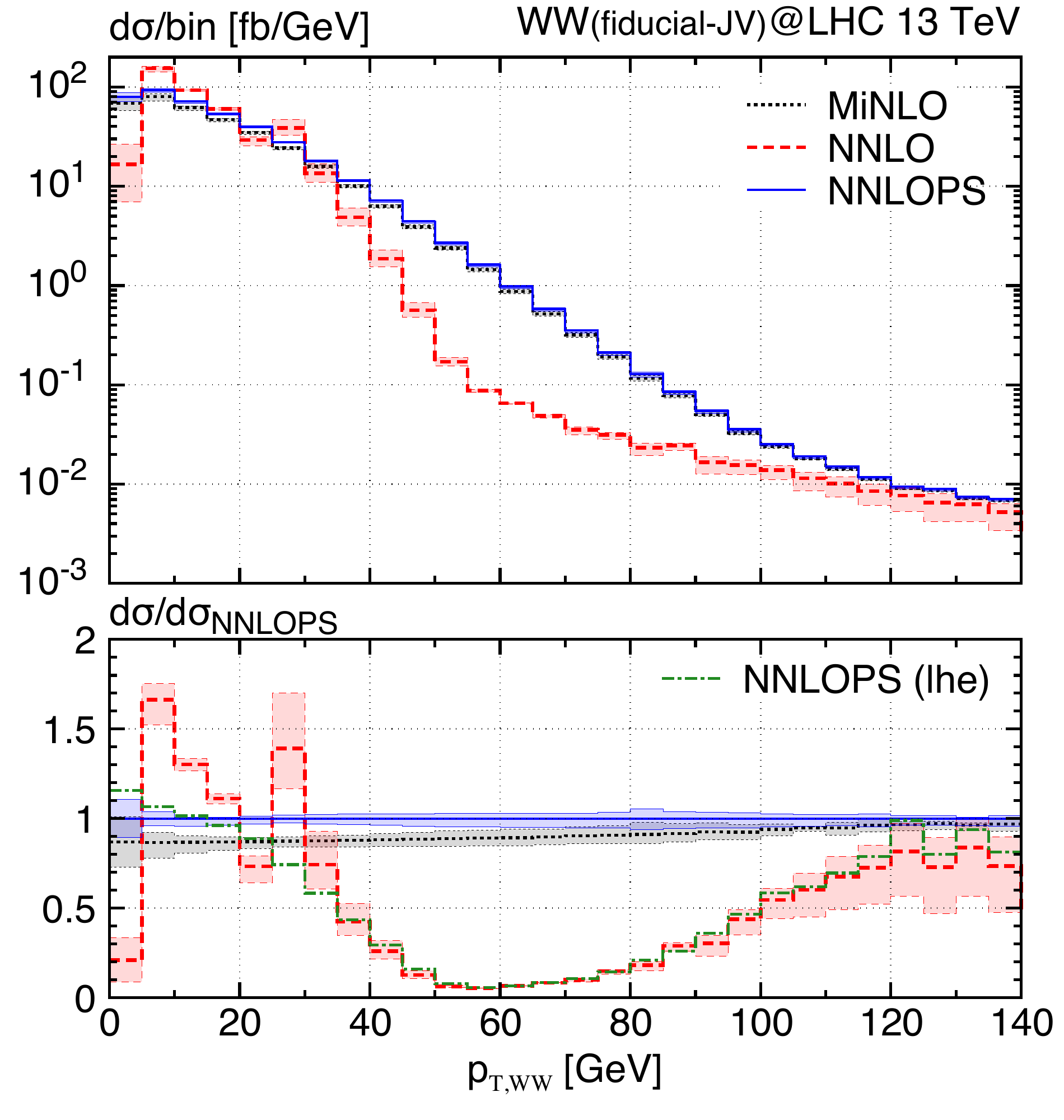} \\[-1em]
\hspace{0.6em} (a) & \hspace{1em}(b)
\end{tabular}
\caption[]{\label{fig:pTWW}{Same as \fig{fig:leadingjet}, but for the distribution in the transverse momentum of the colourless final-state system; (a) {\tt fiducial-noJV} and (b) {\tt fiducial-JV} phase space.}}
\end{center}
\end{figure}

We continue by showing in \fig{fig:pTWW} the transverse momentum of
the colourless final-state (diboson) system (\ptww{}).\footnote{Note that 
we performed a qualitative comparison of the inclusive \ptww{} distribution
with the analytically resummed results of \citeres{Grazzini:2015wpa}, and we
found remarkable agreement in terms of shape between NNLOPS and NNLL+NNLO.}
The {\tt fiducial-noJV} setup in the left panel of that figure reveals no
surprises: the NNLO curve diverges at small transverse momenta, which
is cured in the NNLOPS prediction by the parton shower. The general
behaviour is very similar to the \pt{} distribution of the leading
jet. The \ptww{} distribution in the {\tt fiducial-JV} setup (right
panel of \fig{fig:pTWW}), on the other hand, shows some quite
prominent effects: in the intermediate \pt{} region
($40$\,GeV\,$\lesssim\ptww\lesssim100$\,GeV) NNLO and NNLOPS results
can differ by more than one order of magnitude, while at low \pt{}
NNLO shows the typical unphysical behaviour, and at high \pt{} the two
predictions become similar again.  It is interesting to notice that
the NNLOPS result before showering (see the green, dash-dotted result
at LHE level in the ratio) follows closely the NNLO curve in the
intermediate \pt{} range. Hence, this large gap is filled up in the
NNLOPS prediction entirely by soft radiation due to the parton shower.
This can be understood as follows: beyond the region where jet-veto
requirements are applied ($25$\,GeV and $30$\,GeV respectively) the
NNLO \ptww{} distribution drops significantly as a substantial
fraction of events with a hard jet recoiling against the \ww{} system
is removed. In fact, disregarding high-rapidity jets which escape the
jet veto requirements, the NLO distribution has a boundary at the jet
veto cut. Thus, the NNLO result is effectively only LO accurate at
larger \ptww{} values, and only configurations with two jets of
transverse momentum less than $25$ or $30$\,GeV can contribute to this
region.  Eventually, the shower generates additional configurations
where three and more jets recoil against the \ww{} system, so that the
intermediate \pt{} region gets filled up and a smooth and physical
distribution is obtained.
Since in the intermediate transverse momentum region
($40$\,GeV\,$\lesssim\ptww\lesssim100$\,GeV) the distribution is
mainly built up by the colour singlet recoiling against soft jets from
the parton shower, in this region the distribution is particularly
sensitive to the modeling of soft radiation in the parton shower.
Accordingly, this distribution seems particularly suited to tune the
parton shower inputs of the NNLOPS generator, both the perturbative
components as well as the handling of non-perturbative hadronization
effects.
Compared to other measurements that enter tunes of parton shower it is
interesting to note that the \ww{} system that is measured is in fact
relatively hard.
We finally note that the step in the NNLO curve around
$\ptww{}=25$\,GeV is a perturbative instability from an integrable
logarithmic singularity \cite{Catani:1997xc} caused by the boundary in
the NLO $\ptww{}$ distribution due to the jet-veto cut.

\begin{figure}
\begin{center}
\begin{tabular}{cc}
\hspace*{-0.17cm}
\includegraphics[trim = 7mm -7mm 0mm 0mm, width=.33\textheight]{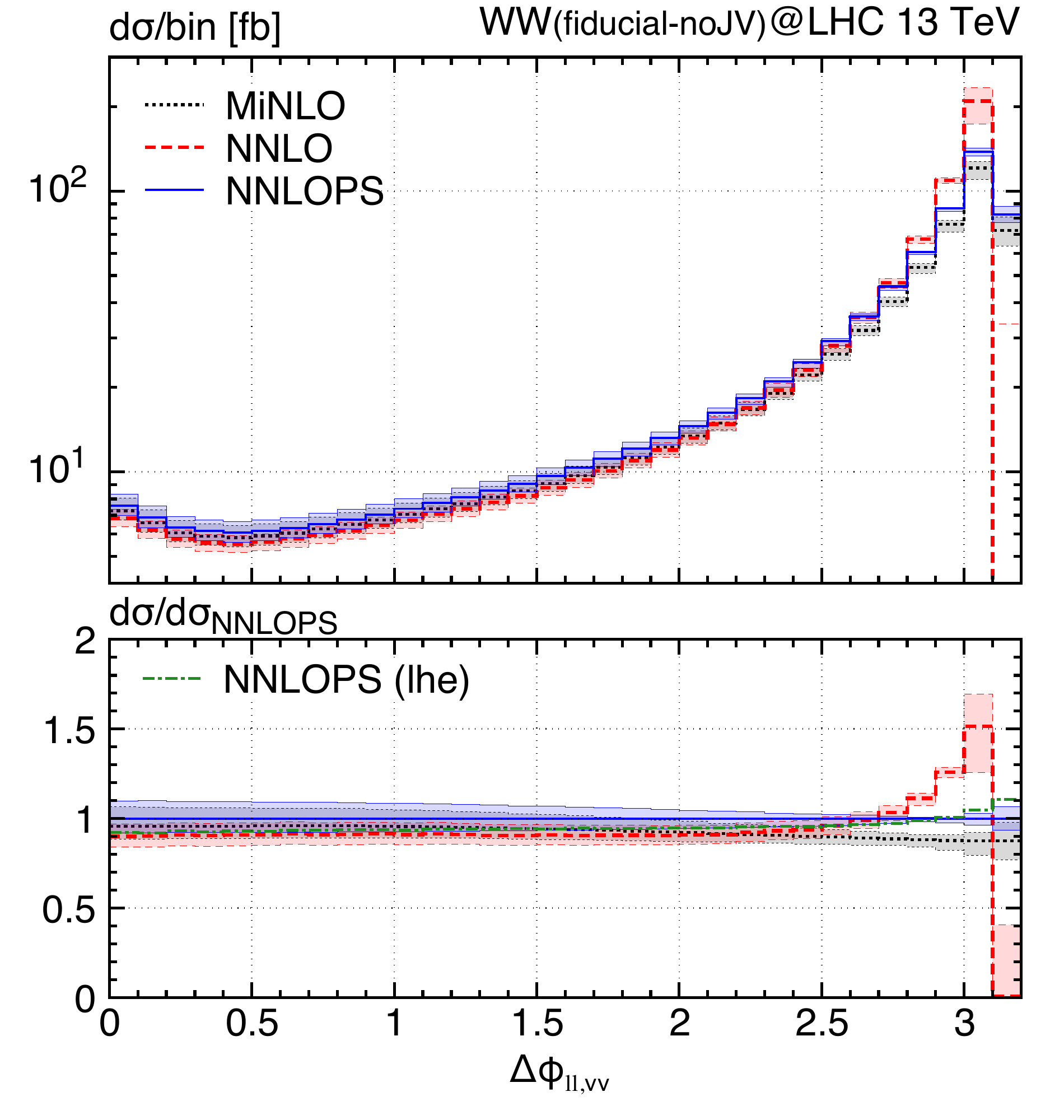} &
\includegraphics[trim = 7mm -7mm 0mm 0mm, width=.33\textheight]{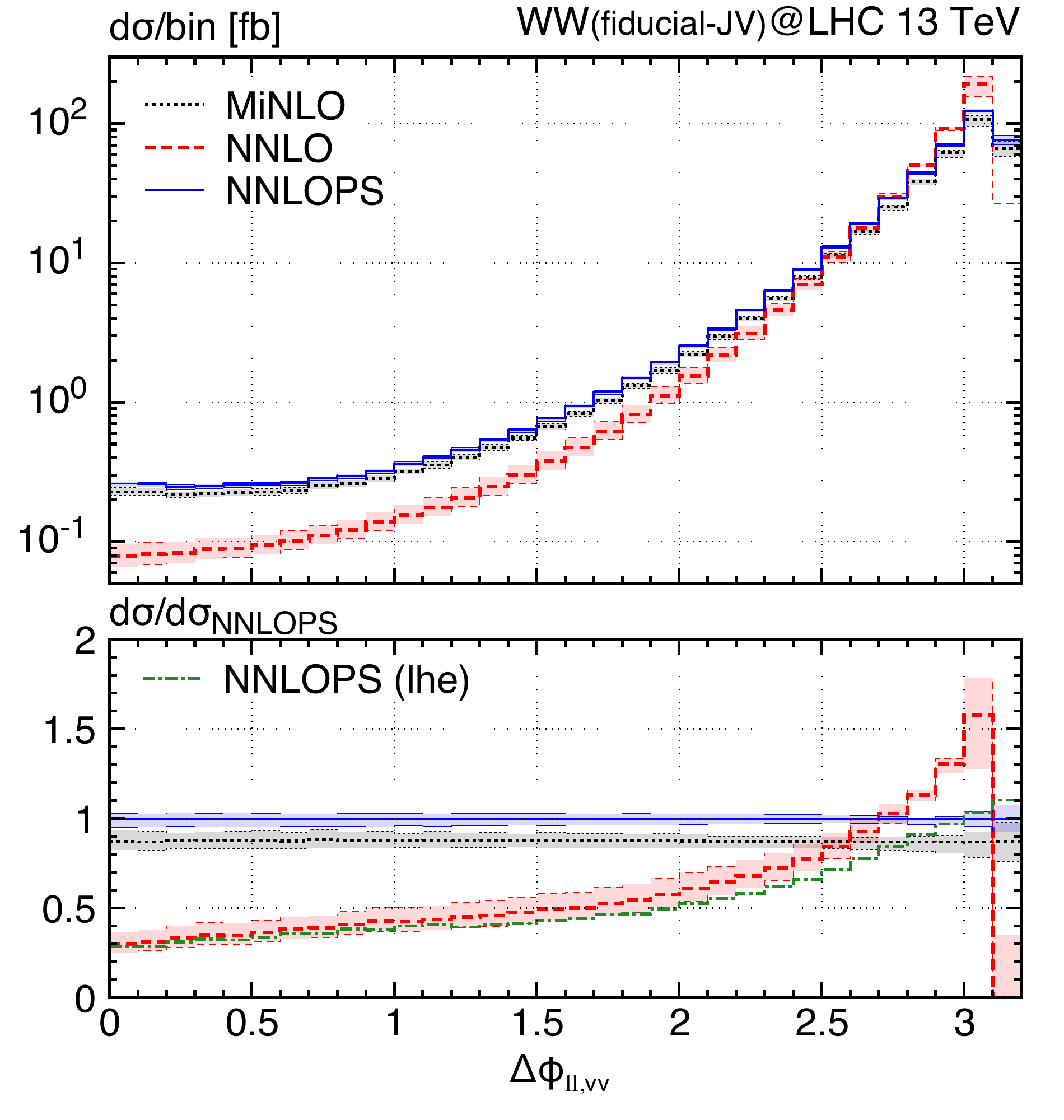} \\[-1em]
\hspace{0.6em} (a) & \hspace{1em}(b)
\end{tabular}
\caption[]{\label{fig:dphillnunu}{Same as \fig{fig:leadingjet}, but
    for the distribution in the azimuthal angle between the dilepton
    system and the two neutrinos; (a) {\tt fiducial-noJV} and (b) {\tt
      fiducial-JV} phase space.}}
\end{center}
\end{figure}

Also the distribution in the azimuthal angle between the dilepton
system and the missing transverse-momentum vector (\dphillnunu{}) is
sensitive to soft-gluon effects: since the two vectors are
back-to-back at LO, values different from $\pi$ are filled only upon
inclusion of higher-order corrections.  The \dphillnunu{} distribution
is shown in \fig{fig:dphillnunu} and develops the expected
Sudakov-like behaviour at large separation angles. While it is clear
that in this region only the showered results provide a proper
prediction, at small separation angles \MINLO{}, NNLO and NNLOPS in
the {\tt fiducial-noJV} setup (left panel of that figure) all have the
same formal fixed-order accuracy and agree within their respective uncertainty
bands. Looking at small \dphillnunu{} angles in the setup with a
jet-veto ({\tt fiducial-JV}) in the right panel of that figure, on the
other hand, we observe a very strong suppression of the NNLO cross
section with respect to the NNLOPS one. As in the \ptww{} case the
green, dash-dotted LHE result in the ratio is very close to NNLO in
that region.  The explanation follows the same logic as for \ptww{}
above: the jet-veto suppresses small \dphillnunu{} separations as they
require the \ww{} system to recoil against hard jet radiation.  The
shower reshuffles events such that more of such configurations are
generated and increases the cross section at small
\dphillnunu{}. Being dominated by corrections from the shower, also
this observable may serve as a way to tune parton showers and as a
probe of non-perturbative models in the parton shower Monte Carlo.

\begin{figure}[tp]
\begin{center}
\hspace*{-0.15cm}
\begin{tabular}{ccc}
\includegraphics[trim = 7mm -7mm 0mm 0mm, width=.226\textheight]{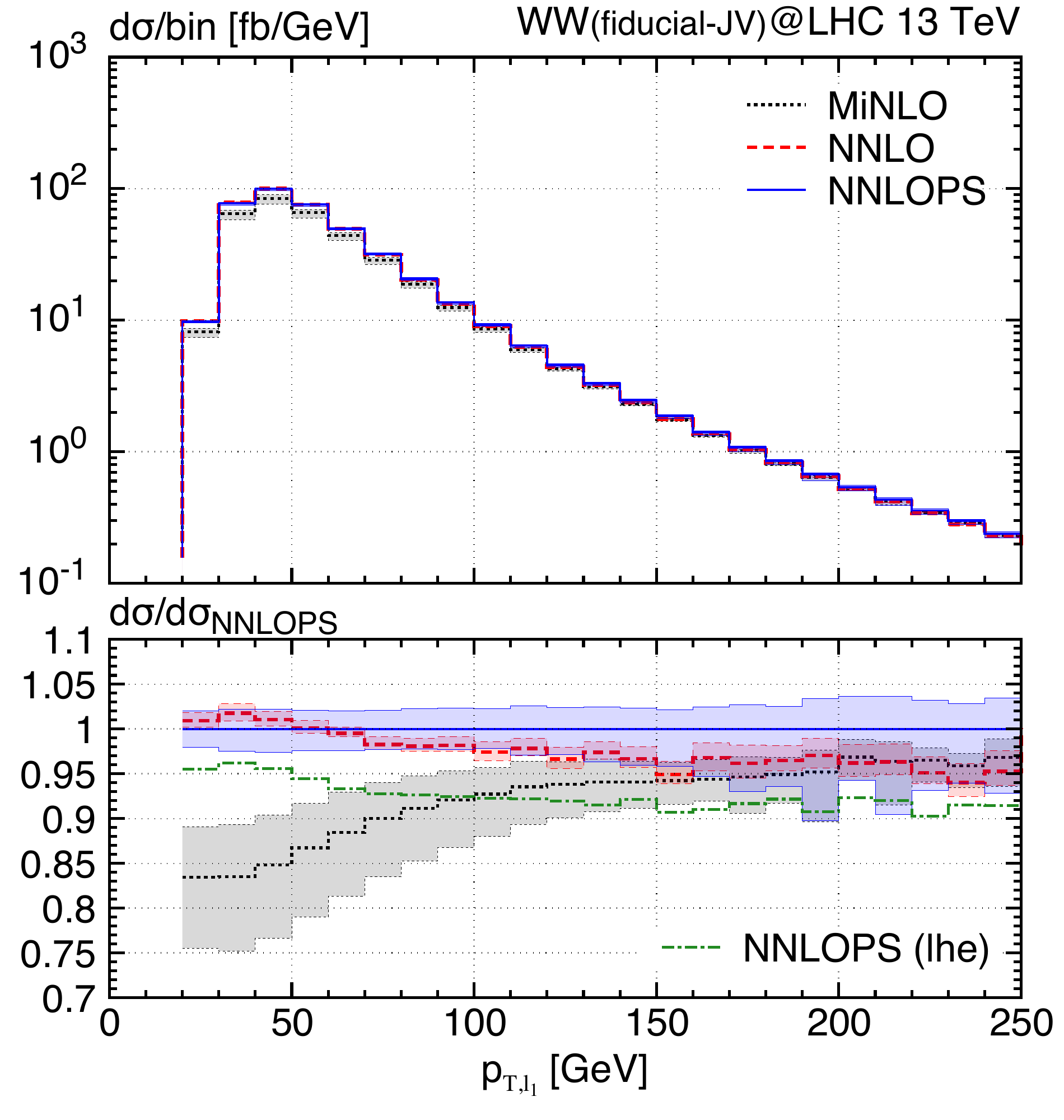} &\hspace{-0.6cm}
\includegraphics[trim = 7mm -7mm 0mm 0mm, width=.226\textheight]{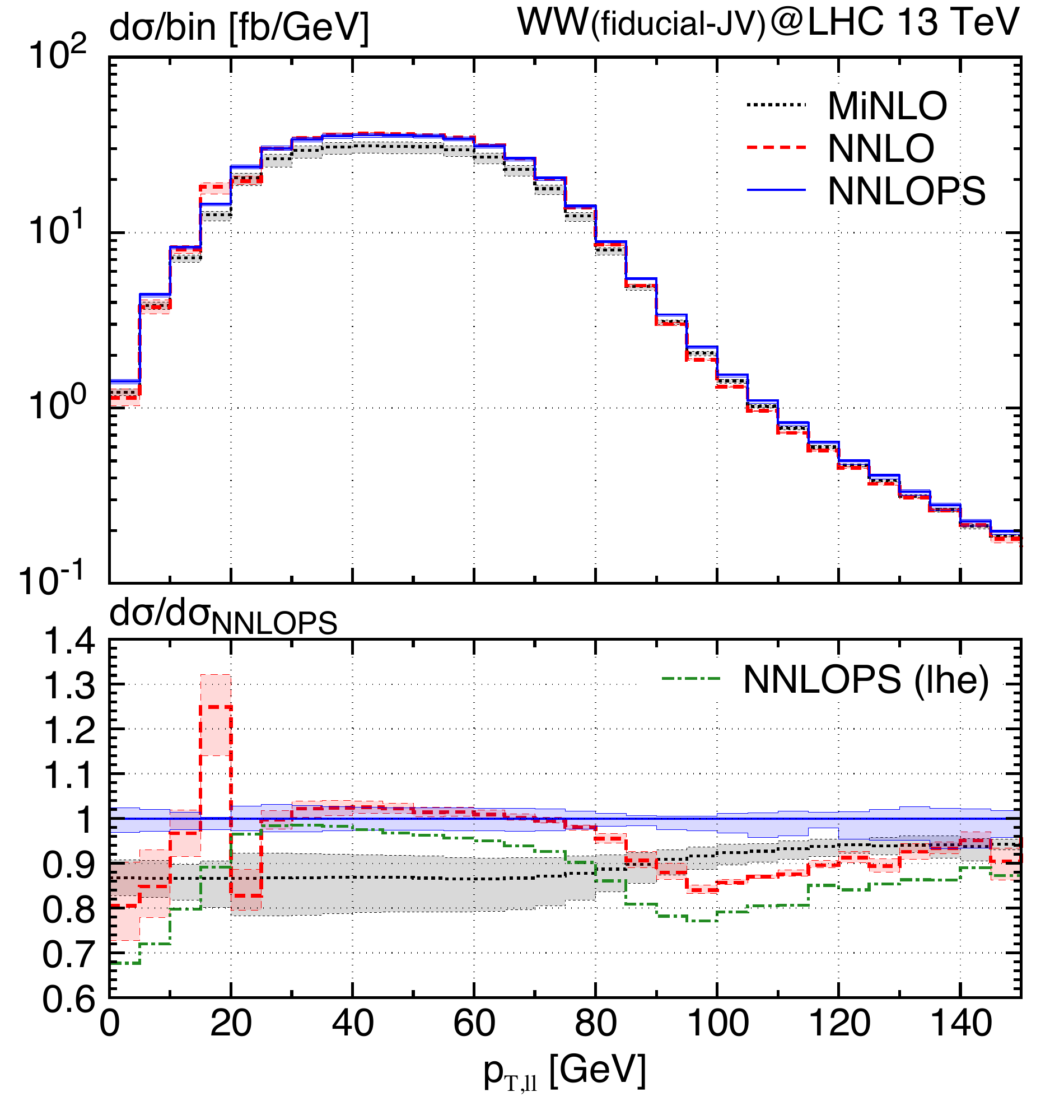} &\hspace{-0.6cm}  \includegraphics[trim = 7mm -7mm 0mm 0mm, width=.226\textheight]{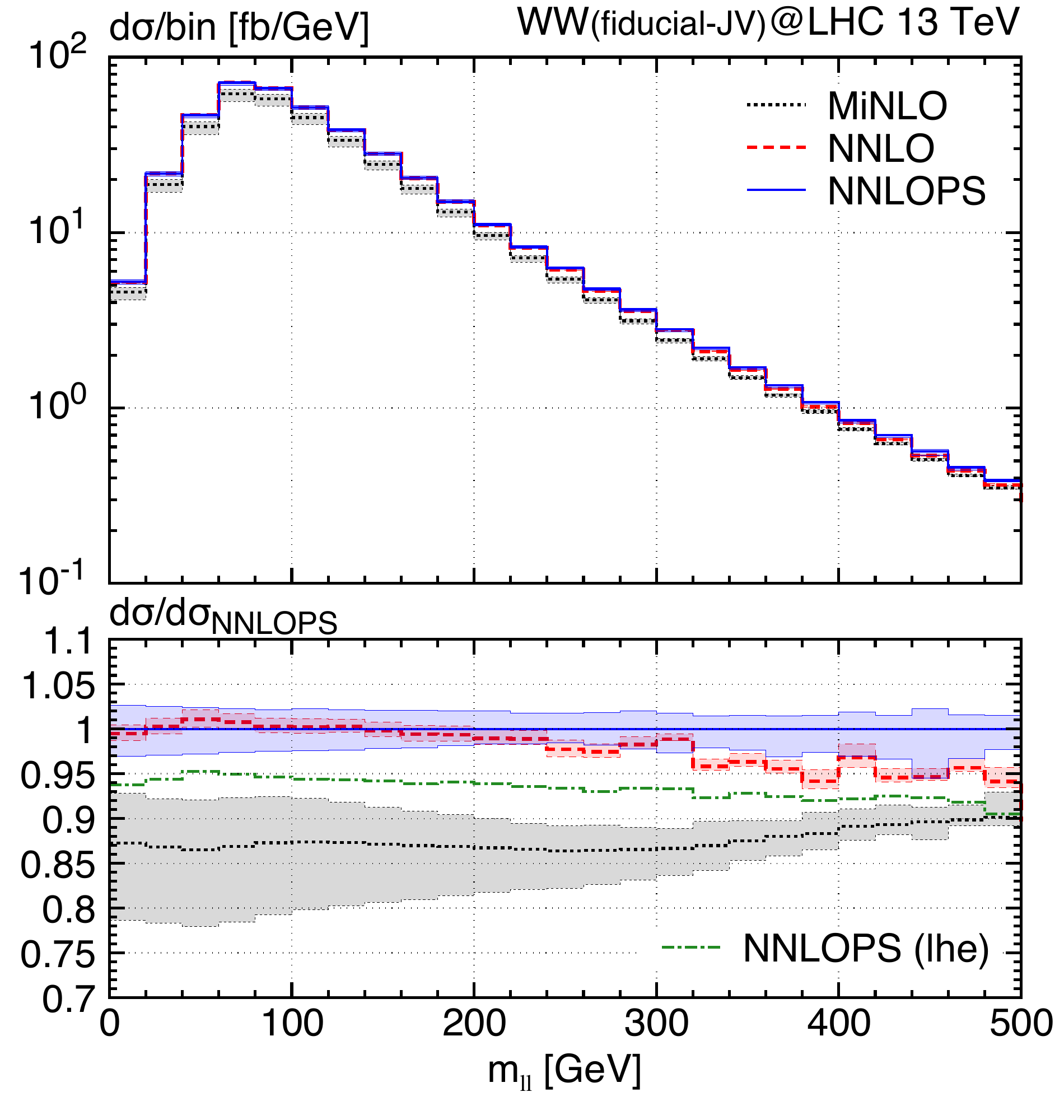} \\[-1em]
 (a) & \hspace{-1em}(b) & \hspace{-1em}(c)\\[0.3cm]
 \includegraphics[trim = 7mm -7mm 0mm 0mm, width=.226\textheight]{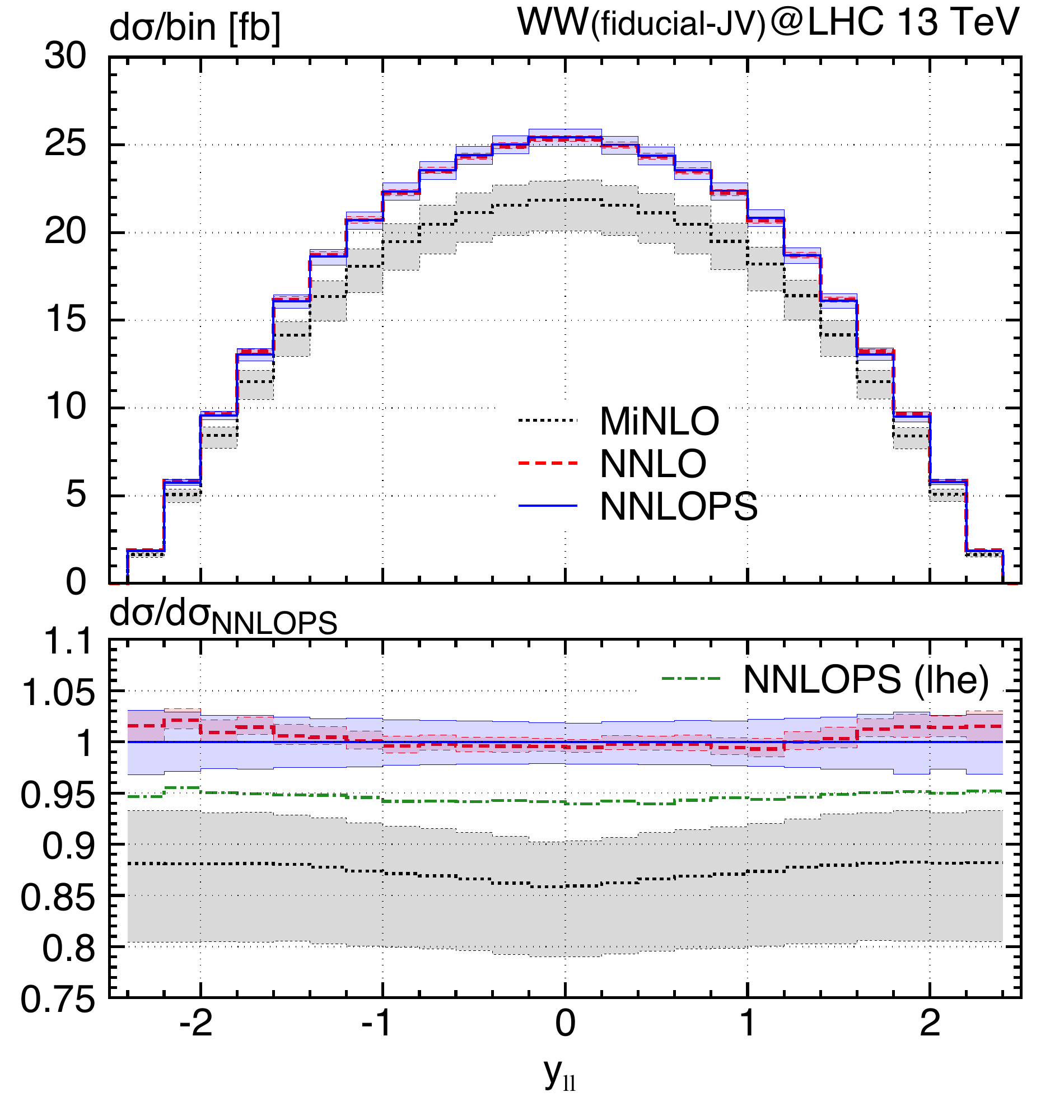} &\hspace{-0.6cm}
\includegraphics[trim = 7mm -7mm 0mm 0mm, width=.226\textheight]{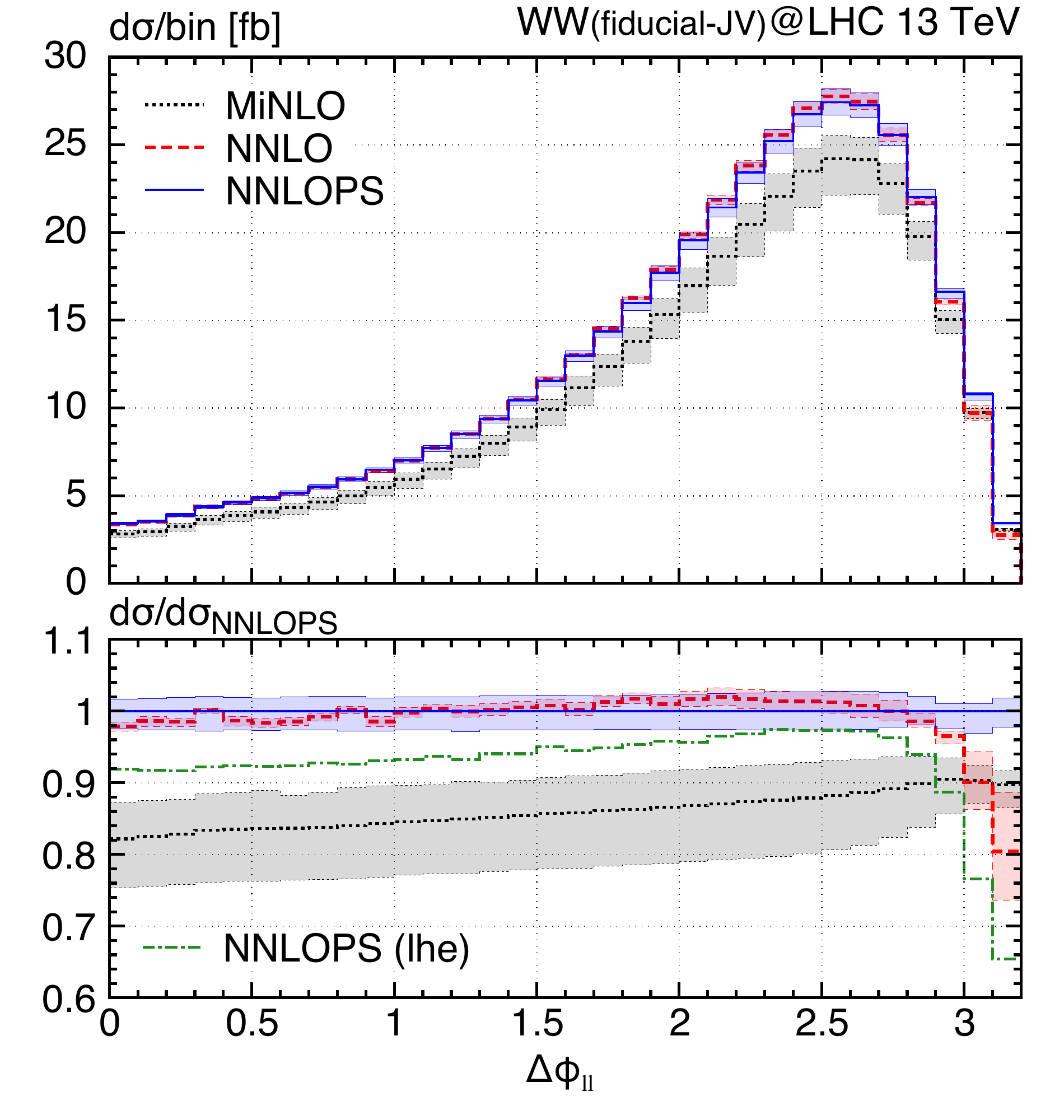} &\hspace{-0.6cm}  \includegraphics[trim = 7mm -7mm 0mm 0mm, width=.226\textheight]{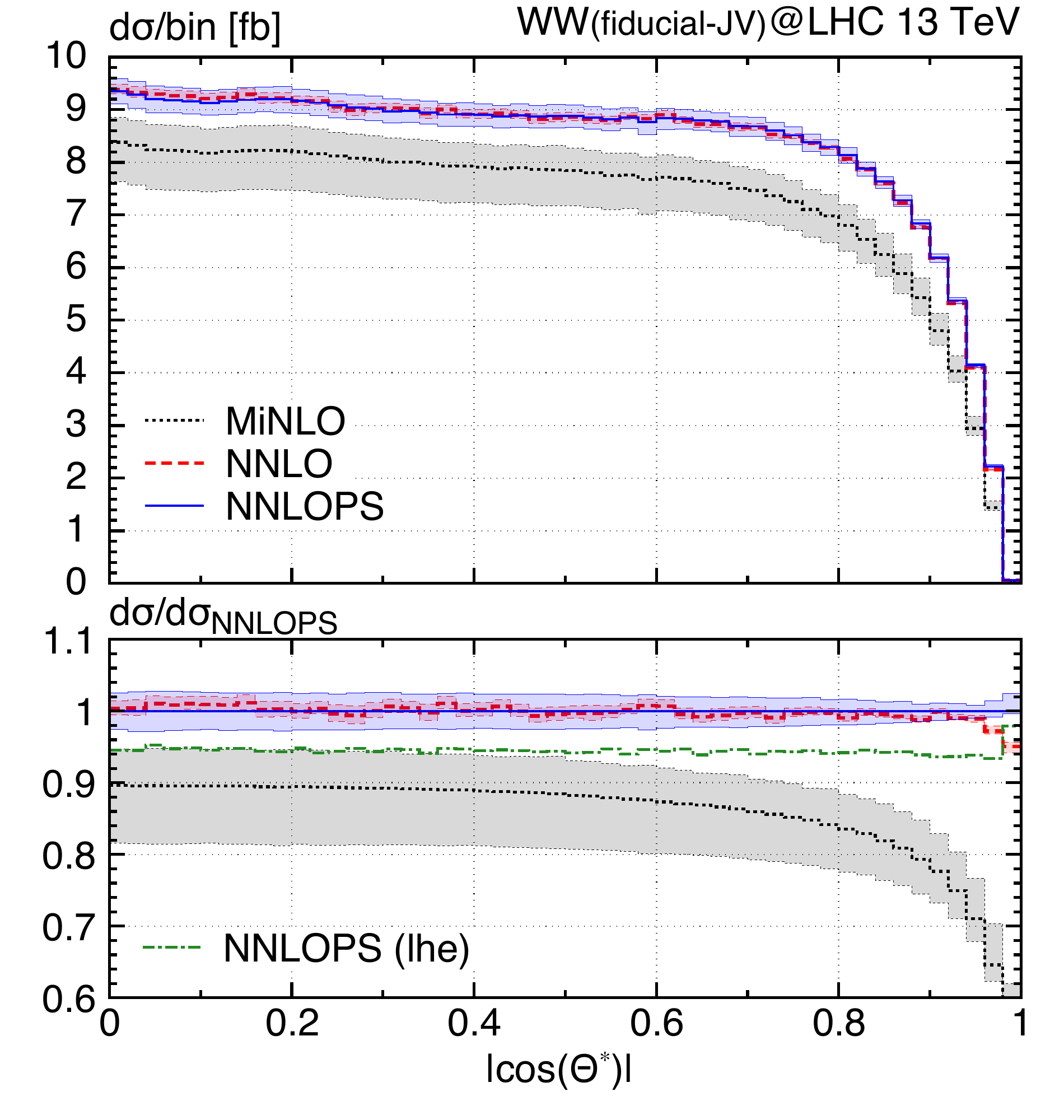} \\[-1em]
 (d) & \hspace{-1em}(e) & \hspace{-1em}(f)
\end{tabular}
\caption[]{\label{fig:allfid}{Same as \fig{fig:leadingjet}, but for
    various distributions in the fiducial phase space measured in the
    8 TeV analysis by ATLAS \cite{Aad:2016wpd}: (a) transverse
    momentum of the leading lepton $\ptlepone$ (b) transverse momentum
    \ptll{}, (c) invariant mass \mll{} and (d) rapidity of the
    dilepton pair, (d) azimuthal lepton separation $\dphill$, and (e)
    $\left|\cos(\theta^\star)\right|$ defined in
    \eqn{eq:costhetastart}.}}
\end{center}
\end{figure}

To conclude our analysis of differential observables in the fiducial
phase, we consider a set of distributions in \fig{fig:allfid} which
have been unfolded in the 8 TeV measurement done by ATLAS in
\citere{Aad:2016wpd}. They involve the leading lepton \pt{}, the
transverse momentum, invariant mass and rapidity of the dilepton
system, the separation in the azimuthal angle of the two leptons as
well as an observable sensitive to new physics searches which is
defined through the separation in $\eta{}$ of the two leptons:
\begin{align}\label{eq:costhetastart}
\left|\cos(\theta^\star)\right|=\left| \tanh\left(\Delta\eta_{\ell\ell}/2\right) \right|\,.
\end{align}
It is nice to see that, on the one hand, the inclusion of NNLO
corrections on top of the \MINLO{} generator is crucial not only for
the correct normalization, but in many cases also to capture relevant
shape effects. On the other hand, the impact of the parton shower on
top of the NNLO predictions is moderate in many phase space regions,
but absolutely vital in cases where the perturbative prediction fails
due to soft radiation effects, as we have already seen in
\figs{fig:leadingjet}$-$\ref{fig:dphillnunu}. Moreover, even in some
of the distributions where the NNLO prediction is not challenged by
large logarithms, the shower induces shape effects at the $5\%$ level,
see \fig{fig:allfid}\,(a) and (c) for example.

The two distribution which require some additional discussion in
\fig{fig:allfid} are \ptll{} and \dphill{}. We note at this point that
in the fiducial phase space the LHE-level NNLOPS result before shower,
which is shown only in the ratio frame, has a different normalization
(by about $-5\%$) than after shower.  This is due to the jet-veto
requirements and does not appear in the inclusive nor the {\tt
  fiducial-noJV} phase space. It can be understood by realizing that
the LHE-level results are unphysical in regions sensitive to
soft-gluon radiation where large logarithmic contributions are
resummed by the shower. In other phase-space regions LHE-level results
coincide with the respective fixed-order result.  Since among the
fiducial cuts only the jet-veto requirements are subject to effects
from soft gluons, large differences between LHE-level and showered
results appear in the {\tt fiducial-JV} setup primarily.

The \ptll{} distribution in \fig{fig:allfid}\,(b) shows some
interesting features: at $20$\,GeV the NNLO curve develops some
perturbative instability.  The integrable logarithmic singularity
\cite{Catani:1997xc} is caused by the fiducial $\ptmiss{} > 20$\, GeV
cut, which at LO implies that the cross section below $\ptll = 20$\,
GeV vanishes.  The reduced formal accuracy of the NNLO calculation in
this region is also evident from the larger uncertainty band.  As
expected, this effect is absent in the NNLOPS result already before
the shower.  Both the NNLO and the LHE-level NNLOPS curve show a very
similar shape at larger \ptll{}. Around $\ptll{}=100$\,GeV a dip
appears in the ratio to the showered NNLOPS prediction. The reason for
this dip is the following: emissions from the parton shower can modify
\ptll{} because of recoil effects. Accordingly events can migrate to a
different bin. The largest impact of this migration will be right
after the point of inflection, which for \ptll{} is at around 100 GeV.

Also for the \dphill{} distribution in \fig{fig:allfid}\,(e) the
parton shower induces some prominent shape differences in the NNLOPS
result. The NNLO and NNLOPS result at LHE level are very similar
shape-wise: their curves relative to the NNLOPS one increase slightly
with \dphill{} up to $\dphill{}\sim 2.5$, after which they drop off
significantly towards configurations where the two leptons are
back-to-back. This behaviour is caused by the fiducial lepton cuts and
is absent in the fully inclusive case.
In particular the cut on $\ptmiss{} > 20$\, GeV suppresses the region
where the two leptons are back to back in the azimuthal
plane. Accordingly, the cross section drops sharply just before
$\dphill{} = \pi$. Because the cross section drops very fast, a small
change in \dphill{} due to the parton shower will have a large effect
in the ratio plot.
This effect is similar to the one observed around $\ptll{} = 90-100$\,
GeV, see \fig{fig:allfid}\,(b).
In particular the effect of the parton shower is to partially
re-populate this region since \ptmiss{} can recoil against extra soft
radiation from the parton shower, hence the region close to $\dphill=
\pi$ is less suppressed.

In summary, the importance of NNLOPS accurate predictions for \ww{}
production has been demonstrated in the fiducial phase space. Besides
IR-sensitive observables which require parton-shower resummation
already in the inclusive phase space, also other distributions in the
fiducial phase space exhibit sizeable corrections from the parton
shower, which cures perturbative instabilities caused by fiducial cuts
and provides a more reliable description of jet-veto logarithms
present in the fiducial cross section. The relevance of NNLO accuracy
beyond the \MINLO{} description is evident in essentially every
observable, irrespective of inclusive or fiducial, integrated or
differential.

\subsection{Charge asymmetry in \ww{} production}\label{sec:chargeasymmetry}

\begin{figure}[h]
\begin{center}
\begin{tabular}{cc}
\hspace*{-0.17cm}
\includegraphics[trim = 7mm -7mm 0mm 0mm, width=.33\textheight]{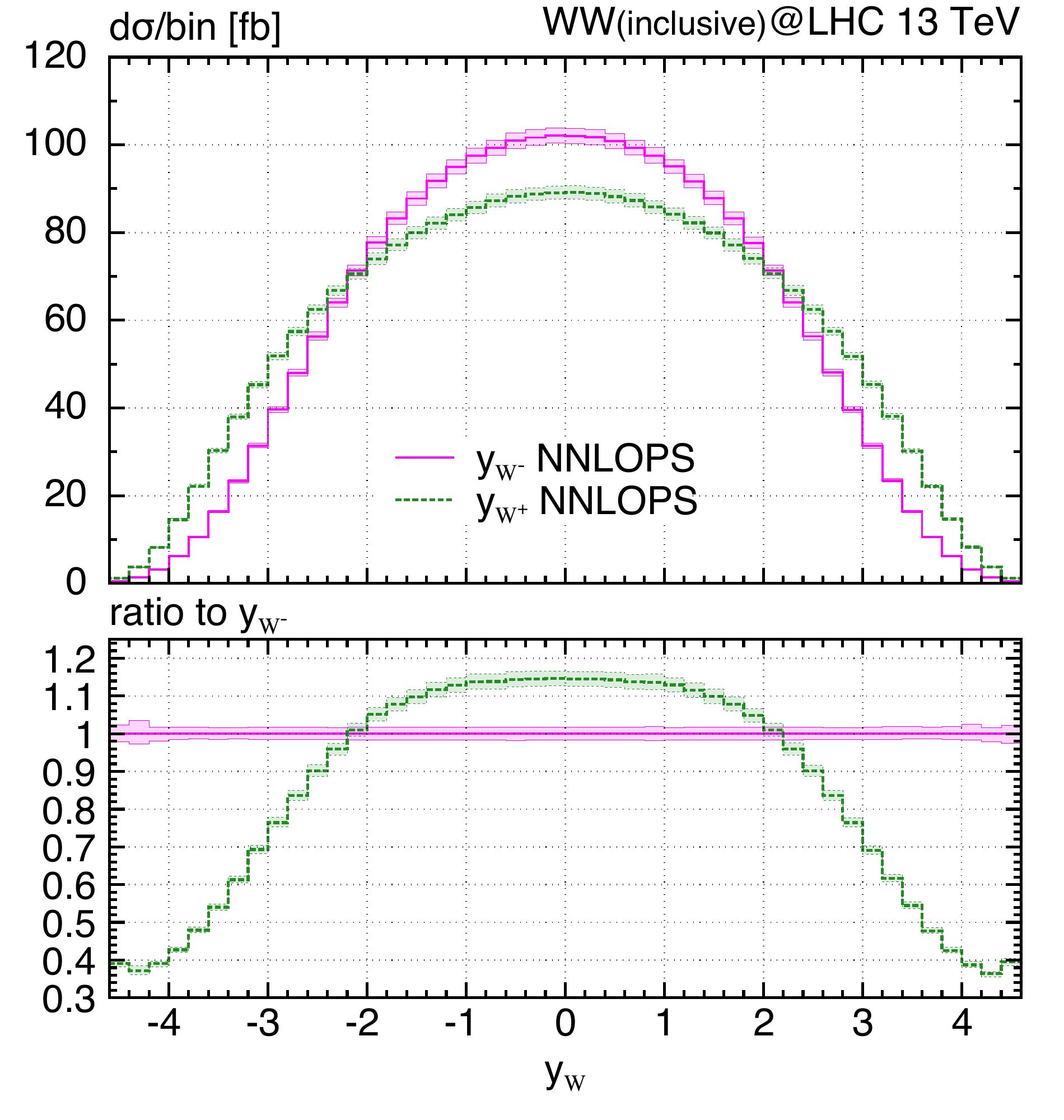} &
\includegraphics[trim = 7mm -7mm 0mm 0mm, width=.33\textheight]{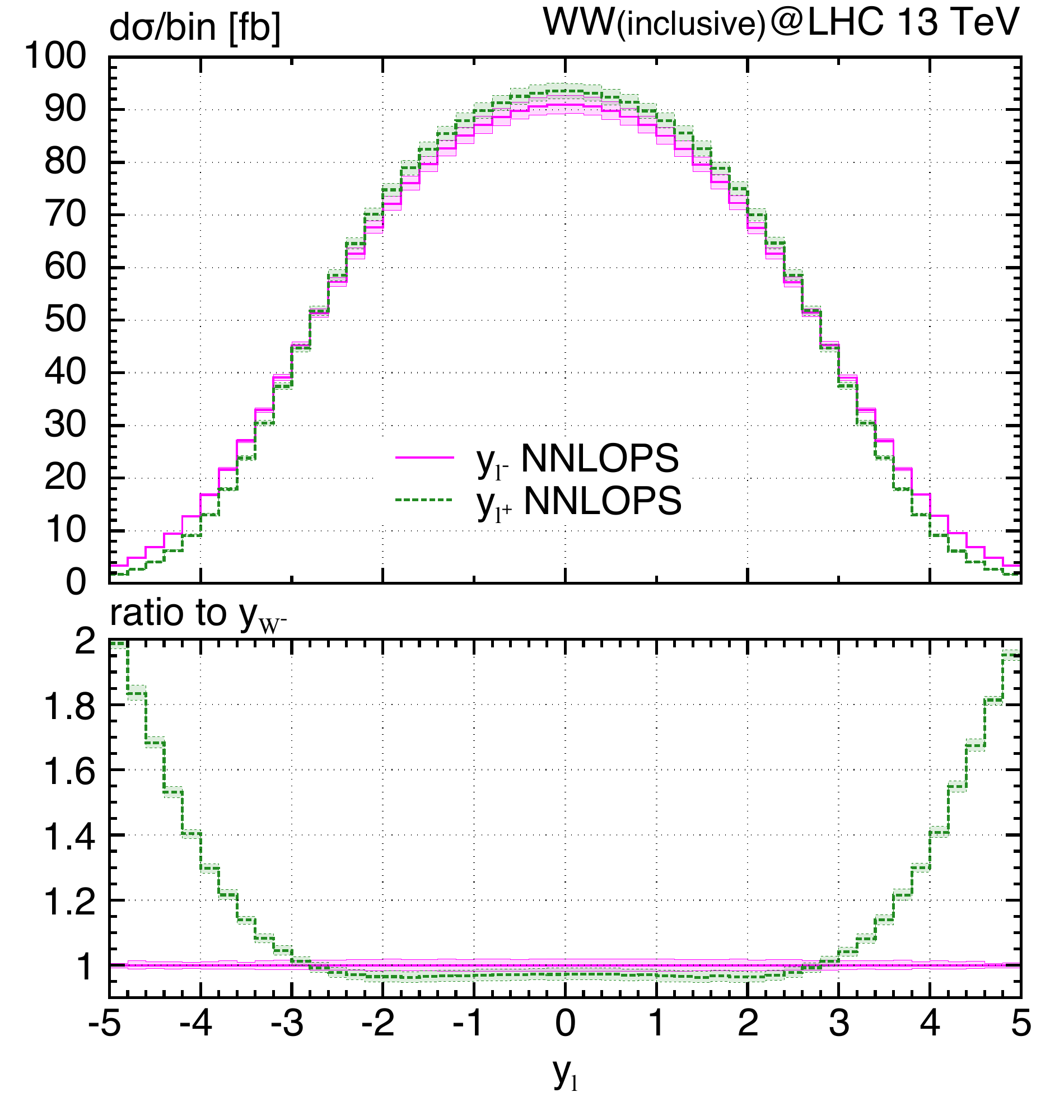} \\[-1em]
\hspace{0.6em} (a) & \hspace{1em}(b)
\end{tabular}
\caption[]{\label{fig:rap}{Comparison of rapidity distributions of negatively (magenta, solid) and positively (green, dotted) charged particles 
at NNLOPS for (a) the two $W$ bosons and (b) the two leptons.}}
\end{center}
\end{figure}

To complete our presentation of phenomenological results, we turn to
discussing the definition of a charge asymmetry in the \ww{}
production process at the LHC.  Similar to $t\bar{t}$ production, the
two $W$ bosons in \ww{} production exhibit an asymmetry.
This is caused by the fact that \ww{} is mainly produced through
$t$-channel $u\bar u$ or $d \bar d$ scattering ($s$-channel production
does not create an asymmetry). Since up quarks in the protons are
faster than down quarks and since the $W$-bosons tend to move in the
same direction as the incoming quark, i.e. $W^+$ ($W^-$) bosons tend
to follow the up (down) quarks, the $W^+$-bosons tend to be more
forward. 
This asymmetry manifests itself in the rapidity distributions of the
positively and negatively charged $W$ bosons as shown in
\fig{fig:rap}\,(a): $W^+$ bosons are generally more forward, while
$W^-$ bosons are situated more at central rapidity. 
However, since the $W$-boson
momenta of the \ww{} final state are not accessible in the measurement
due to the two neutrinos (not even under the assumption that they are
on-shell), one may wonder whether this asymmetry persists in the case
of the leptons.
Indeed, \fig{fig:rap}\,(b) shows a similar, but less
pronounced behaviour for the leptons. In fact, the asymmetry is
reversed with respect to the charges in this case with the $\ell^+$
being more central and the $\ell^-$ more forward.

We can now use the previous observation to define a charge asymmetry
in \ww{} production for the $W$ bosons:
\begin{align}
A_C^W=\frac{\sigma( |y_{W^+}|>|y_{W^-}|)-\sigma(|y_{W^+}|<|y_{W^-}|)}{\sigma(|y_{W^+}|>|y_{W^-}|)+\sigma(|y_{W^+}|<|y_{W^-}|)}, 
\end{align}
as well as for the leptons:\footnote{Note that for the leptons, since
  they are massless, the rapidity entering the asymmetry and the pseudo-rapidity used to define the fiducial cuts coincide ($y_\ell
  \equiv \eta_\ell$).}
\begin{align}
A_C^\ell=\frac{\sigma( |y_{\ell^+}|>|y_{\ell^-}|)-\sigma(|y_{\ell^+}|<|y_{\ell^-}|)}{\sigma(|y_{\ell^+}|>|y_{\ell^-}|)+\sigma(|y_{\ell^+}|<|y_{\ell^-}|)}. 
\end{align}
This allows us to express the size of the asymmetry by a single
number. It is zero if there is no asymmetry, positive if the
positively-charged particle is more forward, and negative
otherwise. Note that the denominator simply corresponds to the
integrated cross section, within the considered cuts.

\renewcommand\arraystretch{1.5}
\begin{table}[t]
\begin{center}
\begin{tabular}{c | c c}
\toprule
NNLOPS 
& inclusive phase space
& fiducial phase space\\
\midrule
$A_C^W$  & $\phantom{-}0.1263(1)_{-1.8\%}^{+2.1\%}$ & $\phantom{-}0.0726(3)_{-2.6\%}^{+2.0\%}$\\
$A_C^\ell$& $-[0.0270(1)_{-6.4\%}^{+5.0\%}]$ & $-[0.0009(4)_{-87\%}^{+72\%}]$\\
\bottomrule
\end{tabular}
\end{center}
\renewcommand{\baselinestretch}{1.0}
\caption{\label{tab:asymmetry} NNLOPS predictions for the charge
  asymmetry for $W$-bosons and charged leptons in \ww{}
  production. The fiducial volume is defined in \tab{tab:cuts} (including the jet-veto requirement).}
\end{table}
\renewcommand\arraystretch{1}

\tab{tab:asymmetry} summarizes the NNLOPS predictions for $A_C^W$ and
$A_C^\ell$ in the inclusive and in the fiducial
phase. The uncertainties are obtained by computing a $7$-point variation in the numerator 
and dividing by the central cross section in the denominator. This choice is motivated 
by the fact that fully correlated uncertainties in the ratio lead to too small uncertainties 
for $A_C^W$.
The $W$-boson asymmetry in the
inclusive phase space is pretty large and positive, as one could
expect from \fig{fig:rap}\,(a). It is significantly reduced by the
fiducial cuts, but still clearly different from zero. Also the leptons
yield a charge asymmetry at inclusive level, which, however, is
smaller than for $W$ bosons and negative.
Unfortunately, once lepton cuts are applied in the fiducial volume
$A_C^\ell$ becomes almost compatible with zero within both
perturbative and numerical uncertainties.
This again is due to the left-handed nature of the $W$-boson
interactions: in the case of the $W^+$ decay, the neutrino tends to
follow the $W$ direction while the positive charged lepton will mostly
end up in the central region of the detector. When the $W^-$ decays,
it is instead the negative charged lepton that tends to follow the $W$
boson in the forward region, while the neutrino ends up in the central
region of the detector compared to the $W$ boson rapidity. This effect
ends up fully compensating the $W$-boson charge asymmetry and even
causing the leptons to have an asymmetry that is reversed in
sign.
The relative importance of these two effects depends on kinematics of
the leptons and can then be altered by probing different kinematic
regions.
For instance, it is clear from the plots that by widening the rapidity
requirements on the leptons and measureing them further into the
forward region (beyond $|\etal| = 2.4$) a non-zero charge asymmetry
could be measured by the experiments.
It would be interesting to see whether such measurement
can be performed at LHCb, which already measured a lepton charge
asymmetry in Drell-Yan production~\cite{Aaij:2016qqz}.
Furthermore, we verified explicitly that the lepton asymmetry
increases when going to a boosted regime of the $W$ bosons, due to its
sensitivity to the $W$-boson polarizations. 
In this region, BSM effects that alter the $W$-polarization
composition could considerably modify the value of $A_C^\ell$, 
so that the lepton asymmetry can be used as a probe of new
physics. A more detailed analysis of such effects is, however, beyond
the scope of the present paper.

\section{Summary}
\label{sec:summary}

In this paper we presented NNLO-accurate parton-shower predictions for
the production of \ww{} pairs at hadron colliders. We include
off-shell effects and spin correlations by considering the full
leptonic process with two charged leptons and the two corresponding
neutrinos in the final state ($\ell\nu_\ell\ell'\nu_{\ell'}$). For the
first time NNLO QCD corrections have been consistently matched to
parton showers for a $2\to4$ process. Our calculation has been
extensively validated by the excellent agreement of the NNLOPS Les
Houches events with NNLO predictions for Born-level observables.

We have studied the impact of including NNLO corrections on top of the
\MINLO{} generator and of including the parton shower on top of NNLO
predictions on rates and distributions in both inclusive and fiducial
phase spaces. Integrated cross sections predicted by our NNLOPS
computation are virtually identical with the NNLO cross section and in
good agreement with cross-section measurements for \ww{} production by
ATLAS and CMS.  The relevance of the parton shower to resum jet-veto
logarithms beyond the ones present at NNLO is moderate: down to
jet-veto cuts of $15$\,GeV NNLO agrees with NNLOPS to better than
about $2\%$, but becomes unreliable below such values.

The importance of NNLOPS predictions compared to fixed order becomes
most apparent in differential distributions which are
sensitive to soft-gluon effects. In these phase-space regions the
validity of QCD perturbation theory breaks down due to the presence of
large logarithmic contributions, but matching to the parton shower
recovers physical predictions by the NNLOPS computation for all
observables.  Even observables which develop no logarithmic
divergences at inclusive level can feature perturbative instabilities
as soon as fiducial cuts are applied.  Hence, also in such cases
NNLOPS matching can induce substantial effects beyond NNLO as far as
distributions in the fiducial phase space are concerned.  Several
examples have been presented where fiducial cuts, primarily the
jet-veto requirements, but also lepton cuts, challenge fixed order
predictions and cause an improved description by NNLOPS.  Moreover, we
found NNLO-corrections to have a significant impact beyond the \MINLO{}
computation throughout: by and large NNLOPS and \MINLO{} show
differences at the $10\%$-level and higher.

We reckon that the NNLOPS calculation and the results presented in
this paper will be highly valuable for experimental measurements,
which feature \ww{} final states as signal or background. The
computation is publicly available within the \POWHEGBOX{} framework
and allows for fully-exclusive hadron-level event generation, which
can be combined by the experiments with their detector
simulation.

\section*{Acknowledgments}
We thank Fabrizio Caola, Massimiliano Grazzini, Uli Haisch, Keith
Hamilton, Stefan Kallweit, Pier Francesco Monni, Paolo Nason and
Andrea Wulzer for many stimulating and helpful discussions.
We are grateful to Kristin Lohwasser, Philip Sommer and Jochen Meyer
for discussion on the experimental setup and details of the
measurements.
We are also indebted to Massimiliano Grazzini, Keith Hamilton and Pier
Francesco Monni for comments on the manuscript.
This work was supported in part by ERC Consolidator Grant HICCUP
(No.\ 614577). The work of ER is supported by a Marie Skłodowska-Curie
Individual Fellowship of the European Commission's Horizon 2020
Programme under contract number 659147 PrecisionTools4LHC.

\bibliography{ww_nnlops}{}
\bibliographystyle{JHEP}

\end{document}